\documentclass[aps,pre,preprint,groupedaddress,floatfix]{revtex4}
\usepackage{epsfig}
\usepackage{amsmath}
\usepackage{amssymb}
\usepackage{subfigure}
\usepackage{rotate}
\usepackage[american]{babel}
\usepackage{graphics}
\usepackage{afterpage}

\newcommand{\beq}{\begin{eqnarray*}}
\newcommand{\eeq}{\end{eqnarray*}}
\newcommand{\beqn}[1]{\begin{eqnarray}\label{#1}}
\newcommand{\eeqn}{\end{eqnarray}}

\newcommand{\nz}{{\if mm {\rm I}\mkern -3mu{\rm N}\else \leavevmode
\hbox{I}\kern -.17em \hbox{N} \fi}}
\newcommand{\calP}{{\cal P}}
\newcommand{\calQ}{{\cal Q}}
\newcommand{\calR}{{\cal R}}
\newcommand{\calS}{{\cal S}}

\newcommand{\id}{\hspace*{0.2cm} {\if mm {\rm I}\mkern - 11.8mu{\rm 1}\else \leavevmode
\hbox{I}\kern -.17em \hbox{1} \fi}\hspace*{1mm}_\nu \hspace*{0.1cm} }

\newcommand{\vphi}{\mbox{\boldmath{$\phi$}}}
\newcommand{\vc}{\mbox{\boldmath{$c$}}}

\begin{document}


\title{Theoretical and numerical study of lamellar eutectic three-phase growth in ternary alloys}


\author{Abhik Choudhury}
\email[1]{abhik.choudhury@hs-karlsruhe.de}
\affiliation{Karlsruhe Institute of Technology (KIT)\\ Haid-und-Neu-Str.7,
76131 Karlsruhe, Germany}

\author{Mathis Plapp}
\email[1]{mathis.plapp@polytechnique.fr}
\affiliation{Physique de la Mati\`{e}re Condens\'ee,\\
Ecole Polytechnique, CNRS, 91128 Palaiseau, France}

\author{Britta Nestler}
\email[1]{britta.nestler@kit.edu}
\affiliation{Karlsruhe Institute of Technology (KIT), Institute of Applied Materials (IAM)\\Haid-und-Neu-Str.7,
76131 Karlsruhe, Germany}


\date{\today}

\begin{abstract}
We investigate lamellar three-phase patterns that form during the directional solidification of ternary eutectic alloys in thin samples. A distinctive feature of this system is that many different geometric arrangements of the three phases are possible, contrary to the widely studied two-phase patterns in binary eutectics. Here, we first analyze the case of stable lamellar coupled growth of a symmetric model {\em ternary} eutectic alloy, using a Jackson-Hunt type calculation in thin film morphology, for arbitrary configurations, and derive expressions for the front undercooling as a function of velocity and spacing. Next, we carry out phase-field simulations to test our analytic predictions and to study the instabilities of the simplest periodic lamellar arrays. For large spacings, we observe different oscillatory modes that are similar to those found previously for binary eutectics and that can be classified using the symmetry elements of the steady-state pattern. For small spacings, we observe a new instability that leads to a change in the sequence of the phases. Its onset can be well predicted by our analytic calculations. Finally, some preliminary phase-field simulations of three-dimensional growth structures are also presented.
\end{abstract}


\maketitle

\section{Introduction}
Eutectic alloys are of major industrial importance because
of their low melting points and their interesting mechanical
properties. They are also interesting for physicists because of 
their ability to form a large variety of complex patterns, which 
makes eutectic solidification an excellent model system
for the study of numerous nonlinear phenomena.

In a binary eutectic alloy, two distinct solid phases co-exist
with the liquid at the eutectic point characterized by the
eutectic temperature $T_{\rm E}$ and the eutectic concentration 
$C_{\rm E}$. If the global sample concentration is close to 
the eutectic concentration, solidification generally results
in composite patterns: alternating lamellae of the two solids,
or rods of one solid immersed in a matrix of the other, grow
simultaneously from the liquid. The fundamental understanding
of this pattern-formation process was established by
Jackson and Hunt (JH) \cite{JH}. They calculated approximate
solutions for spatially periodic lamellae and rods that grow
at constant velocity $v$, and established that the average
front undercooling, that is, the difference between the
average front temperature and the eutectic temperature,
follows the relation
\begin{equation}
\Delta T = K_1 v \lambda + \frac{K_2}{\lambda},
\label{JHlaw}
\end{equation}
where $\lambda$ is the width of one lamella pair (or the distance
between two rod centers), $v$ is the velocity of the 
solidification front, and $K_1$ and $K_2$ are constants 
whose value depends on the volume fractions of the two solid 
phases and various materials parameters \cite{JH}. The two
contributions in Eq.~(\ref{JHlaw}) arise from the redistribution 
of solute by diffusion through the liquid and the curvature of 
the solid-liquid interfaces, respectively.

The front undercooling is minimal for a characteristic spacing
\begin{equation}
\lambda_{JH}=\sqrt{\frac{K_2}{K_1v}}. 
\label{lambdamin}
\end{equation}
The spacings found in experiments in massive 
samples are usually distributed in a narrow range around
$\lambda_{JH}$ \cite{Trivedi91}. However, other spacings can be 
reached in directional solidification experiments by imposing a 
solidification velocity that varies with time. In this way, the 
stability of steady-state growth can be probed \cite{Akamatsu}. 
In agreement with theoretical expections \cite{Manneville}, 
steady-state growth is stable over a range of spacings
that is limited by the occurrence of dynamic instabilities. For
low spacings, a large-scale lamella (or rod) elimination instability
is observed \cite{Faivre}. For high spacings, the type of 
instability that can be observed depends on the sample geometry. 
For thin samples, various oscillatory instabilities
and a tilt instability can occur, depending on the alloy phase diagram 
and the sample concentration. Beyond the onset of these instabilities,
stable tilted patterns as well as oscillatory limit cycles can be
observed in both experiments and simulations \cite{Akamatsu,Sarkissian}. 
For massive samples, a zig-zag 
instability occurs for lamellar eutectics \cite{Akamatsu04,Parisi08}, 
whereas rods exhibit a shape instability \cite{Parisi10}.

In summary, pattern formation in binary eutectics is fairly
well understood. However, most materials of practical importance
have more than two components. Therefore, eutectic solidification
in multicomponent alloys has received increasing attention in
recent years. A particularly interesting situation
arises in alloy systems that exhibit a ternary eutectic point,
at which four phases (three solids and the liquid) coexist.
At such a quadruple point, three 
binary ``eutectic valleys'', that is, monovariant lines
of three-phase coexistence, meet. The existence of three
solid phases implies that there is a far greater variety
of possible structures, even in thin samples. Indeed, for 
two solids $\alpha$ and $\beta$, an array $\alpha\beta\alpha\beta\ldots$ 
is the only possibility for a composite pattern 
in a thin sample; the only remaining degree of 
freedom is the spacing. With an additional $\gamma$ solid, 
an infinite number of distinct periodic cycles with different 
sequences of phases are possible. The simplest cycles are
$\alpha\beta\gamma\alpha\beta\gamma\ldots$ and
$\alpha\beta\alpha\gamma\alpha\beta\alpha\gamma\ldots$ and 
permutations. Clearly, cycles of arbitrary length, and even 
non-periodic configurations are possible. An interesting 
question is then which configurations, if any, will be favored.

In preliminary works, the occurrence of lamellar structures has 
been reported in experiments in massive samples 
\cite{Kerr, Durand, Cooksey, Holder, Rinaldi, Ruggiero, McCartney}.
The spatio-temporal evolution in ternary eutectic systems
was observed in thin samples (quasi-2D experiments) 
in both metallic \cite{In-Bi-Sn} and organic systems \cite{AMPDDCNPGSCN}. 
In both cases, the simultaneous growth of three distinct solid phases
from the liquid with a
$\left(\alpha \beta \alpha \gamma\right)$, (named ABAC in 
Ref.~\cite{AMPDDCNPGSCN}) stacking was observed. Measurements in 
both cases revealed that $\lambda^{2}v$ was approximately constant, 
in agreement with the JH scaling of Eq.~(\ref{lambdamin}).

On the theoretical side, models that extend the JH analysis 
from binary to ternary eutectics for three different growth 
morphologies (rods and hexagon, lamellar, and semi-regular brick 
structures) were proposed by Himemiya {\it et al.} \cite{TPEG}. The
relation between front undercooling and spacing is still of
the form given by Eq.~(\ref{JHlaw}), with constants $K_1$ and
$K_2$ that depend on the morphology. The differences between
the minimal undercoolings for different morphologies were found
to be small. No direct comparison to experiments was given. 

Finally, ternary eutectic growth has also been investigated
by phase-field methods in Refs.~\cite{Apel,Hecht}, who have studied 
different stacking sequences formed by $\alpha$ = Ag$_{2}$Al, 
$\beta$ = ($\alpha$ Al) and $\gamma$ = Al$_{2}$Cu in the ternary 
system Al-Cu-Ag, while transients in the ternary eutectic solidification
of a transparent In-Bi-Sn alloy were studied both by phase field modeling
and experiments \cite{In-Bi-Sn}.

The purpose of the present paper is to carry out a more systematic
investigation of lamellar ternary eutectic growth. The main questions
we wish to address are (i) can an extension of the JH theory adequately
describe the properties of ternary lamellar arrays and reveal the
differences between cycles of different stacking sequences, and (ii) what
are the instabilities that can occur in such patterns. To answer
these questions, we develop a generalization of the JH theory to
ternary eutectics which is capable of describing the front
undercoolings of periodic lamellar arrays with arbitrary stacking 
sequence. Its predictions are systematically compared to phase-field
simulations. We use a generic thermodynamically consistent
phase-field model \cite{Garcke,Stinner}. While this model is
known to exhibit several thin-interface effects which limit
its accuracy \cite{Karma96, Almgren99, Kim98, Karma+01, Eschebaria+04}, 
we show here that we can obtain a very satisfying agreement between
theory and simulations if the solid-liquid interfacial free energy 
is evaluated numerically. In particular, the minimum-undercooling 
spacings are accurately reproduced for all stacking sequences 
that we have simulated.

The model is then used to systematically investigate the instabilities
of lamellar arrays, in particular for large spacings. We find that, as
for binary eutectics, the symmetry elements of the steady-state array
determine the possible instability modes. Whereas the calculation
of a complete stability diagram is not feasible due to the large number
of independent parameters, we find and characterize several new 
instability modes. Besides these oscillatory modes that are
direct analogs of the ones observed in binary eutectics, we also
find a new type of instability which occurs at small spacings: 
cycles in which the same phase appears more than once can 
undergo an instability during which one of these lamellae 
is eliminated; the system therefore transits to
a different (simpler) cycle. Furthermore, we also find that the
occurrence of this type of instability can be well predicted by our
generalized JH theory.

The remainder of the paper is organized as follows. In Sec. 2,
we develop the generalized JH theory for ternary eutectics and
calculate the undercooling-spacing relationships for several
simple cycles. In Sec. 3, the phase-field model is outlined
and its parameters are related to the ones of the theory.
Sec. 4 presents the simulation results concerning both steady-state 
growth and its instabilities. In Sec. 5, we briefly discuss questions
related to pattern selection and present some preliminary
simulations in three dimensions. Sec. 6 concludes the paper.

\section{\label{sec:theory}Theory}
We consider a ternary alloy system consisting of components
$A,B$ and $C$, which can form three solid phases $\alpha, 
\beta$, and $\gamma$ upon solidification from the liquid $l$. 
The concentrations of the components (in molar fractions) are 
denoted by $c_A, c_B$ and $c_C$ and fulfill the constraint 
\begin{equation}
c_A + c_B + c_C = 1.
\label{Eqn-Constraint}
\end{equation}
This obviously implies that there are only two independent
concentration fields.

\begin{figure}[htpb]
 \begin{center}
  \includegraphics[width=8cm]{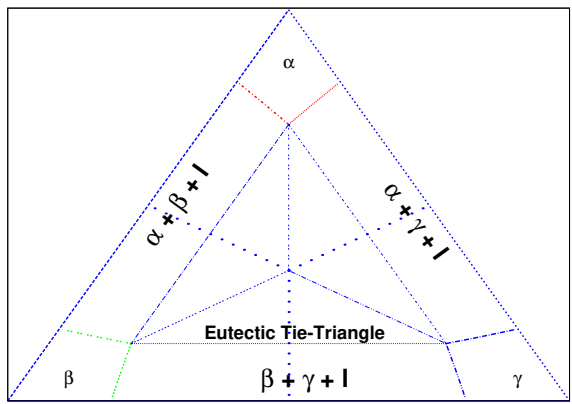}
\caption{(Color Online) Projection of the ternary phase diagram for a model symmetric ternary
eutectic system on the Gibbs simplex. The triangle at the center is the
tie-triangle at the eutectic temperature where four phases 
$\alpha,\beta,\gamma$, and $l$ are in equilibrium. The diagram also 
contains the information on three-phase equilibria. The liquidus lines 
corresponding to each of these equilibria (``eutectic valleys'') are 
shown by dotted lines which meet at the center of the simplex, which 
is also the concentration of the liquid at which all the three solid 
phases and the liquid are at equilibrium.}
\label{Figure1}
\end{center}
\end{figure}

As is customary, isothermal sections of
the ternary phase diagram can be conveniently displayed in
the Gibbs simplex. We are interested in alloy systems that
exhibit a ternary eutectic point: four-phase
coexistence between three solids and the liquid. The isothermal 
cross-section at the ternary eutectic temperature is displayed in 
Figure \ref{Figure1}, here for the particular example
of a completely symmetric phase diagram.

The concentration
of the liquid is located in the center of the simplex ($c_A=c_B=c_C=1/3$),
and the three solid phases are located at the corners
of the eutectic tie triangle. For higher temperatures,
no four-phase coexistence is possible, but each pair
of solid phases can coexist with the liquid (three-phase
coexistence). Each of these three-phase equilibria is a
eutectic, and the loci of the liquid concentrations at 
three-phase coexistence as
a function of temperature form three ``eutectic valleys''
that meet at the ternary eutectic point. On each of the
sides of the simplex (with the temperature as additional
axis), a binary eutectic phase diagram is found.

The key point for the following analysis is the temperature
of solid-liquid interfaces, which depends on the liquid 
concentration, the interface curvature, and the interface velocity.
The dependence on the concentration is described by the liquidus
surface, which is a two-dimensional surface over the Gibbs
simplex. This surface can hence be characterized by two independent
liquidus slopes at each point. For each phase $\nu$
($\nu=\alpha,\beta,\gamma$), we choose the two liquidus
slopes with respect to the minority components. Thus, for
the $\alpha$ phase, the interface temperature is given by the
generalized Gibbs-Thomson relation,
\begin{equation}
T_{\rm int}^\alpha-T_{\rm E} = m_B^\alpha(c_B-c_B^E) + m_C^\alpha(c_C-c_C^E)
-\Gamma_\alpha \kappa - \frac{v_n}{\mu_{\rm int}^\alpha},
\label{GibbsThomson}
\end{equation}
where $c_B$ and $c_C$ are the concentrations in the liquid
adjacent to the interface, $c_B^E$ and $c_C^E$ their values
at the ternary eutectic point, and $m^\alpha_B = \left. \frac{d T_\alpha}{d c_B}
\right|_{c_C={\rm const}}$ and $m^\alpha_C = \left. \frac{d T_\alpha}{d c_C}
\right|_{c_B={\rm const}}$ the liquidus slopes taken at the ternary eutectic 
point. Furthermore, $\Gamma_\alpha=\tilde\gamma_{\alpha l} T_{\rm E}/L_\alpha$
is the Gibbs-Thomson coefficient, with $\tilde\gamma_{\alpha l}$
the solid-liquid surface tension and $L_\alpha$ the latent
heat of fusion per unit volume, and $\mu_{\rm int}^\alpha$ is the
mobility of the $\alpha$-liquid interface. For the typical (slow) 
growth velocities that can be attained in directional solidification
experiments, the last term, which represents the kinetic undercooling 
of the interface, is very small. It will therefore be neglected
in the following. The expression for the other solid 
phases are obtained by cyclic permutation of the indices.

In the spirit of the original Jackson-Hunt analysis, for the
calculation of the diffusion field in the liquid, the 
concentration differences between solid and liquid phases
are assumed to be constant and equal to their values at the
ternary eutectic point. Since we are interested in ternary
coupled growth, which will take place at temperatures close
to $T_E$, this should be a good approximation. Thus, we define   
\[
\Delta c_{j}^{\nu} =  c_{j}^{l} - c_{j}^{\nu}\qquad \mbox{with} \qquad j=A,B,C \qquad 
\mbox{and}  \qquad \nu = \alpha, \beta, \gamma.
\]
In this approximation, the Stefan condition at a $\nu$-$l$ 
interface, which expresses mass conservation upon solidification, reads
\begin{equation}
\partial_n c_j = -\frac{v_n}{D} \Delta c_{j}^{\nu}, 
\label{Eqn-Stefan}
\end{equation}
where $\partial_n c_j$ denotes the partial derivative of $c_j$ in 
the direction normal to the interface, $v_n$ is the normal velocity 
of the interface (positive for a growing solid), and 
$D$ is the chemical diffusion coefficient, for simplicity
assumed to be equal for all the components.

We consider a general periodic lamellar array with $M$ repeating 
units consisting of phases $(\nu_0, \nu_1, \nu_2, \ldots, \nu_{M-1})$
where each $\nu_{i}$ represents the name of one solid phase 
$\left(\alpha, \beta , \gamma\right)$ in the sequence, with
a repeat distance (lamellar spacing) $\lambda$.
The width of the $j$-th single solid phase region is 
$\left(x_{j}-x_{j-1}\right)\lambda$, with $x_{0}=0$ and 
$x_{M}=1$, and the sum of all the widths corresponding to any 
given phase is its volume fraction $\eta_{\nu}$. 
The eutectic front is assumed to grow in the $z$ direction
with a constant velocity $v$. 

\begin{figure}[ht]
\centerline{\epsfxsize=15.5cm \epsfbox{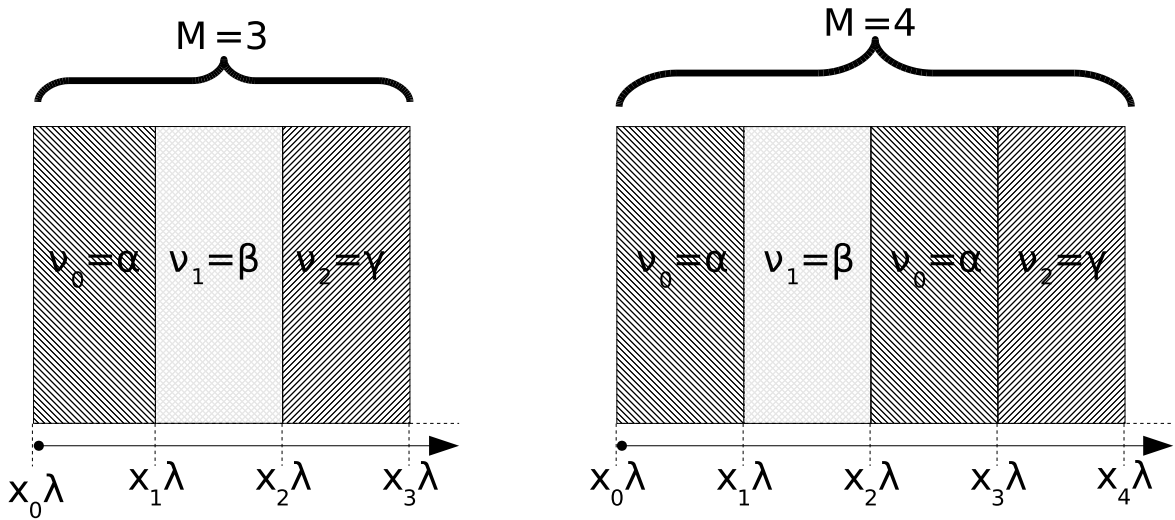}}
\caption{\label{Figure2}
Two examples for periodic lamellar arrays with $M=3$ and $M=4$ units.}
\end{figure}

\subsection{Concentration fields}
First, we consider the diffusion fields of the components
$A,B,C$ ahead of a growing eutectic front. For the calculation
of the concentration fields, the front is supposed to be planar,
as in the sketches of Figure \ref{Figure2}.
We make the following Fourier series expansion for $c_A$ and $c_B$ 
\begin{equation}
c_X = \sum_{n=-\infty}^{\infty} X_n e^{ik_n x-q_nz} + c_X^\infty, \quad 
X=A,B.
\label{Eqn-Series_cX} 
\end{equation}
The third concentration $c_C$ follows 
from the constraint of Eq. (\ref{Eqn-Constraint}). 
In Eq. ~(\ref{Eqn-Series_cX}), $k_n=2\pi n/\lambda$ are wave numbers 
and $q_n$ can be determined from the solutions of the stationary diffusion
equation
\[
v \partial_z c_X + D \nabla^2 c_X = 0,
\]
which yields
\[
q_n = \frac{v}{2D} + \sqrt{k_n^2+\left(\frac{v}{2D}\right)^2}.
\]
For all the modes $n\neq 0$, we thus have $q_n \simeq  |k_n|$ for
small Peclet number ${\rm Pe} = \lambda/\ell \ll 1$ with 
$\ell=2D/v_n$, which will always be the case for slow growth.
The mode $n=0$ describes the concentration boundary layer which is
present at off-eutectic concentrations, and which has a characteristic 
length scale of $\ell$.

To determine the coefficients $X_n$ in the above Fourier series,
we assume the eutectic front to be at the $z=0$ position.
Using the Stefan condition in Eq. (\ref{Eqn-Stefan}) and 
taking the derivative 
of $c_X$ with respect to  the $z$-coordinate
\[
\partial_z c_X|_{z=0} = \sum_{n=-\infty}^\infty - q_n X_n e^{ik_n x},
\]
integration across one lamella period 
$\lambda$ of arbitrary partitioning of phases gives
\begin{eqnarray}
q_n X_n \delta_{nm} \lambda &=& \frac{2}{\ell} 
\sum_{j=0}^{M-1} \int_{x_j \lambda}^{x_{j+1}\lambda}
e^{-ik_m x} \Delta c_{X}^{\nu_j}\: dx, 
\label{ftransform}
\end{eqnarray}
so that the coefficients $X_n, n\in \nz$ in the series ansatz, 
Eq. (\ref{Eqn-Series_cX}) follow
\begin{eqnarray}
\qquad X_n &=& \frac{4}{\ell q_n \lambda k_n} \sum_{j=0}^{M-1}
\Delta c_{X}^{\nu_j} e^{-ik_n \lambda(x_{j+1}+x_j)/2} \sin(k_n \lambda (x_{j+1}+ x_j)/2).
\end{eqnarray}
Applying symmetry arguments for the sinus and cosinus functions, we 
can formulate real combinations of these coefficients if we additionally 
take the negative summation indices into account. We obtain 
\[
X_n + X_{-n} = \frac{8}{\ell q_n \lambda k_n} \sum_{j=0}^{M-1}
\Delta c_{X}^{\nu_j} \cos(k_n \lambda(x_{j+1}+x_j)/2) \sin(k_n \lambda (x_{j+1}+ x_j)/2),
\]
\[
i(X_n - X_{-n}) = \frac{8}{\ell q_n \lambda k_n} \sum_{j=0}^{M-1}
\Delta c_{X}^{\nu_j} \sin(k_n \lambda(x_{j+1}+x_j)/2) \sin(k_n \lambda (x_{j+1}-x_j)/2).
\]
Herewith, Eq. ~(\ref{Eqn-Series_cX}) reads 
\begin{eqnarray}
c_X = c_X^\infty + X_0 + \sum_{j=0}^{M-1} \sum_{n=1}^\infty 
\frac{8}{\ell q_n \lambda k_n} 
\cos(k_n \lambda(x_{j+1}+x_j)/2) \sin(k_n \lambda (x_{j+1}+ x_j)/2)\cos(k_n x) \nonumber \\
+ \sum_{j=0}^{M-1} \sum_{n=1}^\infty 
\frac{8}{\ell q_n \lambda k_n} 
\sin(k_n \lambda(x_{j+1}+x_j)/2) \sin(k_n \lambda (x_{j+1}-x_j)/2)\sin(k_n x).
\end{eqnarray}
The general expression for the mean concentration $\langle c_X \rangle_m$ 
ahead of the $m$-th phase of the phase sequence can be calculated to yield
\begin{eqnarray}
\langle c_X \rangle _m &=& \frac{1}{(x_{m+1} -x_m)\lambda} \int_{x_m \lambda}^{x_{m+1}\lambda}
c_X dx \nonumber \\
&=& c_X^\infty + X_0 + \frac{1}{x_{m+1} - x_m} \sum_{n=1}^{\infty} \sum_{j=0}^{M-1}
\Big\{ \frac{16}{\lambda^2 k_n^2 \ell q_n} \Delta c_{X}^{\nu_j} \sin[\pi n(x_{m+1}-x_m)] \times
\nonumber \\
&& \hspace*{1.5cm} \times \: \sin[\pi n(x_{j+1} - x_j)] \cos[\pi n(x_{m+1}+x_m - x_{j+1}-x_j)]
\Big\}.
\label{Eqn-Mean_cX_general}
\end{eqnarray} 
For a repetitive appearance of a phase $\nu$ in the phase sequence,
the mean concentration of component $X$ ahead of this phase follows by
taking the weighted average of all the lamellae of phase $\nu$, 
\begin{equation}
\langle c_X \rangle_{\nu} = \frac{\sum_{m=0}^{M-1} \langle c_X \rangle_m (x_{m+1} - x_m) \delta_{{\nu_m} \nu}}
{\sum_{m=0}^{M-1} (x_{m+1}-x_m) \delta_{{\nu_m} \nu}} \qquad \mbox{with} \qquad 
\delta_{{\nu_m} \nu} = \left\{ 
\begin{array}{cc} 
0 & \mbox{for} \: \nu = \nu_m \\
1 & \mbox{for} \: \nu \ne \nu_m.
\end{array}
\right.
\end{equation}

\subsection{Average front temperature}
The average front temperature is now found by taking the average
of the Gibbs-Thomson equation along the front, separately for each
phase ($\alpha, \beta$ and $\gamma$): 
\begin{eqnarray}
\Delta T_\nu &=& T_E - T_\nu = -m_B^\nu (\langle c_B \rangle_\nu - c_B^E)
- m_C^\nu (\langle c_C \rangle_\nu - c_C^E) + \Gamma_\nu \langle \kappa \rangle_\nu, \label{Eqn-undercooling}
\end{eqnarray}
for $\nu = \alpha, \beta, \gamma$. Here, $\langle \kappa \rangle_\nu$ 
is the average curvature of the solid-liquid interface which can be
evaluated by exact geometric relations to be
\begin{eqnarray*}
\langle \kappa \rangle_\nu &=& \frac{\sum_{m=0}^{M-1}\langle \kappa \rangle_m (x_{m+1}-x_m) \delta_{{\nu_m}\nu}}{\sum_{m=0}^{M-1} 
(x_{m+1}-x_m)\delta_{{\nu_m}\nu}}
\end{eqnarray*}
and 
\[
\langle \kappa \rangle _m = \frac{\sin \theta_{\nu_{m}\nu_{m+1}} + \sin \theta_{\nu_m \nu_{m-1}}}
{(x_{m+1} - x_m) \lambda}.
\]
Here, $\theta_{\nu_m \nu_{m-1}}$ are the contact angles that are obtained 
by applying Young's law at the trijunction points. More precisely, 
$\theta_{\nu_{m}\nu_{m+1}}$ is the angle, at the triple point (identified 
by the intersection of the two solid-liquid interfaces and the solid-solid one), 
between the tangent to the $\nu_{m}-l$ interface and the horizontal (the $x$ 
direction). For a triple point with the phases $\nu_{m},\nu_{m+1}$ and liquid, 
the two contact angles $\theta_{\nu_{m} \nu_{m+1}}, \theta_{\nu_{m+1}\nu_{m}}$ 
satisfy the following relations, obtained from Young's law,
\begin{equation}
 \frac{\tilde{\gamma}_{\nu_{m+1}l}}{\cos(\theta_{\nu_{m} \nu_{m+1}})}=\frac{\tilde{\gamma}_{\nu_{m}l}}{\cos(\theta_{\nu_{m+1} \nu_{m}})}=\frac{\tilde{\gamma}_{\nu_{m} \nu_{m+1}}}{\sin(\theta_{\nu_{m}\nu_{m+1}} + \theta_{\nu_{m+1}\nu_{m}})}.
\end{equation}
Note that, in general, $\theta_{\nu_{m} \nu_{m+1}} \neq \theta_{\nu_{m+1}\nu_{m}}$.

A short digression is in order to motivate the closure of our
system of equations. Although we have not given the explicit 
expressions, the coefficients $A_0$ and $B_0$ can be simply
calculated by using Eq.~(\ref{ftransform}) with $n=0$. However,
to carry out this calculation, the width of each lamella has
to be given. If these widths are chosen consistent with
the lever rule, that is, the cumulated lamellar width of phase
$\nu$ corresponds to the nominal volume fraction of phase $\nu$
for the given sample concentration $c_A^\infty$, $c_B^\infty$, 
and $c_C^\infty$, the use of Eq.~(\ref{ftransform}) yields 
$X_0=c_X^E-c_X^\infty$ ($X=A,B,C$).
However, this result is incorrect: the concentrations of
the solids are not equal to the equilibrium concentrations at the
eutectic temperature because solidification takes place at a
temperature below $T_{\bf E}$. Therefore, the true volume fractions
depend on the solidification conditions. Their determination
would require a self-consistent calculation
which is exceedingly difficult. Therefore, we will take the same
path as Jackson and Hunt in their original paper \cite{JH}: we
will assume that the volume fractions of the three phases are
fixed by the lever rule at the eutectic temperature, but we will
treat the amplitudes of the two boundary layers, $A_0$ and $B_0$,
as unknowns. As in Ref.~\cite{JH}, one can expect that the difference
to the true solution is of order $\rm Pe$ and therefore
small for slow solidification.

With this assumption, the equations developed above 
can now be used in two ways. For {\em isothermal solidification}, 
the temperatures of all interfaces must be equal to
the externally set temperature, and the three equations
$\Delta T_\nu = \Delta T$ for $\nu = \alpha,\beta,\gamma$, 
can be used to determine the three unknowns $A_0, B_0$  
and the velocity $v$ of the solid-liquid front. All of these
quantities will be a function of the lamellar spacing $\lambda$.
In {\em directional solidification}, the growth velocity in
steady state is fixed and equal to the speed with
which the sample is pulled from a hot to a cold region.
The third unknown is now the total front undercooling.
In the classic Jackson-Hunt theory for binary eutectics,
the system of equations is closed by the hypothesis that
the average undercoolings of the two phases are equal.
This is only an approximation which is quite accurate
for eutectics with comparable volume fractions of the
two solids, but becomes increasingly inaccurate when the
volume fractions are asymmetric \cite{Sarkissian}.
We will use the same approximation for the ternary case
here, and set $\Delta T_\alpha=\Delta T_\beta = \Delta T_\gamma=\Delta T$.
This then leads to expressions for $\Delta T$ as a function
of the growth speed $v$ and the lamellar spacing $\lambda$.

\subsection{Examples}

\subsubsection{Binary systems}

As a benchmark for both our calculations and simulations,
we consider binary eutectic systems with components $A$ and 
$B$ and with three phases: $\alpha, \beta$, and liquid.

\begin{figure}[ht]
\centerline{\epsfxsize=5cm \epsfbox{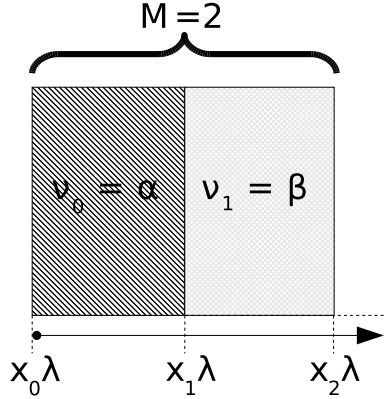}}
\caption{\label{Figure3}
Sketch of a lamellar structure in a binary eutectic system with period length $M=2$. 
$\nu_{i}$ denotes a phase in the sequence $(\alpha \beta)$.}
\end{figure}

Setting $x_0=0, x_1 = \eta_\alpha$, $x_2 = 1$, and applying Eq. (\ref{Eqn-Mean_cX_general}) gives
\begin{eqnarray}
\langle c_X \rangle_\alpha &=& c_X^\infty + X_0 + \frac{1}{\eta_\alpha} \sum_{n=1}^{\infty} 
\Big\{ \frac{16}{\lambda^2 k_n^2 \ell q_n} \left(\Delta c_{X}^{\alpha} - \Delta c_{X}^{\beta} \right) 
\sin^2(\pi n\eta_\alpha) \Big\} \\
&\cong &  c_X^\infty + X_0 + \frac{2 \lambda}{\eta_\alpha \ell} {\cal P}(\eta_\alpha) \Delta c_X 
\qquad \mbox{and} \\
\langle c_X \rangle_\beta &=& c_X^\infty + X_0 - \frac{2 \lambda}{(1-\eta_\alpha) \ell} {\cal P}(1-\eta_\alpha) 
\Delta c_X 
\label{mean_cA_binary}
\end{eqnarray} 
with $k_n = 2 \pi n/\lambda, q_n \approx k_n, \lambda / \ell \ll 1, 
\Delta c_X = \Delta c_{X}^{\alpha} - \Delta c_{X}^{\beta}$, and the
dimensionless function 
\begin{eqnarray}
{\cal P}(\eta) &=& \sum_{n=1}^\infty \frac{1}{(\pi n)^3}\sin^2(\pi n \eta)
\label{Eqn_Pdef}
\end{eqnarray}
which has the properties ${\cal P}(\eta) = {\cal P} (1-\eta)={\cal P}(\eta -1)$.

Furthermore, Eq.~(\ref{Eqn-undercooling}) together with $\ell = 2D/v$ leads to
\begin{eqnarray}
 \Delta T_\alpha = -m_B^\alpha B_0 - \frac{\lambda v}{\eta_\alpha D}\calP(\eta_\alpha)m_B^\alpha\Delta c_B + \Gamma_\alpha \langle\kappa\rangle_\alpha,\\
 \Delta T_\beta = -m_A^\beta A_0 - \frac{\lambda v}{\eta_\beta D}\calP(\eta_\beta)m_A^\beta\Delta c_A + \Gamma_\beta \langle\kappa\rangle_\beta,
\end{eqnarray}
where $\langle\kappa\rangle_\alpha = 2 \sin \theta_{\alpha \beta}/(\eta_\alpha \lambda)$ and $\langle\kappa\rangle_\beta = 2 \sin \theta_{\beta \alpha}/(\eta_\beta \lambda)$. In addition, for a binary alloy $B_0=-A_0$. The unknown $A_{0}$ and the global front undercooling are determined using the assumption of equal interface undercoolings, $\Delta T_\alpha=\Delta T_\beta$. The result is identical to the one of the Jackson-Hunt analysis.


\subsubsection{Ternary Systems}
\begin{figure}[ht]
\centerline{\epsfxsize=6cm \epsfbox{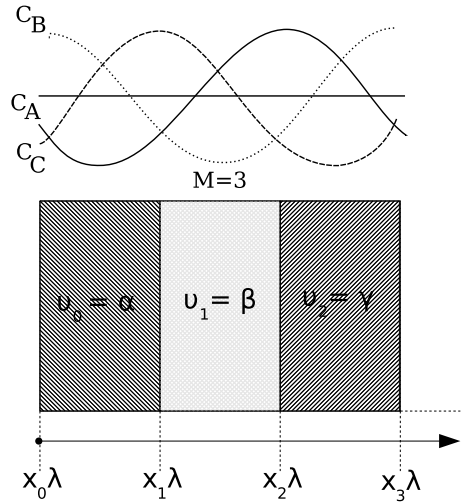}}
\caption{\label{Figure4}Sketch of a ternary stacking 
order $\left(\alpha\beta\gamma\right)$ with period length $M=3$.}
\end{figure}

Next, we study ternary systems with three components ($A, B, C$) and 
four phases ($\alpha, \beta, \gamma$ and liquid). We start with the 
configuration $\left(\alpha\beta\gamma\alpha\beta\gamma \ldots\right)$, 
sketched in Figure \ref{Figure4}.

We set $x_0=0, x_1 = \eta_\alpha, x_2=\eta_\alpha + \eta_\beta = 1 - \eta_\gamma$ 
and $x_3 = 1$ and apply Eq. (\ref{Eqn-Mean_cX_general}). This yields
\begin{eqnarray}
\label{abc_caalpha}
\langle c_X \rangle_\alpha &=& c_X^\infty + X_0 + 
\frac{2 \lambda}{\eta_\alpha \ell} \Big( {\cal P}(\eta_\alpha) \Delta c_{X}^{\alpha} 
+ {\cal Q} (\eta_\alpha,\eta_\beta) \Delta c_{X}^{\beta} + 
{\cal Q}(\eta_\alpha, \eta_\gamma) \Delta c_{X}^{\gamma} \Big)\\
\langle c_X \rangle_\beta &=& c_X^\infty + X_0 + 
\frac{2 \lambda}{\eta_\beta \ell} \Big( {\cal Q}(\eta_\beta, \eta_\alpha) 
\Delta c_{X}^{\alpha} 
+ {\cal P} (\eta_\beta) \Delta c_{X}^{\beta} + 
{\cal Q}(\eta_\beta, \eta_\gamma) \Delta c_{X}^{\gamma} \Big) \\
\label{abc_cabeta}
\langle c_X \rangle_\gamma &=& c_X^\infty + X_0 + 
\frac{2 \lambda}{\eta_\gamma \ell} \Big( {\cal Q}(\eta_\gamma, \eta_\alpha) 
\Delta c_{X}^{\alpha} 
+ {\cal Q} (\eta_\gamma, \eta_\beta) \Delta c_{X}^{\beta} + 
{\cal P}(\eta_\gamma) \Delta c_{X}^{\gamma} \Big). 
\label{abc_cagamma}
\end{eqnarray}
Here, we have used $X=A, B, C$ and $\calP$ is the function defined in 
Eq.~(\ref{Eqn_Pdef}), and
\begin{eqnarray}
\calQ(\eta_{\nu_i}, \eta_{\nu_j}) &=& \sum_{n=1}^\infty \frac{1}{(\pi n)^3}
\sin(\pi n \eta_{\nu_i}) \sin(\pi n \eta_{\nu_j}) \cos[\pi n (\eta_{\nu_i} + \eta_{\nu_j})]
\end{eqnarray}
$\calP(\eta_{\nu_i})$ and $\calQ(\eta_{\nu_i},\eta_{\nu_j})$ fulfill the  properties   
$\calP(\eta_{\nu_i}) = -\calQ(\eta_{\nu_i}, -\eta_{\nu_i})$
and $\calQ(\eta_{\nu_i},\eta_{\nu_j}) = \calQ(\eta_{\nu_j},\eta_{\nu_i})$.\\

For simplicity, we now consider a completely symmetric ternary eutectic
configuration: a completely symmetric ternary phase diagram (that is,
any two phases can be exchanged without changing the phase diagram)
and equal phase fractions $\eta_\alpha = \eta_\beta = \eta_\gamma=\frac{1}{3}$,
which implies $c_X^\infty = c_X^E$. As a consequence, $X_0 = 0$, and 
Eq. (\ref{abc_caalpha}) simplifies to
\begin{eqnarray}
\langle c_A \rangle_\alpha - c_A^E &=&  \frac{2 \lambda}{\eta_\alpha \ell} \calP(\eta_\alpha)
(\Delta c_{A}^{\alpha} - \Delta c_{A}^{\beta}) \\
\langle c_B \rangle_\alpha - c_B^E &=& \frac{ \lambda {\cal P}(\eta_\alpha) }{\eta_\alpha \ell} \Big(\Delta c_{B}^{\alpha} 
- \Delta c_{B}^{\beta}\Big) \\
\langle c_C \rangle_\alpha - c_C^E &=& \frac{\lambda {\cal P}(\eta_\alpha) }{\eta_\alpha \ell} \Big(\Delta c_{C}^{\alpha} 
- \Delta c_{C}^{\gamma} \Big),
\end{eqnarray}

for the three components. Since, in this case, all phases have the same undercooling by symmetry,
the front undercooling is simply given by
\begin{equation}
\Delta T = - \frac{2 \lambda v}{\eta_\alpha D} \calP(\eta_\alpha) 
m_B^\alpha \Delta c_B + \Gamma_\alpha \langle\kappa\rangle_\alpha
\end{equation}
where $\langle\kappa\rangle_\alpha = \tfrac{2}{\eta_\alpha \lambda}(\sin \theta_{\alpha \beta} + \sin \theta_{\alpha \gamma})$. The terms $\Delta c_{B}^{\alpha} - \Delta c_{B}^{\beta}$ and $\Delta c_{C}^{\alpha} - \Delta c_{C}^{\gamma}$ are identical. For convenience, we write the preceding equation using the term we already use for the binaries namely $\Delta c_{B}=\Delta c_{B}^{\alpha} - \Delta c_{B}^{\beta}$. 

Next, we discuss again a ternary eutectic alloy with three components 
and four phases, but now for the phase cycle 
$\left(\alpha\beta\alpha\gamma\alpha\beta\ldots \right)$. 
\begin{figure}[h]
\centerline{\epsfxsize=7cm \epsfbox{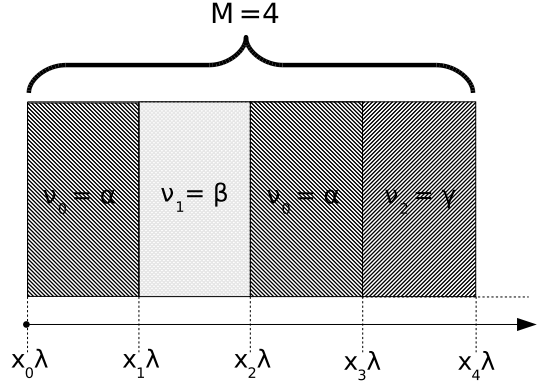}}
\caption{\label{Figure5} Schematic drawing of a ternary eutectic system with a configuration $\left(\alpha\beta\alpha\gamma\alpha\beta\ldots\right)$ of periodic length $M=4$.}
\end{figure}
Furthermore, we suppose that the two lamellae of the $\alpha$ phase
have equal width $\lambda\eta_\alpha/2$.
The average concentrations $\langle c_{X}\rangle_{m}$ are deduced 
from the general expression in Eq.\ref{Eqn-Mean_cX_general} and read
\begin{eqnarray}
\langle c_X \rangle_\alpha &=& c_X^\infty + X_0 + 
\frac{2 \lambda}{\tfrac{\eta_\alpha}{2} \ell} \Big( 
\calS(\eta_\alpha, \eta_\beta) \Delta c_{X}^{\alpha} 
+ {\cal Q} (\tfrac{\eta_\alpha}{2},\eta_\beta) \Delta c_{X}^{\beta} + 
{\cal Q}(\tfrac{\eta_\alpha}{2}, \eta_\gamma) \Delta c_{X}^{\gamma} \Big) \\
\label{abac_caalpha}
\langle c_X \rangle_\beta &=& c_X^\infty + X_0 + 
\frac{2 \lambda}{\eta_\beta \ell} \Big( 2 {\cal Q}(\eta_\beta, \tfrac{\eta_\alpha}{2}) 
\Delta c_{X}^{\alpha} 
+ {\cal P} (\eta_\beta) \Delta c_{X}^{\beta} + \calR(\eta_\beta, \eta_\gamma) 
\Delta c_{X}^{\gamma} \Big) \\
\label{abac_cabeta}
\langle c_X \rangle_\gamma &=& c_X^\infty + X_0 + 
\frac{2 \lambda}{\eta_\gamma \ell} \Big( 2 {\cal Q}(\eta_\gamma, \tfrac{\eta_\alpha}{2}) 
\Delta c_{X}^{\alpha} + \calR(\eta_\gamma, \eta_\beta) \Delta c_{X}^{\beta} + 
{\cal P}(\eta_\gamma) \Delta c_{X}^{\gamma} \Big), 
\label{abac_cagamma}
\end{eqnarray}
where $X=A, B, C$. Furthermore, we have introduced the short notations
\begin{eqnarray}
\calR(\eta_{\nu_i},\eta_{\nu_j}) &=& \sum_{n=1}^\infty \frac{1}{(\pi n)^3} \sin(\pi n \eta_{\nu_i}) 
\sin(\pi n \eta_{\nu_j}) \cos(\pi n)\\
\calS(\eta_{\nu_i},\eta_{\nu_j}) &=& \sum_{n=1}^\infty \frac{1}{(\pi n)^3} 
\sin^2(\pi n \eta_{\nu_i}/2)
\{1 + \cos(\pi n)\cos[\pi n (\eta_{\nu_j} - \eta_{\nu_i})]\}.
\end{eqnarray}
From the general formulation of the Gibbs-Thomson equation in Eq.~(\ref{Eqn-undercooling}), we determine the undercoolings,
\begin{eqnarray}
\label{abag-1lamella}
\Delta T_\alpha &=& -m_{B}^{\alpha}\left(B_{0}+\frac{4 \lambda}{\eta_\alpha \ell} \Big( 
{\cal S}(\eta_\alpha, \eta_\beta) \Delta c_{B}^{\alpha} 
+ {\cal Q} (\tfrac{\eta_\alpha}{2},\eta_\beta) \Delta c_{B}^{\beta} + 
{\cal Q}(\tfrac{\eta_\alpha}{2}, \eta_\gamma) \Delta c_{B}^{\gamma} \Big)\right) \nonumber \\
&+& -m_{C}^{\alpha}\left(C_{0}+\frac{4 \lambda}{\eta_\alpha \ell} \Big( 
{\cal S}(\eta_\alpha, \eta_\beta) \Delta c_{C}^{\alpha} 
+ {\cal Q} (\tfrac{\eta_\alpha}{2},\eta_\beta) \Delta c_{C}^{\beta} + 
{\cal Q}(\tfrac{\eta_\alpha}{2}, \eta_\gamma) \Delta c_{C}^{\gamma} \Big)\right) \nonumber \\
&+& \Gamma_\alpha \dfrac{2\left(\sin \theta_{\alpha \beta} + \sin \theta_{\alpha \gamma}\right)}{\eta_{\alpha}\lambda}\\
\Delta T_\beta &=& -m_{A}^{\beta}\left(A_{0}+\frac{2 \lambda}{\eta_\beta \ell} \Big( 2 {\cal Q}(\eta_\beta, \tfrac{\eta_\alpha}{2}) 
\Delta c_{A}^{\alpha} 
+ {\cal P} (\eta_\beta) \Delta c_{A}^{\beta} + {\cal R}(\eta_\beta, \eta_\gamma) 
\Delta c_{A}^{\gamma} \Big)\right) \nonumber \\
&+& -m_{C}^{\beta}\left(C_{0}+\frac{2 \lambda}{\eta_\beta \ell} \Big( 2 {\cal Q}(\eta_\beta, \tfrac{\eta_\alpha}{2}) 
\Delta c_{C}^{\alpha} 
+ {\cal P} (\eta_\beta) \Delta c_{C}^{\beta} + {\cal R}(\eta_\beta, \eta_\gamma) 
\Delta c_{C}^{\gamma} \Big)\right) \nonumber \\
&+& \Gamma_\beta \dfrac{2\sin \theta_{\beta \alpha}}{\eta_{\beta}\lambda}\\
\Delta T_\gamma &=& -m_{A}^{\gamma}\left(A_{0}+\frac{2 \lambda}{\eta_\gamma \ell} \Big( 2 {\cal Q}(\eta_\gamma, \tfrac{\eta_\alpha}{2}) 
\Delta c_{A}^{\alpha} + {\cal R}(\eta_\gamma, \eta_\beta) \Delta c_{A}^{\beta} + 
{\cal P}(\eta_\gamma) \Delta c_{A}^{\gamma} \Big)\right) \nonumber \\
&+& -m_{B}^{\gamma}\left(B_{0}+\frac{2 \lambda}{\eta_\gamma \ell} \Big( 2 {\cal Q}(\eta_\gamma, \tfrac{\eta_\alpha}{2}) 
\Delta c_{B}^{\alpha} + {\cal R}(\eta_\gamma, \eta_\beta) \Delta c_{B}^{\beta} + 
{\cal P}(\eta_\gamma) \Delta c_{B}^{\gamma} \Big)\right) \nonumber \\
&+& \Gamma_\gamma \dfrac{2\sin \theta_{\gamma \alpha}}{\eta_{\gamma}\lambda}.
\end{eqnarray}
For a symmetric phase diagram (all slopes equal, $m_{X}^{\nu_{i}}=m$) one can show using the assumption of equal undercooling of all phases that an expression for the global interface undercooling can be derived as $\Delta T = 1/3(\Delta T_{\alpha} + \Delta T_{\beta} + \Delta T_{\gamma})$ by elimination of the constants $A_{0}, B_{0}$ and $C_{0}$ using the relation $(A_{0} + B_{0} + C_{0})=0$.

\subsection{Discussion}
\label{theoretical_discussion}
A point which merits closer attention is the question which of all
the possible steady-state configurations exhibits the lowest undercooling.
Whereas the general idea that a eutectic system will always select the state
of lowest undercooling is wrong (see Sec. \ref{sec_selection} below),
an information about this point constitutes nevertheless a useful
starting point. Whereas the general solution to this problem is
non-trivial, in the following we present some partial insights.

Let us, for the sake of discussion, first compute the average total 
curvature undercooling $\Delta T_\kappa$ of an arbitrary arrangement. 
Consider a configuration of period M having $M_a$ lamella of the $\alpha$ 
phase, $M_b$ lamella of the $\beta$ phase, and $M_c$ lamella of the $\gamma$
phase, where the integers $M_a$, $M_b$, and $M_c$ add up to M. 
In a system where all the solid-liquid and solid-solid surface tensions
are identical, the total average curvature undercooling $\Delta T_{\kappa}^\nu$ 
of each phase $\nu$ is, 
\begin{eqnarray}
\Delta T_{\kappa}^\alpha &=& \Gamma_\alpha \dfrac{2\sin \theta}{\lambda} \dfrac{M_a}{\eta_\alpha}\\
\Delta T_{\kappa}^\beta &=& \Gamma_\beta  \dfrac{2\sin \theta}{\lambda} \dfrac{M_b}{\eta_\beta}\\
\Delta T_{\kappa}^\gamma &=& \Gamma_\gamma \dfrac{2\sin \theta}{\lambda} \dfrac{M_c}{\eta_\gamma}.
\end{eqnarray}
It is remarkable that the average curvature undercooling is independent of 
the individual widths of each lamella, but depends only on the total volume
fraction and the number of lamellae of the specific phase.
Furthermore, it is quite clear from the above examples that the final
expression for the global average interface undercooling can always
be written in the same form as Eq. (\ref{JHlaw}). The second term
of this expression (that is, the one proportional to $1/\lambda$) can
be computed for the case where all Gibbs-Thomson coefficients and 
liquidus slopes are equal, and reads
\begin{eqnarray}
\dfrac{K_2}{\lambda} &=& \dfrac{\Delta T_{\kappa}^\alpha + \Delta T_{\kappa}^\beta + \Delta T_{\kappa}^\gamma}{3} \nonumber \\
&=& \Gamma \dfrac{2\sin \theta}{3\lambda}\left ( \dfrac{M_a}{\eta_\alpha} + \dfrac{M_b}{\eta_\beta} + \dfrac{M_c}{\eta_\gamma}\right).
\label{average_curvature}
\end{eqnarray}
For the special case of a completely symmetric phase diagram and
a sample at the eutectic composition,
Eqn.(\ref{average_curvature}) yields
\begin{eqnarray}
 \dfrac{K_2}{\lambda} &=& \Gamma \dfrac{2\sin \theta}{\lambda}\left(M_a+M_b+M_c\right),
\end{eqnarray}
where we have used the fact that $\eta_\alpha = \eta_\beta = \eta_\gamma =1/3$. 
Using, $M_a+M_b+M_c=M$,
$\dfrac{K_2}{\lambda} = \Gamma \dfrac{2\sin \theta}{\left(\lambda\right/M)}$.
Thus, we see that the magnitude of this term per unit lamella in 
an arrangement is the same for all the possible arrangements, irrespective 
of the individual widths of the lamella and the relative positions of the 
lamellae in a configuration. Moreover, we see that for a general off-eutectic
composition, choosing the number of lamellae in the ratio
$\eta_\alpha:\eta_\beta:\eta_\gamma$ renders the average curvature 
undercoolings of all the three phases equal. This condition is, however,
relevant only for the special case of identical solid-solid and
solid-liquid surface tensions and equal liquidus slopes of the phases.
For the case when the solid-liquid and solid-solid
surface tensions are unequal, the curvature undercooling is
no longer independent of the arrangement of the lamella in the 
configuration. Hence, the problem of determining the minimum undercooling
configuration is complex and no general expression regarding the number, 
position and widths of lamellae can be derived.

Another point is worth mentioning.
Under the assumption that the volume fractions of the solid phases 
are fixed by the lever rule, the width of the three lamellae in the 
$\alpha\beta\gamma$ cycle is uniquely fixed by the alloy concentration.
However, for the $\alpha\beta\alpha\gamma$ cycle, and more generally
for any cycle with $M>3$, this is not the case any more because there
have to be at least two lamellae of the same phase in the cycle.
Whereas the cumulated width of these lamellae is fixed by the global 
concentration, the width of each individual lamella is not. For example, 
in the $\alpha\beta\alpha\gamma$ cycle at the eutectic concentration 
$c_A^\infty=c_B^\infty=c_C^\infty=1/3$, all the configurations
$(\xi,1/3,1/3-\xi,1/3)$ for $0<\xi<1/3$ are admissible, where the
notation $(\cdot,\cdot,\ldots)$ is a shorthand for the list of the lamella
widths $x_{n+1}-x_n$. The number $\xi$ is an internal degree of freedom
that can be freely chosen by the system. With our method, the global
front undercooling can be calculated for any value of $\xi$. For the
$\alpha\beta\alpha\gamma$ cycle, we found that the configuration 
with equal widths of the $\alpha$ phases ($\xi=1/6$) was 
the one with the minimum average front undercooling. This gives a
strong indication that this value is stable, and that perturbations 
of $\xi$ around this value should decay with time. Hence, the 
analytic expressions given above for the $\alpha\beta\alpha\gamma$ cycle, 
which are for $\xi=1/6$, should be the relevant ones.

\section{Phase-field Model}
\subsection{Model}
A thermodynamically consistent phase-field model is used for the present study \cite{Garcke,Stinner}.
The equations are derived from an entropy functional of the form
\begin{eqnarray}
 {\cal
S}\left(e,\vc,\vphi\right)&=&\int_{\Omega}\left(s\left(e,\vc,
\vphi\right)-\left(\epsilon a\left(\vphi,\nabla \vphi\right) +
\dfrac{1}{\epsilon}w\left(\vphi\right)\right)\right)d\Omega,
 \label{functional}
\end{eqnarray}
where $e$ is the internal energy density, $\vc$ = $(c_{i})_{i=1}^{K}$ is a
vector of concentration variables, $K$ being the number of components, and
$\vphi=\left(\phi_{\alpha}\right)_{\alpha=1}^{N}$ is a vector of phase-field
variables, $N$ being the number of phases present in the system. $\vphi$ 
and $\vc$ fulfill the constraints
\begin{equation}
 \sum_{i=1}^{K}c_{i}=1 \qquad \mbox{and} \qquad
\sum_{\alpha=1}^{N}\phi_{\alpha}=1,
\label{constraint}
\end{equation}
so that these vectors always lie in $K-1$- and $N-1$-dimensional planes,
respectively. Moreover, $\epsilon$ is the small length scale parameter related 
to the interface width, $s\left(e,\vc,\vphi\right)$ is the bulk entropy density, 
$a\left(\vphi,\nabla\vphi\right)$ is the gradient entropy density and 
$w\left(\vphi\right)$ describes the surface entropy potential of the 
system for pure capillary-force-driven problems.

We use a multi-obstacle potential for $w\left(\vphi\right)$ of the form
\begin{align}
w\left(\vphi\right)=\left\{
\begin{array}{ll}
\dfrac{16}{\pi^{2}}\displaystyle \sum_{\substack{
\alpha, \beta = 1 \\
(\alpha < \beta)}}^{N, N}
\sigma_{\alpha \beta}\phi_{\alpha}\phi_{\beta}+\displaystyle \sum_{\substack{
\alpha, \beta, \gamma= 1 \\
(\alpha < \beta < \gamma)}}^{N, N, N}
\sigma_{\alpha \beta \gamma}\phi_{\alpha}\phi_{\beta}\phi_{\gamma}, \quad
& \textrm{if} \hspace{0.1cm} \vphi \in \sum\\
\infty, & \textrm{elsewhere}
\\
\end{array}
\right.
\label{FunctionW}
\end{align}
where $\sum=\{ \vphi \, | \sum_{\alpha=1}^{N}\phi_{\alpha} = 1$ and
$\phi_{\alpha} \geq 0 \}$,
$\sigma_{\alpha \beta}$ is the surface entropy density and $\sigma_{\alpha \beta
\gamma}$  is a term added to reduce the presence of unwanted third or higher
order phase at a binary interface (see below for details).

The gradient entropy density $a\left(\vphi,\nabla \vphi\right)$ can be written
as
\begin{align}
 a\left(\vphi,\nabla \vphi\right)=
\begin{array}{ll}
 \displaystyle \sum_{\substack{
\alpha, \beta = 1 \\
(\alpha < \beta)}}^{N,N}
\sigma_{\alpha \beta}\left[a_{c}\left(q_{\alpha \beta}\right)\right]^{2}\lvert
q_{\alpha \beta}\lvert^{2},
\end{array}
\end{align}
where $q_{\alpha \beta}=\left(\phi_{\alpha}\nabla
\phi_{\beta}-\phi_{\beta}\nabla \phi_{\alpha}\right)$ is a vector normal to the
$\alpha \beta$ interface. The function $a_{c}\left(q_{\alpha \beta}\right)$ 
describes the form of the anisotropy of the evolving phase boundary. For the 
present study, we assume isotropic interfaces, and 
hence $a_{c}\left(q_{\alpha \beta}\right) = 1$. Evolution
equations for $\vc$ and $\vphi$ are derived from the entropy functional through
conservation laws and phenomenological maximization of entropy, respectively
\cite{Garcke,Stinner}. A linearized temperature field with positive gradient 
$G$ in the growth direction ($z$ axis) is imposed and moved forward with a
velocity $v$,
\begin{align}
 T = T_{0} + G(z-vt)
\label{Tfield}
\end{align} 
where $T_{0}$ is the temperature at $z=0$ at time $t=0$. The
evolution equations for the phase-field variables read
\begin{align}
\omega \epsilon \partial_{t}\phi_{\alpha}=\epsilon \left(\nabla \cdot a,_{\nabla
\phi_{\alpha}}\left(\vphi,\nabla
\vphi\right)-a,_{\phi_{\alpha}}\left(\vphi,\nabla
\vphi\right)\right)-\dfrac{1}{\epsilon}w,_{\phi_{\alpha}}
\left(\vphi\right)-\dfrac{f,_{\phi_{\alpha}}(\vc,\vphi;T)}{T} - \Lambda,
\label{Evolution}
\end{align}
where $\Lambda$ is the Lagrange multiplier which maintains the constraint of 
Eq.~(\ref{constraint}) for $\vphi$, and the constant $\omega$ is the 
relaxation time of the phase fields. Furthermore,
$a,_{\nabla \phi_{\alpha}}$, $a,_{\phi_{\alpha}}$,
$w,_{\phi_{\alpha}}$ and $f,_{\phi_{\alpha}}$ indicate the derivatives of the
respective entropy densities with respect to $\nabla \phi_\alpha$ and
$\phi_\alpha$.
The function $f(\vc,\vphi;T)$  in Eq. (\ref{Evolution}) describes the
free energy density, and is related to the entropy density $s(\vc,\vphi;T)$,
through the relation $f(\vc,\phi;T) = e(\vc, \phi;T) - Ts(\vc,\vphi;T)$,
where $e(\vc, \phi;T)$ is the internal energy density. The free energy
density is given by the summation over all bulk free energy 
contributions $f_\alpha(\vc;T)$ of the individual phases in the system.
We use an ideal solution model,
\begin{align}
f(\vc,\vphi;T) = \sum_{i=1}^{K}\left(T c_{i}\ln c_{i} +
\sum_{\alpha=1}^{N}c_{i}L_{i}^{\alpha}\dfrac{\left(T-T_{i}^{\alpha}\right)}{T_{i
}^{\alpha}}
h_{\alpha}\left(\vphi\right)\right),
\label{freeen1}
\end{align}
where
\begin{align}
f_{\alpha}(\vc;T)=\sum_{i=1}^{K} \left( T c_{i}\ln c_{i}+ c_{i}L_{i}^{\alpha}\dfrac{\left(T-T_{i}^{\alpha}
\right)}{T_{i}^{\alpha}} \right)
 \label{freeen2}
\end{align}
is the free energy density of the $\alpha$ solid phase, and
\begin{align}
f_{l}(\vc;T)=T\sum_{i=1}^{K}\left(c_{i}\ln\left(c_{i}\right)\right)
\label{freeen3}
\end{align}
is the one of the liquid. The parameters $L_{i}^{\alpha}$ and 
$T_{i}^{\alpha}$ denote the latent heats and the melting temperatures 
of the $i^{th}$ component in the $\alpha$ phase, respectively. We
choose the liquid as the reference state, and hence $L_{i}^{l}=0$.

The function $h_\alpha(\vphi)$ is a weight function which we choose to be of 
the form $h_{\alpha}\left(\vphi\right)=\phi_{\alpha}^{2}\left(3-2\phi_{\alpha}\right)$. 
Thus, $f=f_\alpha$ for $\phi_\alpha=1$. Other interpolation functions involving
other components of the $\vphi$ vector could also be used, but here we restrict
ourselves to this simple choice.

The evolution equations for the concentration fields are derived from 
Eq.~(\ref{functional}),
\begin{align}
\partial_{t}c_{i}=-\nabla \cdot \left(M_{i0}(\vc,
\vphi)\nabla\dfrac{1}{T}+\sum_{j=1}^{K}M_{ij}\left(\vc, \vphi
\right)\nabla\left(\dfrac{1}{T}
\dfrac{\partial f(\vc,\vphi;T)}{\partial c_{j}}\right)\right).
\end{align}
By a convenient choice of the mobilities $M_{ij}\left(\vc, \vphi
\right)$, self- and interdiffusion
in multicomponent systems (including off-diagonal terms of the
diffusion matrix) can be modelled. Here, however, we limit ourselves
to a diagonal diffusion matrix with all individual diffusivities
being equal, which can be achieved by choosing
\begin{align}
M_{ij}(\vc, \vphi)=D_i(\vphi) c_{i}\left(\delta_{ij}-c_{j}\right) \\
M_{i0} (\vc, \vphi)= M_{0i}(\vc, \vphi) =
-\sum_{\alpha=1}^{N}\sum_{j=1}^{K}M_{ji}h_{\alpha}\left(\vphi\right)L_{i}^{
\alpha}.
\end{align}
The terms $M_{i0} (\vc, \vphi)= M_{0i}(\vc, \vphi)$ are the mobilities for the concentration current of the component $i$ due to a temperature gradient. The diffusion coefficient is taken as a linear interpolation between the
phases, $D_i(\vphi) = \sum_{\alpha=1}^{N} D_{i}^{\alpha} \phi_\alpha$, where
$D_{i}^{\alpha}$ is the non-dimensionalized diffusion coefficient of the $i^{th}$
component in the $\alpha$ phase, using the liquid diffusivity $D^l$ as the reference,
where the diffusivities of all the components in the liquid phase are assumed to be equal.
In the simulations we assume zero diffusivity in the solid, and take the effective
diffusivity to be $D_i(\vphi) = D^l \phi_l$.
The quantity $d^{*}= \sigma / \left(R/v_{m}\right)$ is used as the reference 
length scale in the simulations, where the molar volume $v_m$ is assumed
to be independent of the concentration. Here, $\sigma$ is one of the surface entropy density 
parameters introduced in Eq.~(\ref{FunctionW}), and the surface entropies of all 
the phases are assumed to be equal. The reference time scale is chosen to be
$t^* = {d^{*}}^{2}/D^l$. The temperature scale is the eutectic temperature
corresponding to the three phase stability regions at the three edges of 
the concentration simplex and is denoted by $T^*$ while the 
energy scale is given by $RT^*/v_{m}$.

\subsection{Relation to sharp-interface theory}

In order to compare our phase-field simulations to the theory
outlined in Sec.~2, we need to relate the parameters of the 
phase-field model to the quantities needed as input for the theory.
For some, this is straightforward. For example, all the parameters
of the phase diagram (liquidus slopes, coexistence temperatures 
etc.) can be deduced from the free energy densities of Eqs. (\ref{freeen1})--(\ref{freeen3}) 
in the standard way. For others, the correspondence is less immediate.
In the following, we will discuss in some detail two quantities
that are crucial for the theory: the surface free energies and
the latent heats, both needed to calculate the Gibbs-Thomson
coefficients in Eq.~(\ref{GibbsThomson}).

The surface free energy $\tilde\sigma_{\alpha\beta}$ is 
defined as the interface excess of the thermodynamic potential density 
that is equal in two coexisting phases.
For alloys, this is not the free energy, but the grand potential.
Indeed, the equilibrium between two phases is given by $K$ conditions
for $K$ components: $K-1$ chemical potentials (because of the constraint
of Eq.~(\ref{Eqn-Constraint}), only $K-1$ chemical potentials 
are independent) as well as
$f-\sum_{i=1}^{K-1} \mu_ic_i$, which is the grand potential, have to
be equal in both phases. This is the mathematical expression of the
common tangent construction for binary alloys and the common tangent
plane construction for ternary alloys.

The grand potential excess has several contributions. Since $f=e-Ts$,
we need to consider the entropy excess. Both the gradient term in 
the phase fields and the potential $w(\vphi)$ present in the entropy functional
give a contribution inside the interface. If, along an $\alpha\beta$ interface, 
all the other phase fields remain exactly equal to zero, then this contribution 
can be calculated analytically. However, this is generally not the case:
in the interface, the phase fields $\phi_\nu$, $\nu\ne\alpha,\beta$ 
can be different from zero, which corresponds to an ``adsorption'' 
of the other phases. Since the grand potential excess has to be
calculated along the equilibrium profile of the fields, the 
presence of extra phases modifies the value of $\tilde\sigma_{\alpha\beta}$.
The three-phase terms proportional to $\sigma_{\alpha\beta\gamma}$ have been
included in the potential function to reduce (or even eliminate)
the additional phases. However, the total removal of these phases 
requires to choose high values of $\sigma_{\alpha \beta \gamma}$. Such
high values (\textgreater 10 times the binary constant $\sigma_{\alpha\beta}$) 
cause the interface to become steeper near the regions of triple points
and lines in 2D and 3D, respectively, which is a natural consequence of the
fact that the higher order term affects only the points inside the phase-field 
simplex where three phases are present. The thinning of the interfaces 
leads to undesirable lattice pinning, which could only be circumvented
by a finer discretization. This, however, would lead to a large increase 
of the computation times. Therefore, if computations are to remain feasible,
we have to accept the presence of additional phases in the interfaces.

\begin{figure}[htpb]
 \begin{center}
  \includegraphics[width=8cm]{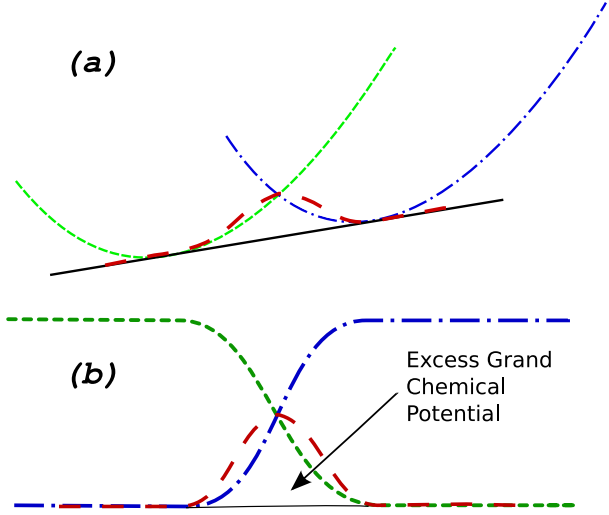}
\caption{(Color Online) Illustration of the existence of an excess interface energy 
contribution from the chemical free energy. Upper panel: the concentration
inside the interfacial region does not necessarily follow the common
tangent line. Here, the two convex curves are the free energy densities of
the individual phases in contact, the straight line is the common tangent,
and the thick non-monotonous line is the concentration along a cut through
the interface. Lower panel: the grand chemical potential in the interface
differs from the one obtained by a weighted sum of the bulk phase free energies,
where the weighting coefficients are the interpolating functions of 
the order parameters.}
\label{Figure6}
 \end{center}
\end{figure}

Furthermore, there is also a contribution due to the chemical part
of the free energy functional. This contribution, identified
for the first time in Ref.~\cite{Kim}, arises from the fact
that the concentrations inside the interface (which are fixed
by the condition of constant chemical potentials) do not, in
general, follow the common tangent plane, as illustrated 
schematically in Figure ~\ref{Figure6}. 

Therefore, there is
a contribution to the surface free energy which is given by the
following expressions.
For binary eutectic systems ($N = 3$ phases, $\vphi=\left(\phi_{\alpha},\phi_{\beta},\phi_{l}\right)$; $K=2$
components $\vc=\left(c_{A},c_{B}\right)$),
the vector $\vc$ is one-dimensional and we define the concentration 
($c_{A}$) to be the independent field $\vc=\left(c_{A},1-c_{A}\right)$. 
Then, we have
\begin{align}
{\Delta f}_{\textrm{chem}}\left(\vphi,\vc;T\right) = f\left(\vphi, \vc;
T\right) - f_{l} - \mu_{A}(T)\left(c_{A}-c^{l}_{A}\right),
\end{align}
where $\mu_{A}\left(T\right)=\dfrac{\partial
f\left(\vphi,\vc;T\right)}{c_{A}}$ is the chemical potential of component A.
For ternary eutectic systems ($N = 4$ phases, $\vphi =
\left(\phi_{\alpha},\phi_{\beta},\phi_{\gamma},\phi_{l}\right)$; $K=3$ components,
$\vc=\left(c_{A},c_{B},c_{C}\right)$),
the vector $\vc$ is two-dimensional and with
the concentrations of $A, B$ as the independent
concentration fields, we get $\vc=\left(c_{A},c_{B},1-c_{A}-c_{B}\right)$ 
and the chemical free energy excess becomes

\begin{multline}
{\Delta f}_{\textrm{chem}}\left(\vphi,\vc;T\right) = f\left(\vphi, \vc;
T\right) - f_{l} - \left(\mu_{A}(T)\right)\left(c_{A}-c^{l}_{A}\right)- \left(\mu_{B}\left(T\right)\right)\left(c_{B}-c^{l}_{B}\right).
\end{multline}

The entire surface excess can thus be written as the following
\begin{eqnarray}
\widetilde{\sigma}_{\alpha l} &=& \int_{x} \Big (T \epsilon a\left(\vphi,\nabla
\vphi\right) + \frac{T}{\epsilon}{w\left(\vphi\right)} + {\Delta f}_{\textrm{chem}}\left(\vphi,\vc;T\right)\Big) dx
\end{eqnarray}
where $x$ is the coordinate normal to the interface, and the
integral is taken along the equilibrium profile $\vphi(x)$, $\vc(x)$.
This integral cannot be calculated analytically. Therefore, we
determine the surface free energy numerically. To this end, we
perform one-dimensional simulations to determine the equilibrium
profiles of concentration and phase fields, and insert the
solution into the above formula to calculate $\tilde\sigma$.
For these simulations, the known bulk values of the concentration
fields are used as boundary conditions. To accurately calculate 
the surface excesses, it is important to include the contribution 
of the adsorbed phases. For this, the above calculations are performed 
by letting a small amount of these phases equilibrate at the interface of 
the major phases. Since the adsorbed phases equilibrate with very different 
concentrations compared to that of the bulk phases, the domain is chosen 
large enough such that the chemical potential change of the bulk phases 
during equilibration is kept negligibly low.

Another important quantity which is required as an input in the theoretical
expressions is the latent heat of fusion $L^{\alpha}$ of the $\alpha$
phase. We follow the thermodynamic definition for the latent heat of transformation $L^{\alpha}$,

\begin{eqnarray}
L^{\alpha} &=& T_{E} \left(s^{l} - s^{\alpha}\right), \\
\textrm{with} \hspace{0.25cm} s &=& -\left(\dfrac{\partial f\left(\vphi,\vc;T\right)}{\partial T}\right) \\
\textrm{and in particular} \hspace{0.25cm} s^{l} &=& \sum_{i=1}^{K}c^{l}_{i}ln\left(c^{l}_{i}\right)\\
\textrm{and} \hspace{0.25cm} s^{\alpha} &=&
\sum_{i=1}^{K}c^{\alpha}_{i}\dfrac{L_{i}^{\alpha}}{T_{i}^{\alpha}} +
c^{\alpha}_{i}ln\left(c^{\alpha}_{i}\right),
\end{eqnarray}
where the concentrations of the phases are taken from the phase diagram at the
eutectic temperature.

Finally, let us give a few comments on the interface mobility $\mu_{\rm int}$
that appears in Eq.~(\ref{GibbsThomson}). In early works {\cite{Caginalp}}, 
it was shown that an expression for this mobility in terms of the phase-field 
parameters can be easily derived in the sharp-interface limit in which the
interface thickness tends to zero. Later on, Karma and Rappel \cite{Karma96} 
proposed the thin-interface limit, in which the interface width remains 
finite, but much smaller than the mesoscopic diffusion length of the 
problem. This limit relaxes some of the stringent requirements of the 
sharp-interface method for the achievement of quantitative simulations. 
Additionally, this method introduces a correction term to the original 
expression for the interface mobility, which makes it possible to carry 
out simulations in the vanishing interface kinetics (infinite interface 
mobility) regime.

Clearly, such modifications of the interface kinetics are also present 
in our model, where they arise both from the presence of adsorbed phases 
in the interface and from the structure of the concentration profile 
through the interface. Furthermore, it is well known that solute 
trapping also occurs in phase-field models of the type used here 
\cite{Ahmad98}. Since the interface profile can only be evaluated 
numerically, and since several phase-field and concentration variables 
need to be taken into account, it is not possible to evaluate 
quantitatively the contribution of
these effects to the interface mobility. However, this lack of knowledge 
does not decisively impair the present study since we are mainly interested 
in undercooling versus spacing curves at a fixed interface velocity. At 
constant velocity, the absolute value of the interface undercooling 
contains an unknown contribution from the interface kinetics, but the 
relative comparison between steady states of different spacings remains 
meaningful. In addition, even though our simulation parameters correspond
to higher growth velocities than typical experiments, it will be seen below
that the value of the kinetic undercooling in our simulations is small.
This indicates once more that our comparisons remain consistent.

\section{Simulation results}

\begin{figure}[htpb]
\begin{center}
\includegraphics[width=8cm]{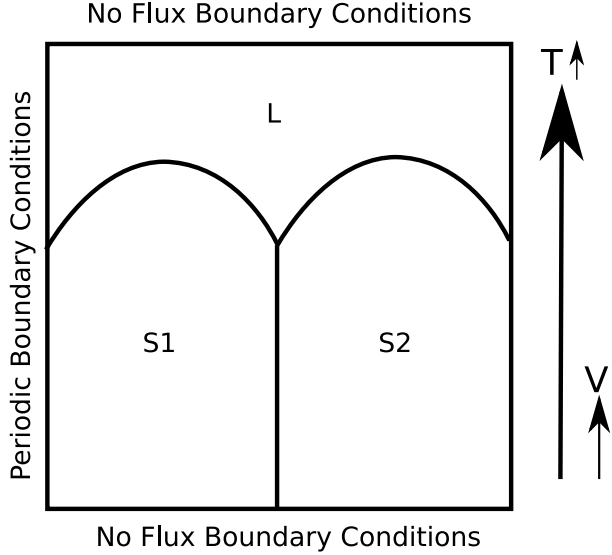}
\end{center}
\caption{Simulation setup for the phase-field simulations of binary 
and ternary eutectic systems. We impose a temperature gradient $G$
along the $z$ direction and move it with a fixed velocity. The 
average interface position follows the isotherms at steady 
state in case of stable lamellar coupled growth.}
\label{Figure7}
\end{figure}

In this section, we compare data extracted from phase-field simulations 
with the theory developed in Sec.\ref{sec:theory}, for the case of coupled 
growth of the solid phases in directional solidification.
The simulation setup is sketched in Figure \ref{Figure7}.
Periodic boundary conditions are used in the transverse direction, while 
no-flux boundary conditions are used in the growth direction. 
The box width in the transverse direction directly controls the
spacing $\lambda$. The box length in the growth direction is 
chosen several times larger than the diffusion length.
The diffusivity in the solid is assumed to be zero. 
A non-dimensional temperature gradient, G is imposed in the 
growth direction and moved with a velocity $v$, such that the
temperature field is given by Eq.~(\ref{Tfield}). 

The outline of this section is as follows: first,
we will briefly sketch how we extract the front undercooling from
the simulation data. Then, this procedure will be validated by
comparisons of the results to analytically known solutions as well
as to data for binary alloys, for which well-established benchmark
results exist. We start the presentation of our results on ternary
eutectics by a detailed discussion of the two simplest possible
cycles, $\alpha\beta\gamma$ and $\alpha\beta\alpha\gamma$. We
compare the data for undercooling as a function of spacing to our
analytical predictions and determine the relevant instabilities
that limit the range of stable spacings.
Finally, we also discuss the behavior of more complicated cycles,
for sequences up to length $M=6$.

\subsection{Data extraction}
\label{Data-extraction}
At steady state, the interface velocity matches the velocity
of the isotherms. The undercooling of the solid-liquid interface 
is extracted at this stage by the following procedure. First, a
vertical line of grid points is scanned until the interface is
located. Then, the precise position of the interface is determined 
as the position of the level line $\phi_\alpha=\phi_\beta$ for an
$\alpha\beta$-interface (and in an analogous way for all the
other interfaces). This is done by calculating the intersection of the 
phase-field profiles of the corresponding phases, which are extrapolated
to subgrid accuracy by polynomial fits. In the presence of adsorbed phases 
at the interface, the two major phases along the scan line are used 
for determining the interface point. The major phases are determined 
from the maximum values that a particular order parameter assumes along 
the scan line. The temperature at a calculated interface point is then 
given by Eq.~(\ref{Tfield}).

\begin{figure}[htbp]
\begin{center}
\parbox{8cm}{
\includegraphics[width=10cm]{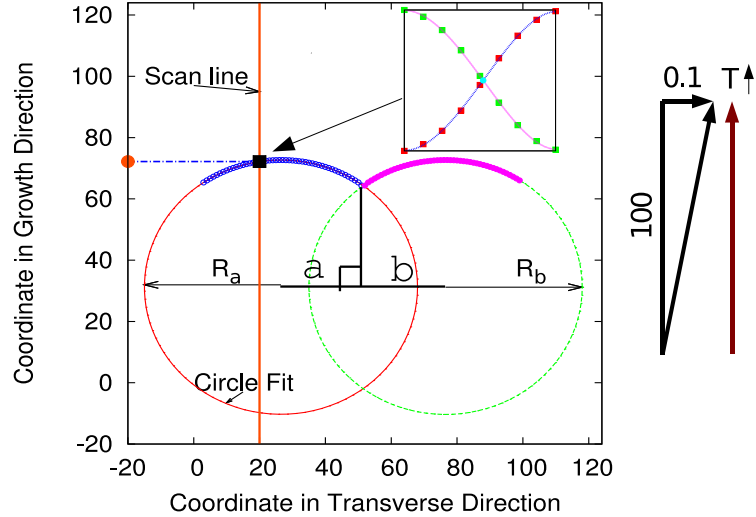}
}
\caption{(Color Online) Procedure to extract the interface points from the simulation data
with sub-grid resolution using higher order interpolation of the phase-field 
profiles. For the evaluation of the equilibrium properties, the solid-liquid 
interface points of each lamella are fitted with a circle which is then
used to measure the radius of curvature of the particular lamella. We also
calculate the triple point angles as the angles between the tangents to 
the circles at one of the points of intersection.}
\label{Figure8}
\end{center}
\end{figure}

In order to test both our data extraction methods and our calculations 
of the surface tensions, we have performed the following consistency
check. For an alloy with a symmetric phase diagram at the eutectic
concentration, a lamellar front has an equilibrium position when a 
small temperature gradient ($G=0.001$) is applied to the system at 
zero growth speed. Since the concentration in the liquid is uniform 
for a motionless front, according to the Gibbs-Thomson relation the 
interface shapes should just be arcs of circles. This was indeed the case
in our simulations, and the fit of the interface shapes with circles
has allowed us to obtain the interface curvature and the contact angles
with very good precision. The extraction of the data is illustrated
in Figure \ref{Figure8}.

We fit the radius and the coordinates of the
circle centers. Then, the angle at the trijunction point $\theta$ is 
deduced from geometrical relations, with
$d = a + b$ and $a = \dfrac{R_a^2 - R_b^2 + d^{2}}{2d}$,
\begin{align*}
b = d-a\\
\theta =
\cos^{-1}\left(\dfrac{a}{R_a}\right)+\cos^{-1}\left(\dfrac{b}{R_b}
\right).
\end{align*}
The meaning of the lengths $a$ and $b$ is given in  
Figure \ref{Figure8}.

\subsection{Validation: Binary Systems}
\begin{figure}[htpb]
\begin{center}
\subfigure[]{
\includegraphics[width=7cm]{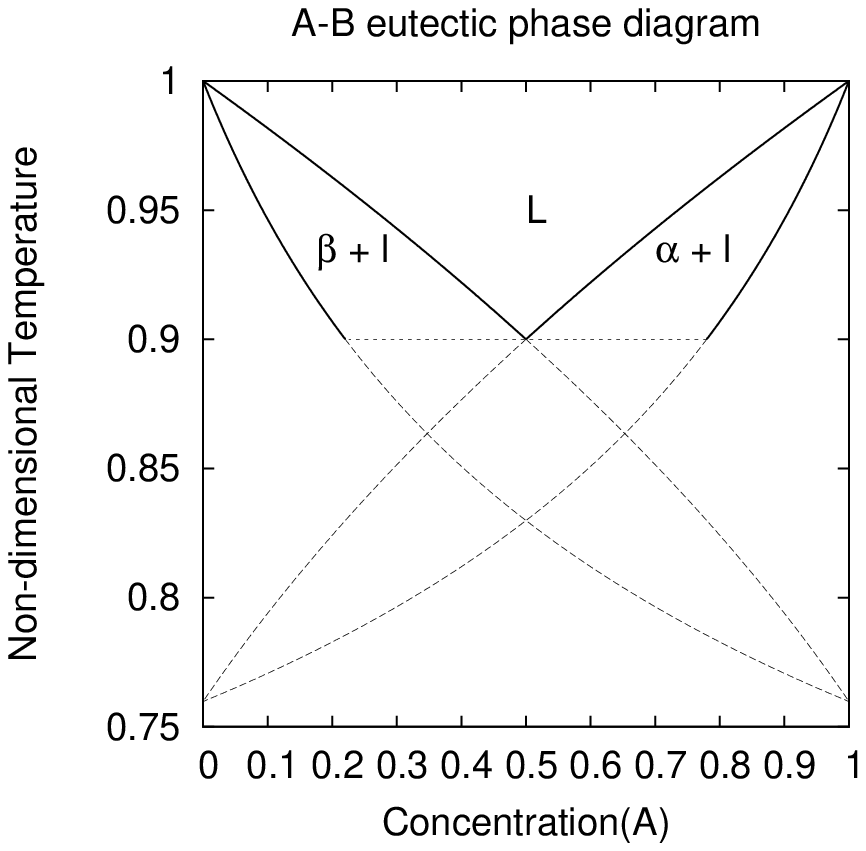}
\label{Figure9a}
}
\subfigure[]{
\includegraphics[width=7cm]{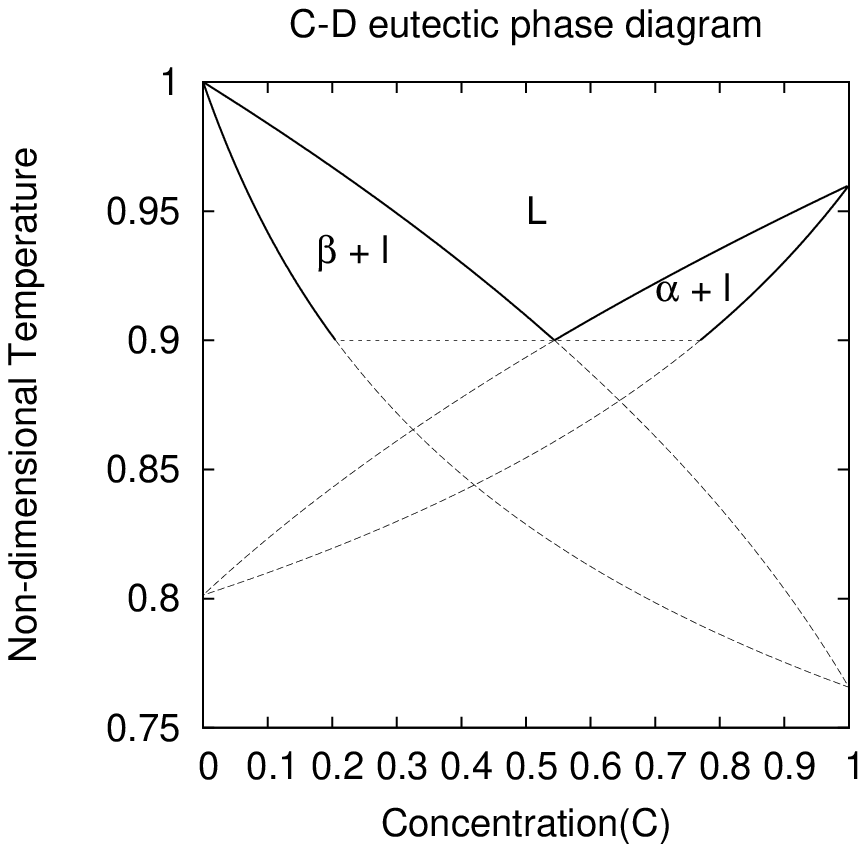}
\label{Figure9b}
}
\caption{Binary eutectic phase diagrams for a model system with stable (solid lines) and metastable (light dashed lines)
extensions of the solidus and the liquidus lines, of (a) a symmetric A-B and (b)
an unsymmetric C-D system.}
\label{Figure9}
\end{center}
\end{figure}

For comparison with the $\Delta T - \lambda$ relationship known
from Jackson-Hunt(JH) theory, we create two binary eutectic systems by
choosing suitable parameters $L_{i}^{\alpha}$ and $T_{i}^{\alpha}$
in the free energy density $f\left(\vphi,\vc;T\right)$.
A symmetric binary eutectic system, shown in Figure \ref{Figure9a},
is created by
\begin{center}
\begin{minipage}{10cm}
\parbox{4cm}{
\begin{displaymath}
L_{i}^{\alpha} = \left(\begin{array}{lll}
       &  A  & B  \\
\alpha & 4.0 & 4.0 \\ 
\beta  & 4.0 & 4.0
\end{array}\right)
\end{displaymath}
}
\hspace{0.2cm}
\parbox{4cm}{
\begin{displaymath}
T_{i}^{\alpha} = \left(\begin{array}{lll}
       & A   & B   \\
\alpha & 1.0 & 0.75980 \\ 
\beta  & 0.75980 & 1.0
\end{array}\right).
\end{displaymath}
}
\end{minipage}
\end{center}

To create an asymmetric binary eutectic system, shown in Figure \ref{Figure9b},
we choose
\begin{center}
\begin{minipage}{10cm}
\parbox{4cm}{
\begin{displaymath}
L_{i}^{\alpha} = \left(\begin{array}{lll}
       &  C  & D  \\
\alpha & 5.0 & 5.0 \\ 
\beta  & 5.0 & 5.0
\end{array}\right)
\end{displaymath}
}
\hspace{0.2cm}
\parbox{4cm}{
\begin{displaymath}
T_{i}^{\alpha} = \left(\begin{array}{lll}
       & C   & D   \\
\alpha & 0.96 & 0.80137 \\ 
\beta  & 0.76567 & 1.0
\end{array}\right)
\end{displaymath}
}
\end{minipage}

\end{center}

The numbers $L_{i}^{\alpha}$, $T_{i}^{\alpha}$ are chosen such that the widths
of each of the (lens-shaped) two-phase coexistence regions remain reasonably broad, 
and that the approximation of using the values of concentration difference between the
solidus and liquidus $\left(\Delta c_{\nu}^{l}\right)$ at the eutectic 
temperature for the theoretical expressions holds for a good range of 
undercoolings. This implies that the value of the $L_{i}^{\alpha}$ should not be 
too small. Conversely, a too high value is also not desirable since for large
values of $L_{i}^{\alpha}$ the chemical contribution to the surface free energy
becomes large, which leads to very steep and narrow interface profiles.

\begin{table}[!ht]
\caption{\label{Jacktab0604}Parameters for the sharp-interface theory, with proper calculation 
of the surface tension in the phase-field simulations for (a) a symmetric 
binary eutectic system with components A and B and (b) for an unsymmetric binary eutectic
system with components C and D.}
\label{JacksonBin}
\begin{center}
(a) \parbox{5cm} {
\begin{tabular}{|c|c|}
\hline
$\widetilde{\sigma}_{\alpha l}$ & 1.01146 \\
\hline
$\widetilde{\sigma}_{\beta l}$ & 1.01146 \\
\hline
$\widetilde{\sigma}_{\alpha \beta}$ & 1.23718 \\
\hline
$\theta_{\alpha \beta}$ & 37.70 \\
\hline
$\theta_{\beta \alpha}$ & 37.70 \\
\hline
$L^{\alpha}$ & 4.0 \\
\hline
$L^{\beta}$ & 4.0 \\
\hline
$m_{B}^\alpha=m_{A}^\beta$ & -0.206975 \\
\hline
\end{tabular}
}
\label{Jacktab0505}
\hspace{0.25cm}
(b) \parbox{5cm} {
\begin{tabular}{|c|c|}
\hline
$\widetilde{\sigma}_{\alpha l}$ & 0.97272 \\
\hline
$\widetilde{\sigma}_{\beta l}$ & 1.07235 \\
\hline
$\widetilde{\sigma}_{\alpha \beta}$ & 1.24836 \\
\hline
$\theta_{\alpha \beta}$ & 33.903 \\
\hline
$\theta_{\beta \alpha}$ & 41.161 \\
\hline
$L^{\alpha}$ & 4.686 \\
\hline
$L^{\beta}$ & 4.711 \\
\hline
$m_{D}^{\alpha}$ & -0.13161 \\
\hline
$m_{C}^{\beta}$ & -0.22138 \\
\hline
\end{tabular}
}
\end{center}
\end{table}

We perform simulations at two different velocities V = 0.01 and V = 0.02,
with a mesh size $\Delta x = 1.0$ and the parameter set 
$ \epsilon = 4.0, D^{l}_{A} = D^{l}_{B} = D^{l}_{C} = D^{l}_{D} = 1.0, 
\sigma_{\alpha \beta} = \sigma_{\alpha l} = \sigma_{\beta l} = 1.0, 
\sigma_{\alpha \beta \gamma} = 10.0$. To give an idea of the order
of magnitude of the corresponding dimensional quantities, we remark that 
if we assume the melting temperatures to be around 1700K and the other 
values to correspond to the Ni-Cu system used in the study of Warren 
{\it et al.} \cite{Warren}, the length scale $d^*$ for the case of the 
binary eutectic system turns out to be around 0.2 nm and the time scale 0.04 ns.

\begin{figure}[htbp]
\begin{center}
\subfigure[]{
\includegraphics[width=6cm]{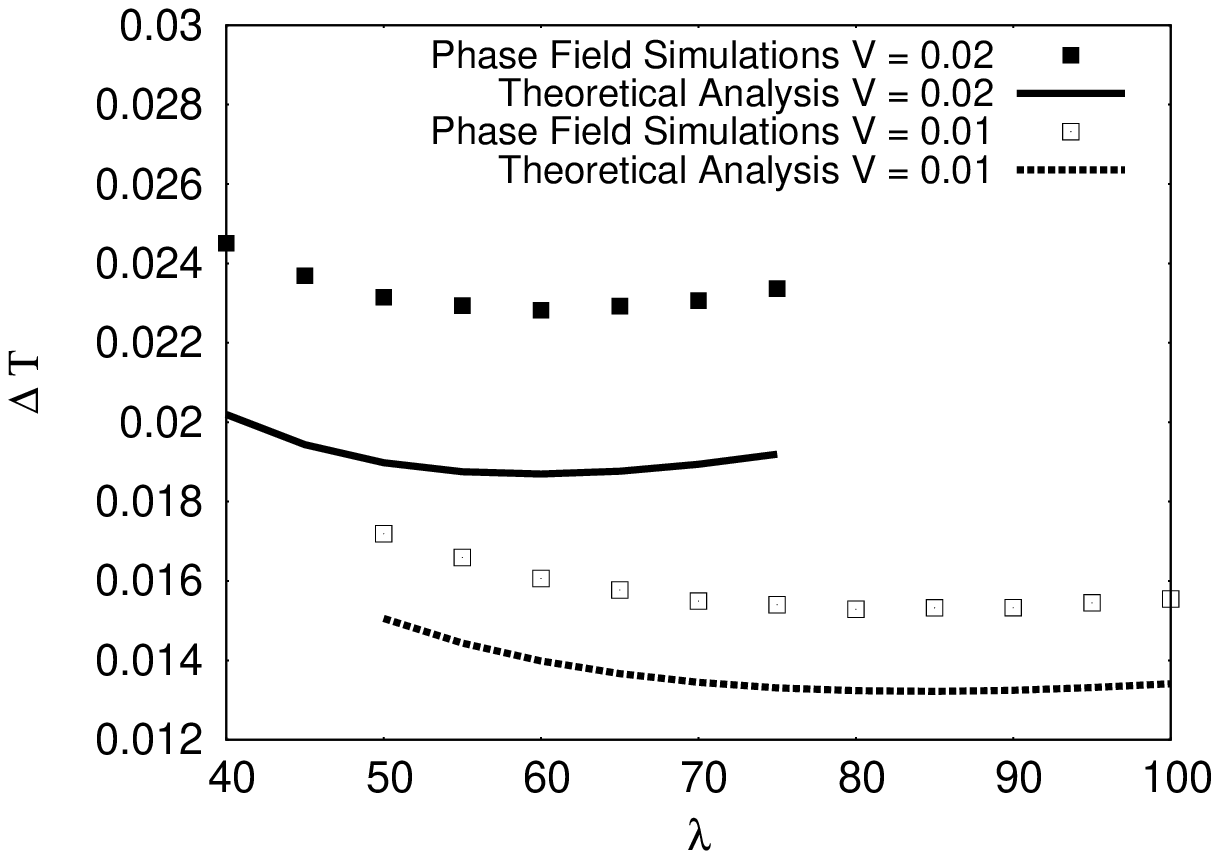}
\label{Figure10a}
}
\subfigure[]{
\includegraphics[width=6cm]{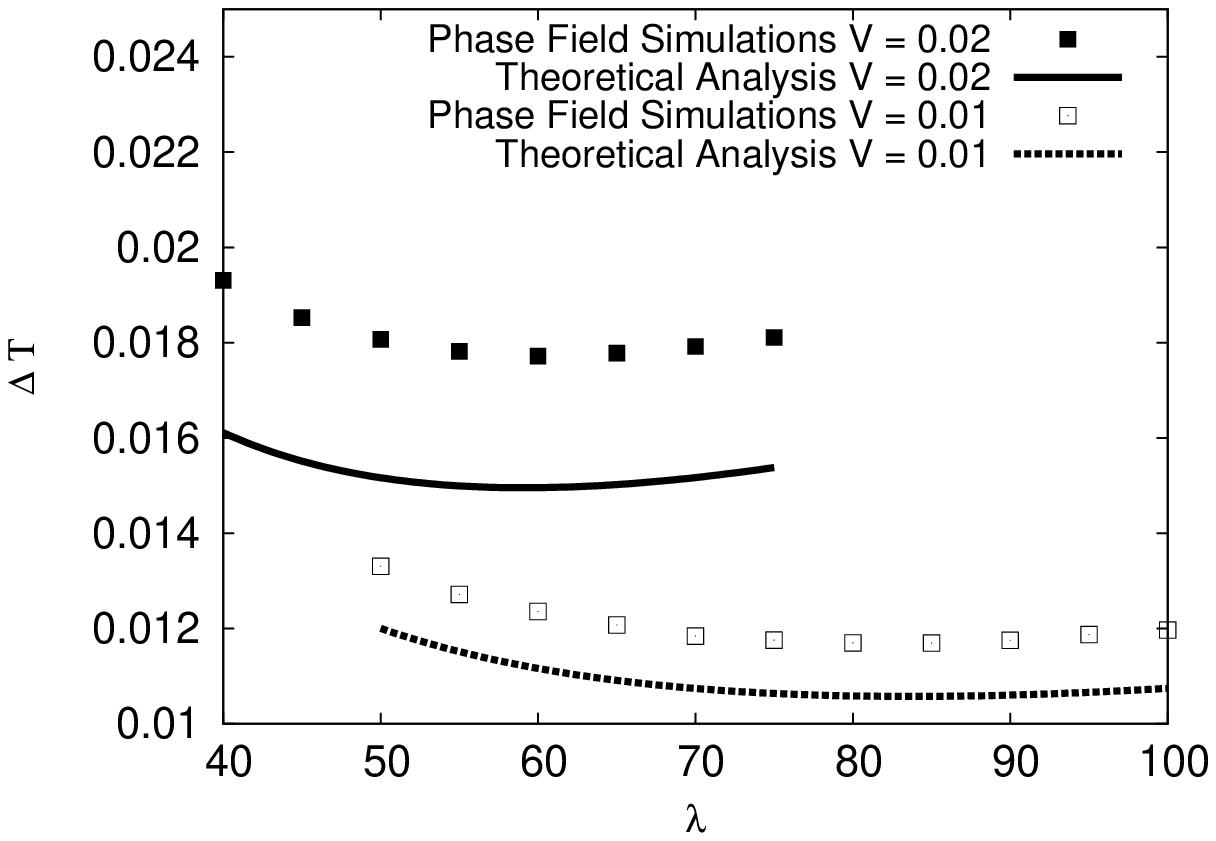}
\label{Figure10b}
}
\caption{Comparison of $\Delta T-\lambda$ relations resulting from the theoretical
analysis and from the phase-field simulations at two different velocities for systems;
(a) symmetric binary eutectic system (A-B) and (b) unsymmetric binary eutectic
system (C-D).}
\label{Figure10}
\end{center}
\end{figure}

The corresponding parameters for the sharp-interface theory are given in
Table~\ref{JacksonBin}. The comparisons between our numerical results and
the analytic theory are shown in Figs.~\ref{Figure10a} and~\ref{Figure10b}.

Consistent differences can be observed in the undercooling values 
between our data and the predictions from JH theory for both systems. 
The difference in undercoolings is smaller at lower velocities, which hints
at the presence of interface kinetics. We find indeed that when 
we change the relaxation constant in the phase-field evolution 
equation by about 50 \%, the difference between the predicted and 
measured undercoolings is removed for the case of the considered 
symmetric binary phase diagram. This clearly shows that the interface
kinetics is not negligible. It seems difficult, however, to obtain a
precise numerical value for its magnitude in the framework of the
present model.

The spacing at minimum undercooling, however, is reproduced to a good 
degree of accuracy (error of 5 \%), while the minimum undercooling has a
maximum error of 10 \%. It should also be noted that the JH theory only
is an approximation for the true front undercooling. Results obtained
both with boundary integral \cite{Sarkissian} and quantitative phase-field methods
\cite{Plapp} have shown that, whereas the prediction for the minimum 
undercooling spacing is excellent, errors of 10 \% for the value of
the undercooling itself are typical. If the JH curve is drawn without 
taking into account the additional chemical contributions to the surface 
tension, a completely different result is obtained, with minimum undercooling
spacings that are largely different from the simulated ones. We can
therefore conclude that we have captured the principal corrections.

In addition, we have performed equilibrium measurements of the angles 
at the trijunction point and of the radius of curvature of the lamellae as
described in the preceding sub-section (\ref{Data-extraction}) 
for the symmetric eutectic system. 
The contact angles differ from the ones 
predicted by Young's equilibrium conditions only by a 
value of 0.2 degrees. 
The theoretical (from the Gibbs-Thomson equation) and measured 
undercoolings differ in the third decimal, with an error of 0.1 \%.

\subsection{Ternary Systems: Parameter set}

\begin{minipage}{16cm}
 \parbox{9cm}{
\begin{displaymath}
L_{i}^{\alpha} = \left(\begin{array}{llll}
       & A         & B          &  C       \\
\alpha & 1.46964038 & 1.0        &  1.0     \\ 
\beta  & 1.0        & 1.46964038 &  1.0     \\
\gamma & 1.0        & 1.0        &  1.46964038
\end{array}\right)
\end{displaymath}
}\\
\parbox{5cm}{
\begin{displaymath}
T_{i}^{\alpha} = \left(\begin{array}{llll}
       & A   & B   &  C       \\
\alpha & 1.5 & 0.5 &  0.5     \\ 
\beta  & 0.5 & 1.5 &  0.5     \\
\gamma & 0.5 & 1.0 &  1.5
\end{array}\right).
\end{displaymath}
}
\end{minipage}

We use a symmetric ternary phase diagram. The following matrices list the parameters $L_{i}^{\alpha}$,$T_{i}^{\alpha}$ in the free energy $f\left(\vphi,\vc;T\right)$ that 
were used to create a symmetric ternary eutectic system, shown in Figure \ref{Figure1}.
We perform simulations with the parameter set $\epsilon = 8.0, \Delta x = 1.0, D^{l}_{A} =
D^{l}_{B} = D^{l}_{C} = 1.0 , \sigma_{\alpha \gamma} = \sigma_{\beta \gamma} =
\sigma_{\gamma \beta} = \sigma_{\alpha l} = \sigma_{\beta l} = \sigma_{\gamma l}
= 1.0, \sigma_{\alpha \beta l} = \sigma_{\alpha \beta \gamma} = 
\sigma_{\alpha \gamma l} = \sigma_{\beta \gamma l} =10.0$ and compare with the 
theoretical expressions using the input parameters listed in
Table~\ref{Jacktabtern}.

\begin{table}[!ht]
\begin{center}
\caption{Input parameters for the theoretical relations 
for the ternary eutectic system.}
\begin{tabular}{|c|c|}
\hline
$\widetilde{\sigma}_{\alpha l} = \widetilde{\sigma}_{\beta l} = \widetilde{\sigma}_{\gamma l}$ & 1.194035 \\
\hline
$\widetilde{\sigma}_{\alpha \beta} = \widetilde{\sigma}_{\alpha \gamma} = \widetilde{\sigma}_{\beta \gamma}$ &
1.430923\\
\hline
$\theta_{\alpha \beta} = \theta_{\beta \alpha} = \theta_{\gamma \alpha}$ = $\theta_{\alpha \gamma}$ & 36.81 \\
\hline
$L^{\alpha} = L^{\beta} = L^{\gamma}$ & 1.33 \\
\hline
$m_{B}^\alpha=m_{C}^{\alpha}$ & -0.91 \\
\hline
$m_{A}^{\beta} = m_{C}^{\beta}$ & -0.91 \\
\hline
$m_{A}^{\gamma} = m_{B}^{\gamma}$ & -0.91 \\
\hline
\end{tabular}
\label{Jacktabtern}
\end{center}
\end{table}

\subsection{Simple cycles: steady states and oscillatory instability}

\begin{figure}[htbp]
\begin{center}
\includegraphics[width=8cm]{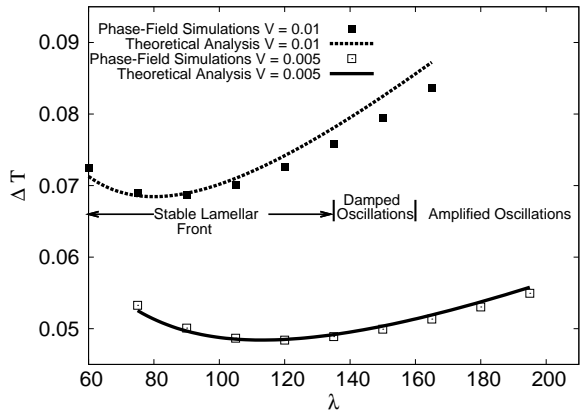}
\caption{Comparison between theoretical analysis and phase-field simulations
at two different velocities for the arrangement $\left(\alpha \beta
\gamma\right)$ of ternary eutectic solids at $V=0.005$ and $V=0.01$. 
The demarcation shows the regions of stable lamellar growth and the 
critical spacing beyond which we observe amplified oscillatory behavior. 
There is a small region named ``Damped Oscillations'', which is a region 
where oscillations occur but die down slowly with time.}
\label{Figure11}
\end{center}
\end{figure}

We first perform simulations to isolate the regime of stable lamellar growth 
for the configuration $\alpha \beta \gamma$. For this regime, we measure the average 
interface undercooling and compare it to our
theoretical predictions. The results are shown in Figure \ref{Figure11}.

The agreement in the undercoolings is much 
better than for the binary eutectic systems, with a smaller dependence of 
undercoolings on the velocities. Consequently, both the spacing at minimum
undercooling (error ~4 \% for V=0.005 and ~6 \% for V=0.01) and the minimum
undercooling (error of 1-2 \%), match very well with the theoretical
relationships, as shown in Figure \ref{Figure11}. The 
equilibrium angles at the triple point also agree with the ones 
predicted from Young's law to within an error of 0.3 degrees, 
while the radius of curvature matches that from the  
Gibbs-Thomson relationship with negligible error (\textless
0.5 \%).

It should be noted that the steady lamellae remain straight, 
contrary to the results of Ref.~\cite{Hecht}, where
a spontaneous tilt of the lamellae with respect to the direction
of the temperature gradient was reported. This difference is due
to the different phase diagrams: we are using a completely
symmetric phase diagram and equal surface tensions for all
solid-liquid interfaces, whereas \cite{Hecht} uses the thermophysical 
data of a real alloy.

\begin{figure}[htbp]
 \begin{center}
\includegraphics[width=8cm]{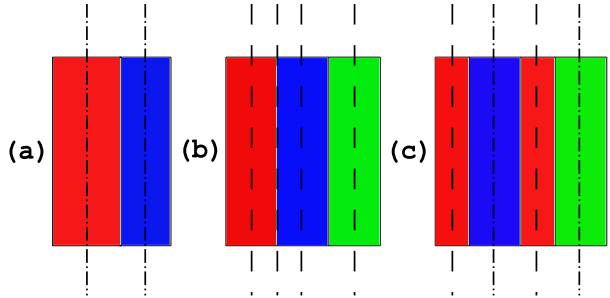}
\caption{(Color Online) In a periodic arrangement of lamellae, we can identify certain 
lines of symmetry, as shown in (a) for a binary eutectic. Similarly, 
for the case of the two simplest configurations, (b) $\alpha \beta \gamma$  
and (c) $\alpha \beta \alpha \gamma$ in a symmetric ternary eutectic 
system, such planes of symmetry exist. While in the case of a binary 
eutectic, the lines are mirror symmetry axes (shown by dash-dotted lines), 
in the special case of a symmetric ternary phase diagram, one can also 
identify quasi-mirror lines (dashed lines) where we retrieve the original 
configuration after a spatial reflection and an exchange of two phases. 
Only quasi-mirror lines exist in the $\alpha \beta \gamma$ arrangement, 
which are shown in (b), while both true- and quasi-mirror planes exist in 
the $\alpha \beta \alpha \gamma$ arrangement as shown in (c).}
\label{Figure12}
 \end{center}
\end{figure}

Next, we are interested in the stability range of three-phase coupled growth.
From general arguments, we expect a long-wavelength lamella elimination
instability (Eckhaus-type instability) to occur for low spacings, as
in binary eutectics \cite{Akamatsu04}. Here, we will focus on oscillatory
instabilities that occur for large spacings. It is useful to first recall 
a few facts known about binary eutectics, where all the instability modes
have been classified \cite{Sarkissian,Akamatsu}. Lamellar arrays in binary
eutectics are characterized (in the absence of crystalline anisotropy) by 
the presence of two mirror symmetry planes that run in the center of each
type of lamellae, as sketched in Figure \ref{Figure12}(a). Instabilities 
can break certain of these symmetries while other symmetry elements 
remain intact \cite{Coullet}. In binary eutectics, the oscillatory 1-$\lambda$-O 
mode is characterized by an in-phase oscillation of the thickness of all 
$\alpha$ (and $\beta$) lamellae; both mirror symmetry planes remain in 
the oscillatory pattern. In contrast, in the 2-$\lambda$-O mode, one 
type of lamellae start to oscillate laterally, whereas the mirror plane 
in the other type of lamellae survives; this leads to a spatial period doubling. 
Finally, in the tilted pattern both mirror planes are lost.

It is therefore important to survey the possible symmetry elements in the
ternary case. At first glance, there seems to be no symmetry plane in the
pattern. However, for our specific choice of phase diagram, new symmetry
elements not present in a generic phase diagram exist:
mirror symmetry planes combined with the exchange of two phases.
Consider for example the $\beta$ phase in the center of Figure \ref{Figure12}(b): 
if the system is reflected at its center, and then the $\alpha$ and $\gamma$
phases are exchanged, we recover the original pattern. At the eutectic
concentration, there are three such symmetry planes running in the center 
of each lamella, and three additional ones running along the three 
solid-solid interfaces. Off the eutectic point, two of these planes 
survive if any two of the three phases have equal volume fractions.

\begin{figure}[htbp]
 \begin{center}
  \subfigure[]{
   \includegraphics[height=3cm]{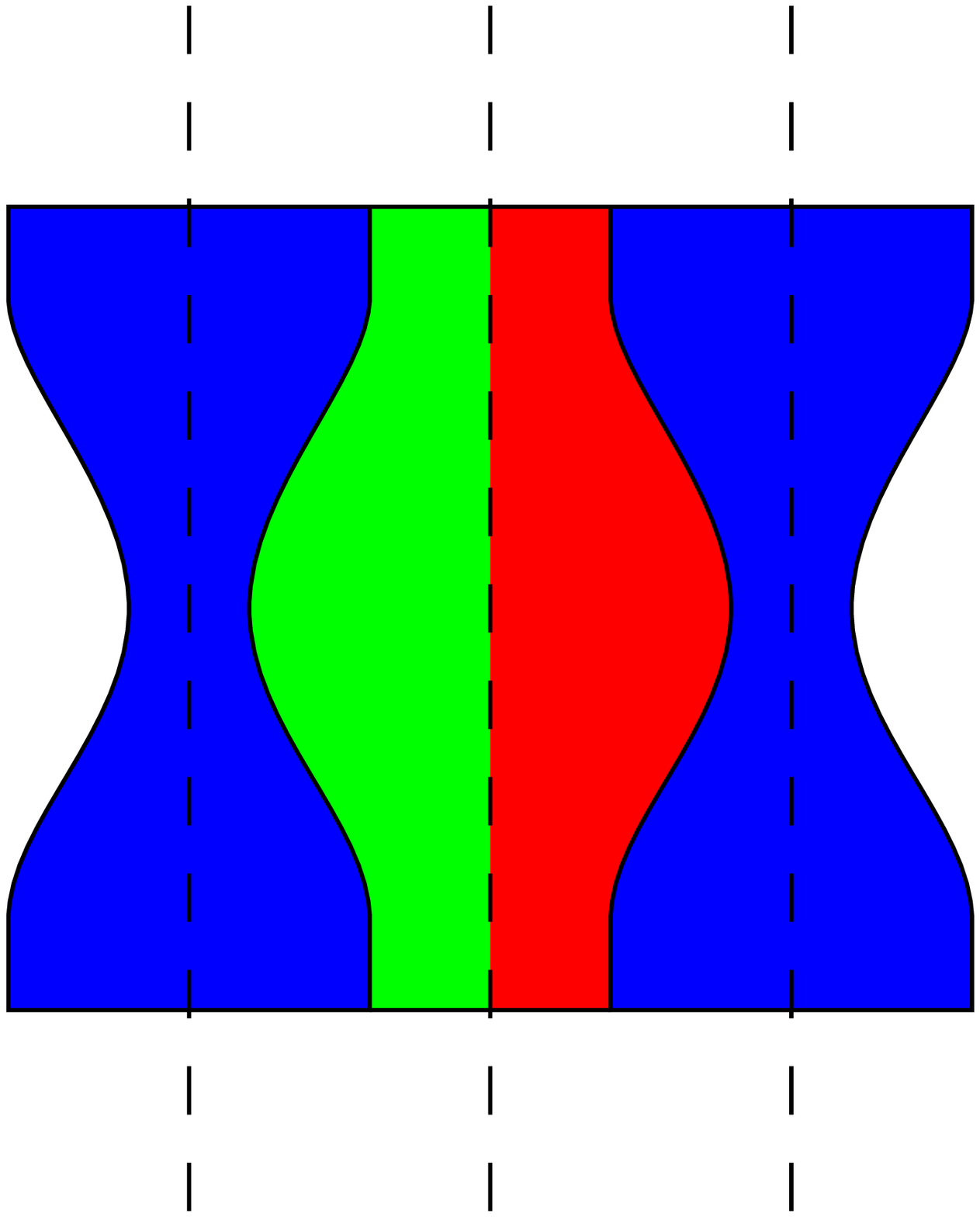}
  \label{Figure13a}
  }
  \subfigure[]{
   \includegraphics[height=3cm]{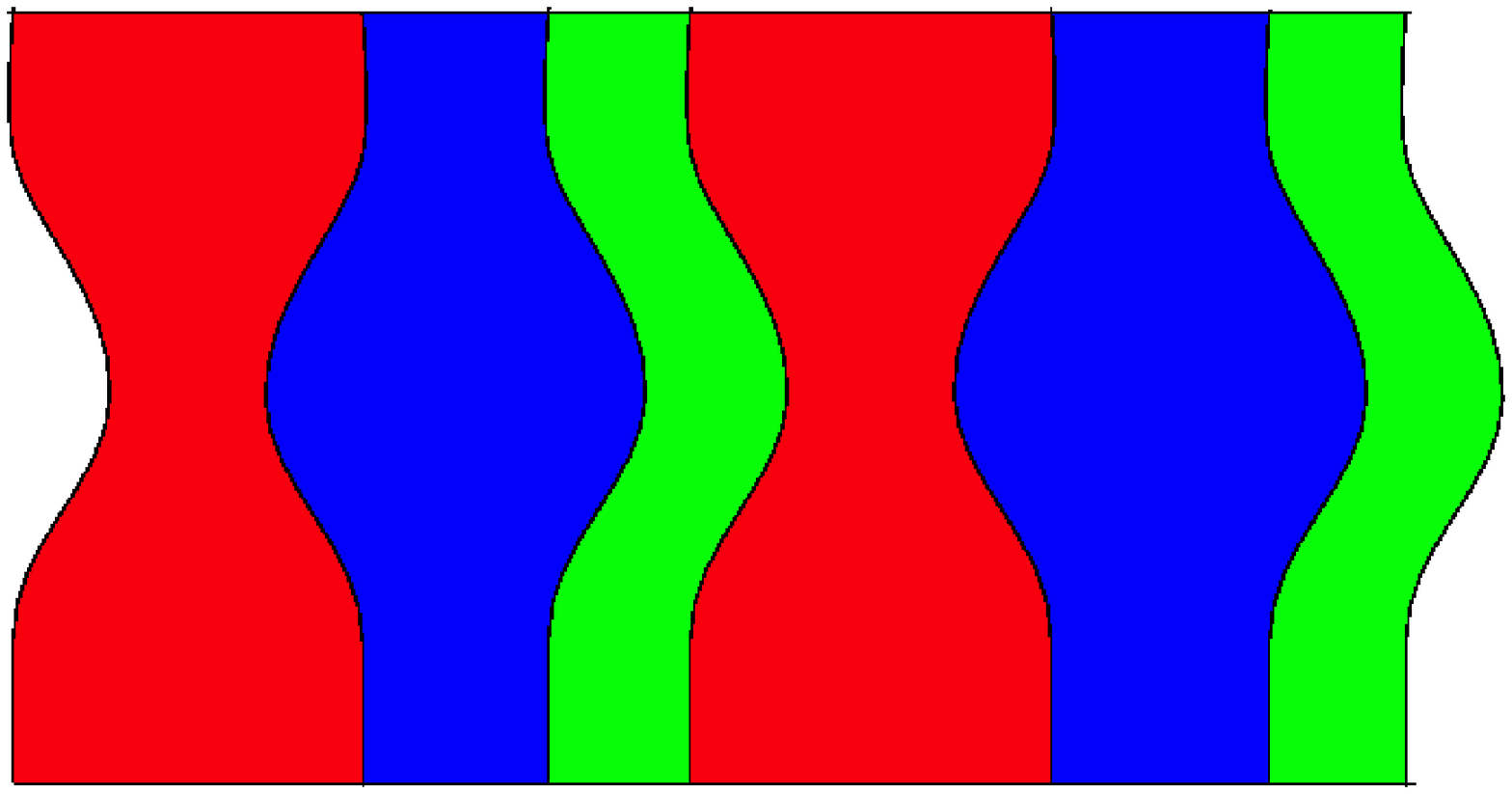}
   \label{Figure13b}
  }
\caption{(Color Online) Guided by the symmetry axes in the $\alpha \beta \gamma$ arrangement, 
one can expect two possible oscillatory modes at off-eutectic concentrations 
along the eutectic groove. The oscillations in (a), which keep all the 
quasi-mirror planes intact, are expected to occur at concentrations towards 
the apex of the simplex along the eutectic groove. Another possibility, shown
in (b) exists in which no symmetry plane remains, which is expected to 
occur at a concentration towards the binary edge of the simplex.}
\label{Figure13}
 \end{center}
\end{figure}

Guided by these considerations, we can conjecture that there are two obvious 
possible instability modes, sketched in Figure \ref{Figure13}.

In the first, called mode 1 in the following, 
two symmetry planes survive: the width of one lamella
oscillates, whereas the two other phases form a ``composite lamella'' that
oscillates in opposition of phase; the interface in the center of this
composite lamella does not oscillate at all and constitutes one of the
symmetry planes. In the second (mode 2), the lateral position of one of 
the lamellae oscillates with time, whereas the other two phases oscillate
in opposition of phase to form a ``composite lamella'' that oscillates
laterally but keeps an almost constant width. There is no symmetry plane 
left in this mode.

\begin{figure}[htbp]
 \begin{center}
  \subfigure[]{
      \includegraphics[height=5cm]{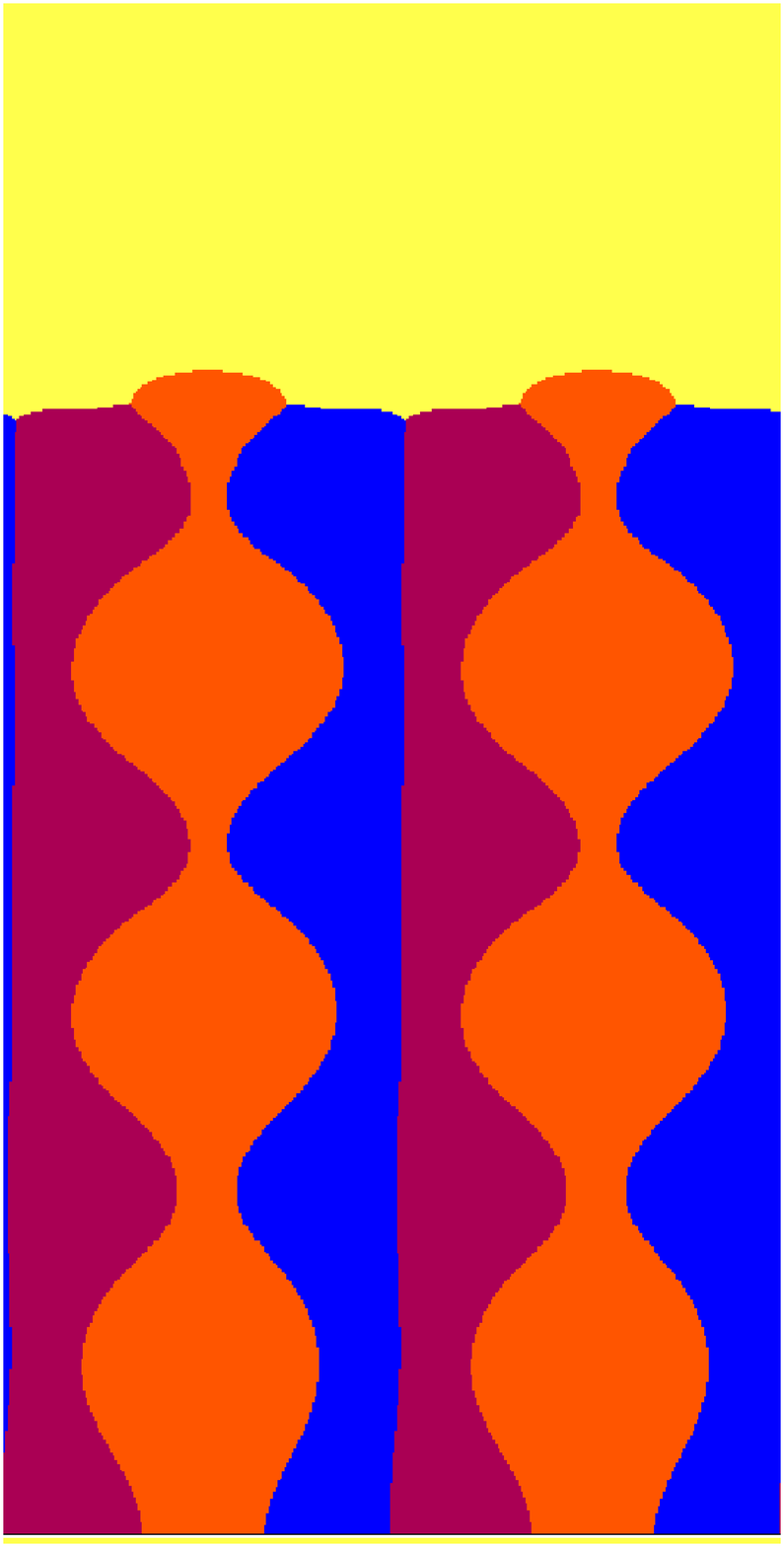}
      \label{Figure14a}
     }
  \subfigure[]{
      \includegraphics[height=5cm]{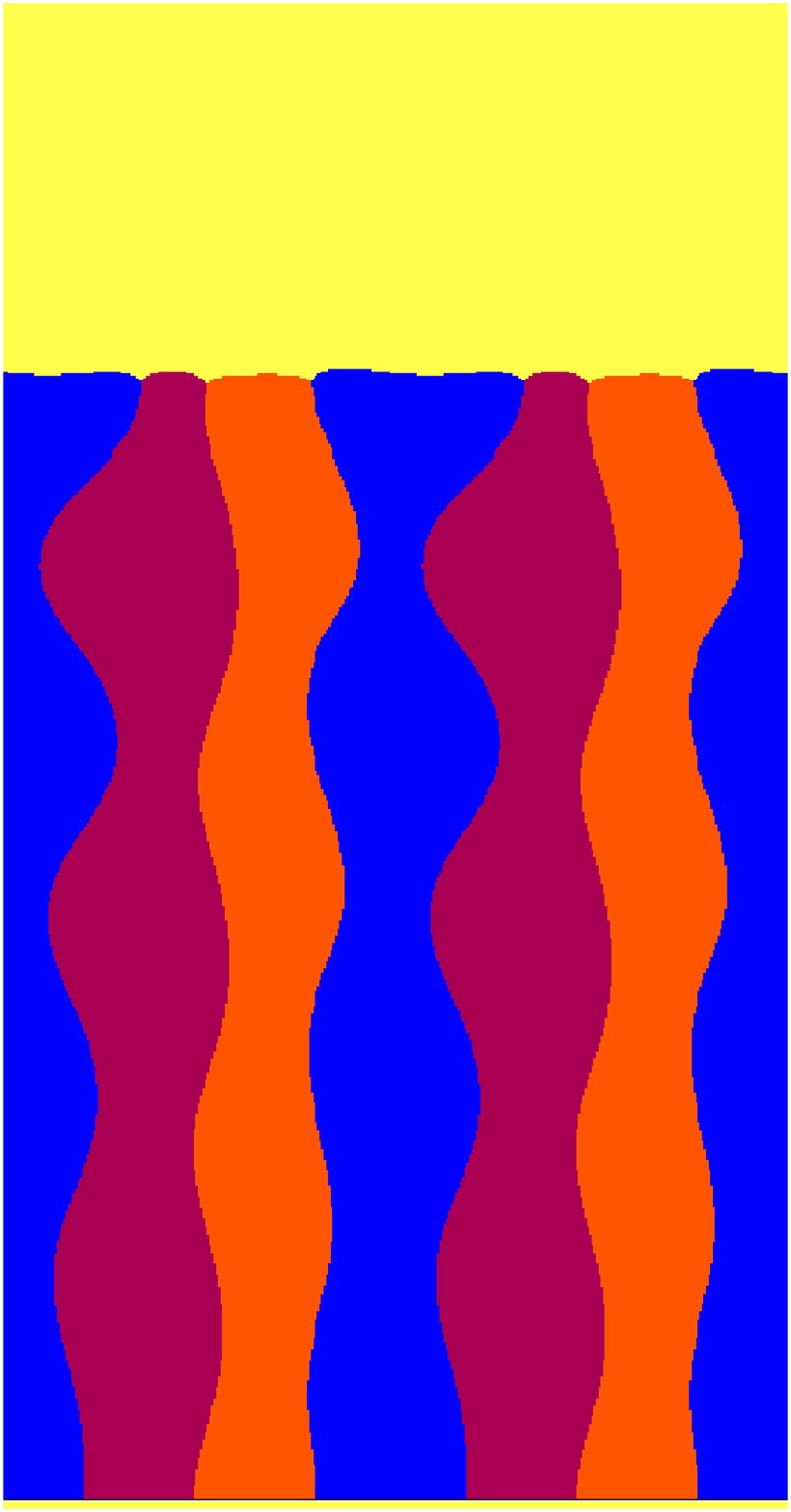}
      \label{Figure14b}
     }
  \subfigure[]{
      \includegraphics[height=5cm]{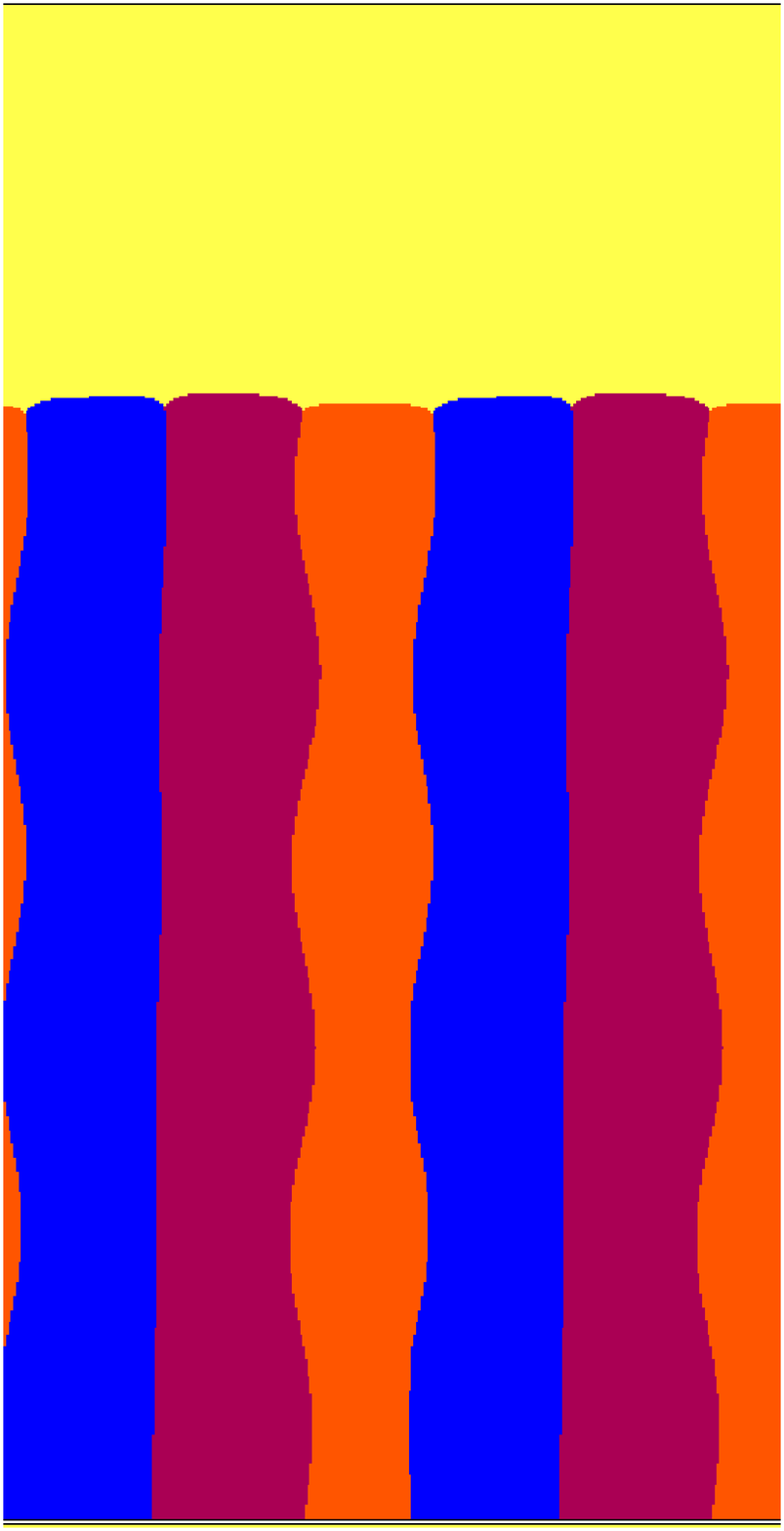}
      \label{Figure14c}
     }
\caption{(Color Online) Oscillatory modes in simulations for the $\alpha \beta \gamma$ configuration at the off-eutectic concentrations $\vc=\left(0.32,0.32,0.36)\right)$ in (a), and $\vc=\left(0.34,0.34,0.32\right)$ in (b) and (c). The spacings are $\lambda=170$ in (a) and (c) and $\lambda=165$ in (b).}
\label{Figure14}
\end{center}
\end{figure}

The stability range of the coupled growth regime of the lamellar 
arrangement is indicated in Figure \ref{Figure11}. Steady 
lamellar growth is stable from below the minimum undercooling 
spacing up to a point where an oscillatory instability occurs.
In the region marked ``damped oscillations'', oscillatory motion
of the interfaces was noticed, but died out with time. Above a threshold in
spacing, oscillations are amplified. We monitored the modes that emerged,
and found indeed good examples for the two theoretically expected
patterns, shown in Figure \ref{Figure14}. 

\begin{figure}[htbp]
 \begin{center}
  \subfigure[]{
   \includegraphics[height=6cm]{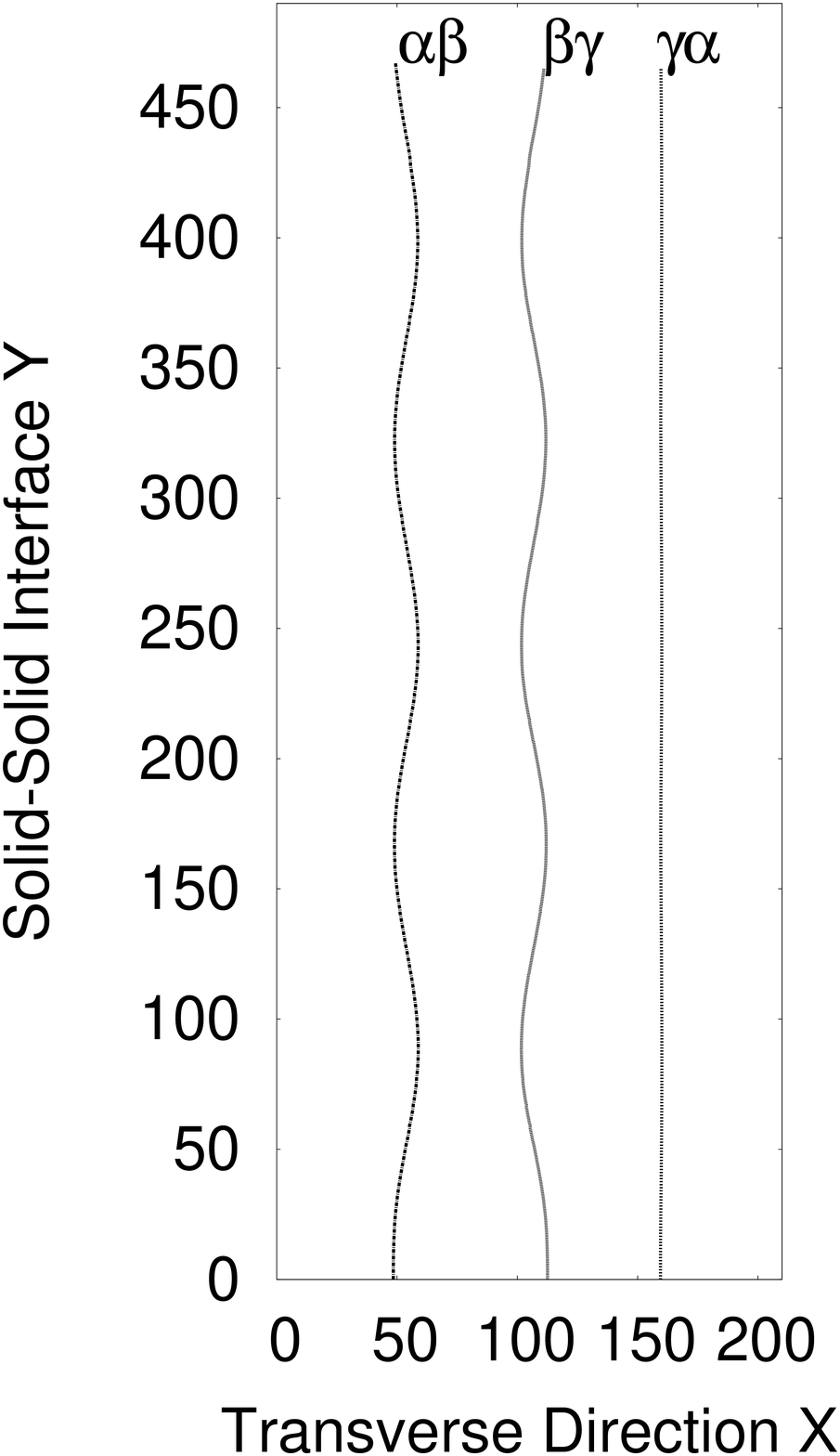}
  \label{Figure15a}
  }
  \subfigure[]{
      \includegraphics[height=6cm]{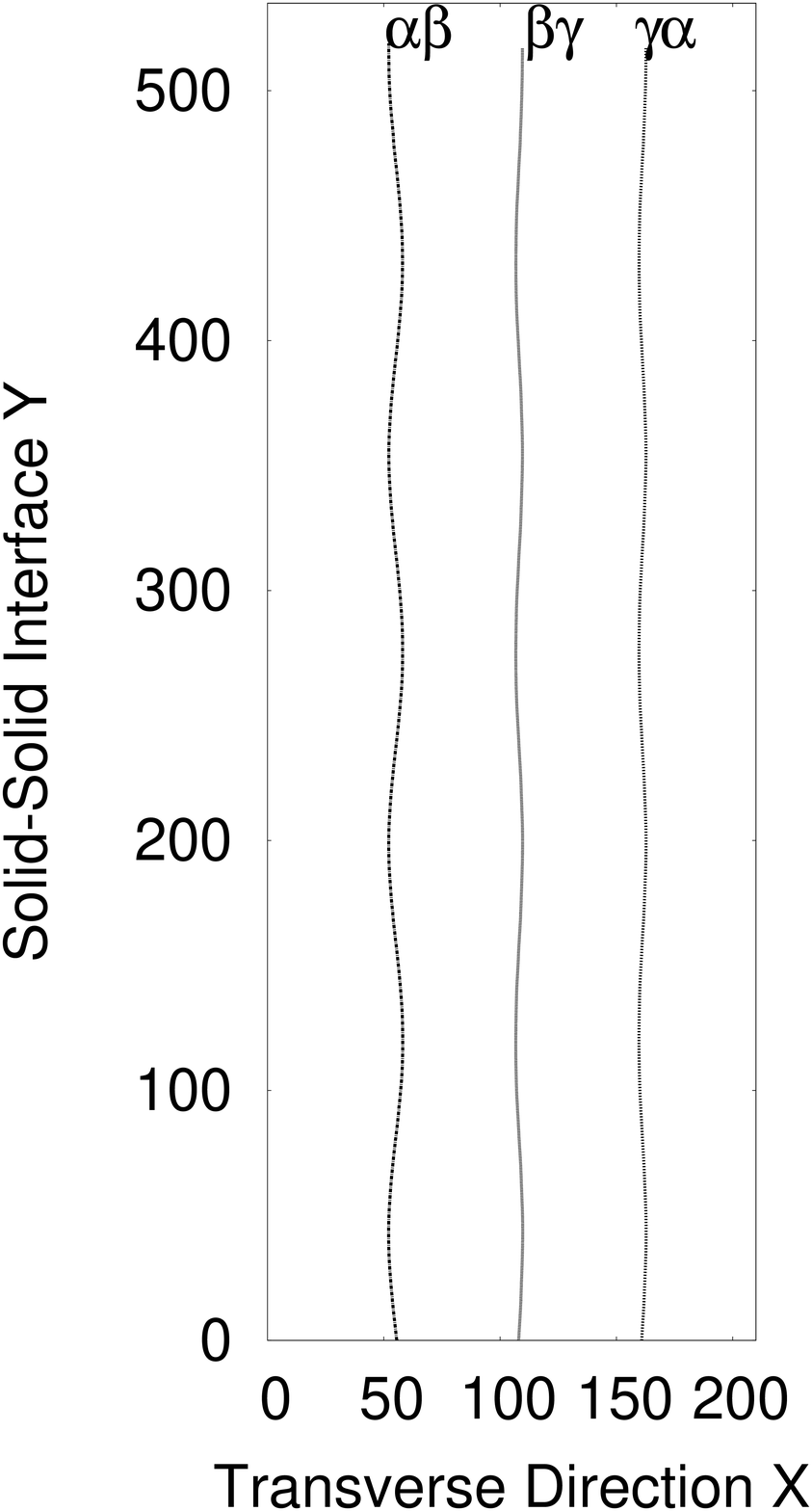}
     \label{Figure15b}
   }
  \subfigure[]{
      \includegraphics[height=6cm]{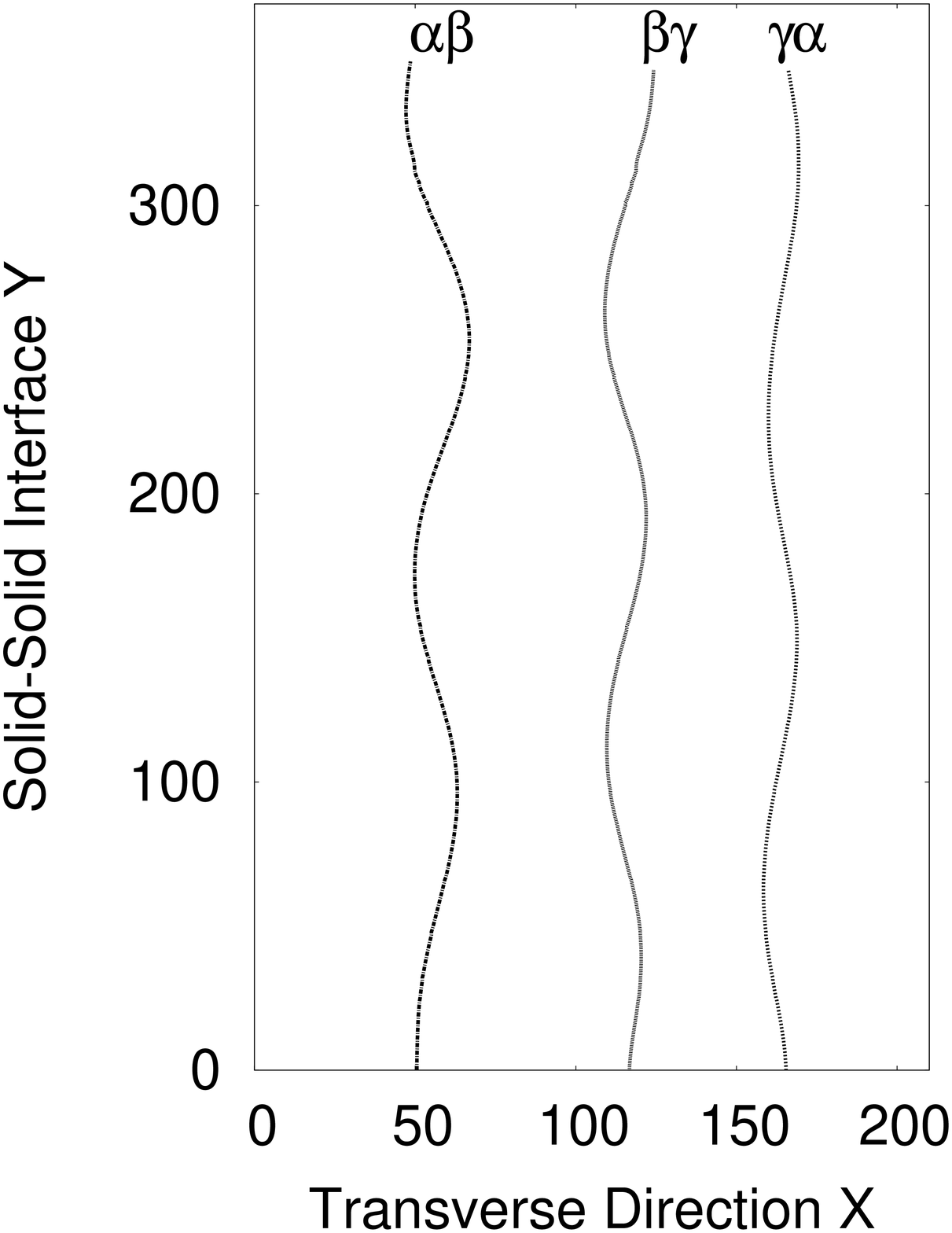}
     \label{Figure15c}
     }
\caption{The plot shows the trace of the triple points  for the $\alpha \beta \gamma$ arrangement. The growth direction is compressed in these plots with respect to the transverse direction in order to better visualize the modes. We get multiple modes at the eutectic concentration for the same spacing $\lambda=159$, shown in (a) and (b). In (a) we get back mode 1 while (b) matches well to our predicted mode 2. A mixed mode (c) is obtained at an off-eutectic concentration $\vc=\left(0.34,0.34,0.32\right)$, at a spacing $\lambda=165$, which is a combination of oscillations in both the width and lateral spacing.}
\label{Figure15}
 \end{center}
\end{figure}

Mode 1 is favored for off-eutectic concentrations in which one of the lamellae is wider 
than the two others, such as $\vc=(0.32,0.32,0.36)$. Indeed, in that case the
(unstable) steady-state pattern exhibits the same symmetry planes
as the oscillatory pattern. This mode can also appear when one 
lamella is {\it smaller} than the two others, see Figure \ref{Figure14c}.
We detect mode 2 at the eutectic concentration, see Figure \ref{Figure15b}.
However, a ``mixed mode'' can also occur, in which no symmetry
plane survives, but the three trijunctions oscillate laterally with
phase differences that depend on the concentration and possibly on the
spacing, see Figs. \ref{Figure14b} and \ref{Figure15c}.

\begin{figure}[htpb]
\begin{center}
\subfigure[]{
\includegraphics[width=8cm]{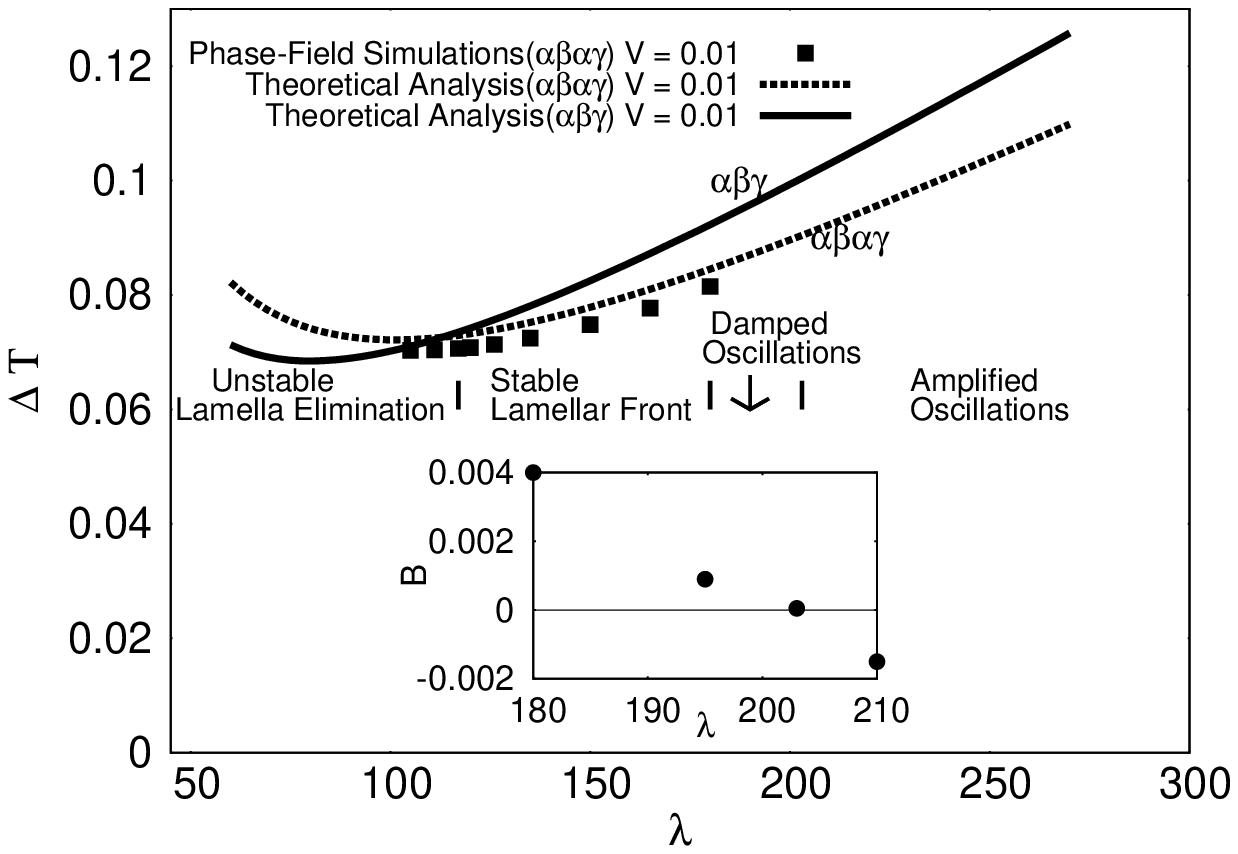}
\label{Figure16a}
}\\
\subfigure[]{
\includegraphics[width=8cm]{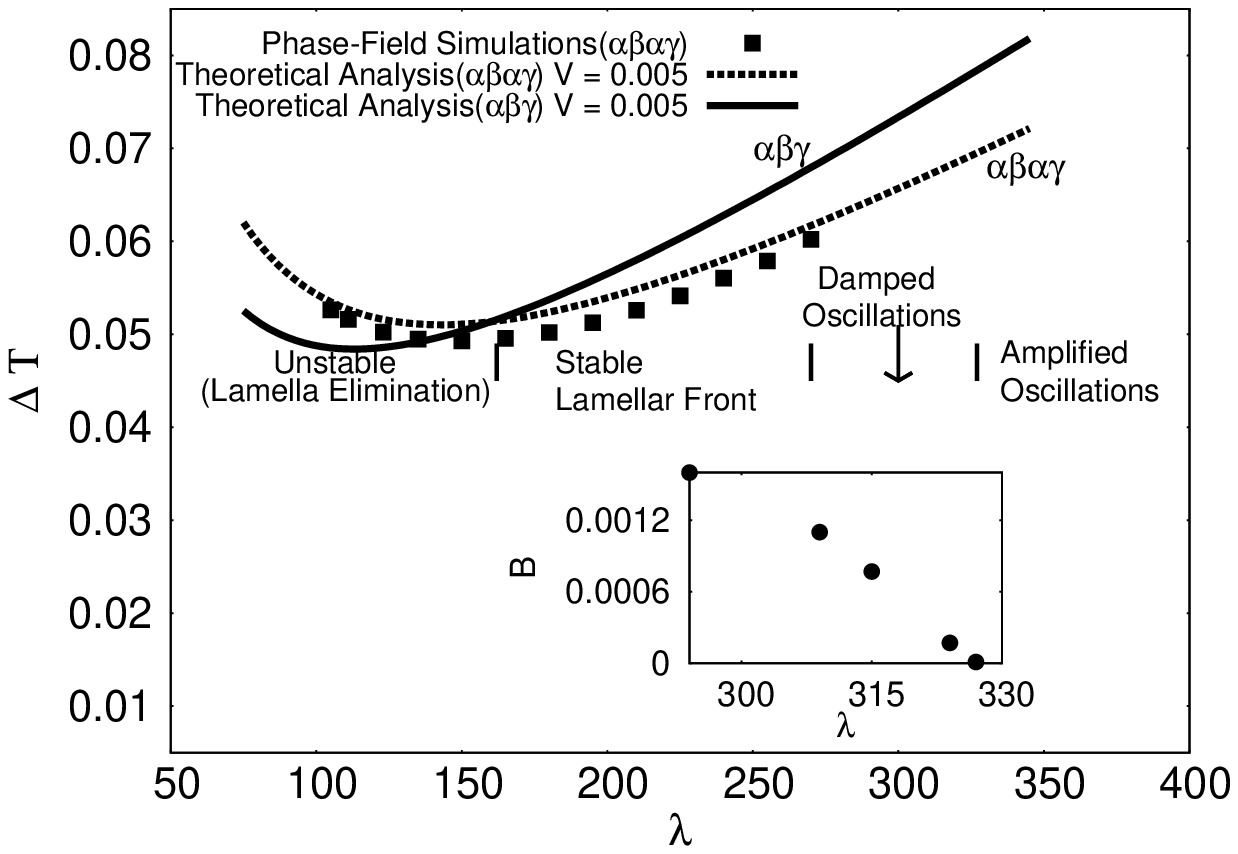}
\label{Figure16b}
}
\caption{Theoretical analysis and phase-field simulations: Comparison between the arrangements $\left(\alpha
\beta \gamma\right)$ and $\left(\alpha \beta \alpha \gamma\right)$ for two
different velocities (a) V = 0.005 and (b) V = 0.01. Plots convey information 
on the stability ranges, and the onset of oscillatory behavior of the 1-$\lambda$-O type.}
\label{Figure16}
\end{center}
\end{figure}

Let us now turn to the $\alpha \beta \alpha \gamma$ cycle.
We perform simulations for two different velocities $V = 0.01$ 
and $V = 0.005$. The comparison of the measurements with the theoretical analysis 
for steady-state growth is shown in Figure \ref{Figure16}. 
For the purpose of analysis, 
predictions from the theory for both arrangements ($\alpha \beta \gamma$ and 
$\alpha \beta \alpha \gamma$) are also shown. Here again, the minimum undercooling 
spacings match those of the theory to a good degree of accuracy (error ~5\%, V=0.005). 
However, the undercooling is lower than the one predicted by JH-theory, with a discrepancy of 4\% 
for the case of $V=0.005$, Figure \ref{Figure16a}.
For $V=0.01$, Figure \ref{Figure16b}, simulations were not possible 
for a sufficient range of $\lambda$ to determine the minimum undercooling, because 
the width of the narrowest lamellae became comparable to the interface 
width $\epsilon \simeq 8.0$ before the minimum was reached. However, the 
general trend of the data follows the predictions of the theory for 
both velocities. This was also the case for simulations carried out
at an off-eutectic concentration $\vc = (0.32, 0.34, 0.34)$ at a 
velocity of $V=0.005$, for the same configuration $\alpha\beta\alpha\gamma$.

\begin{figure}[htbp]
 \begin{center}
  \subfigure[]{
      \includegraphics[height=4cm]{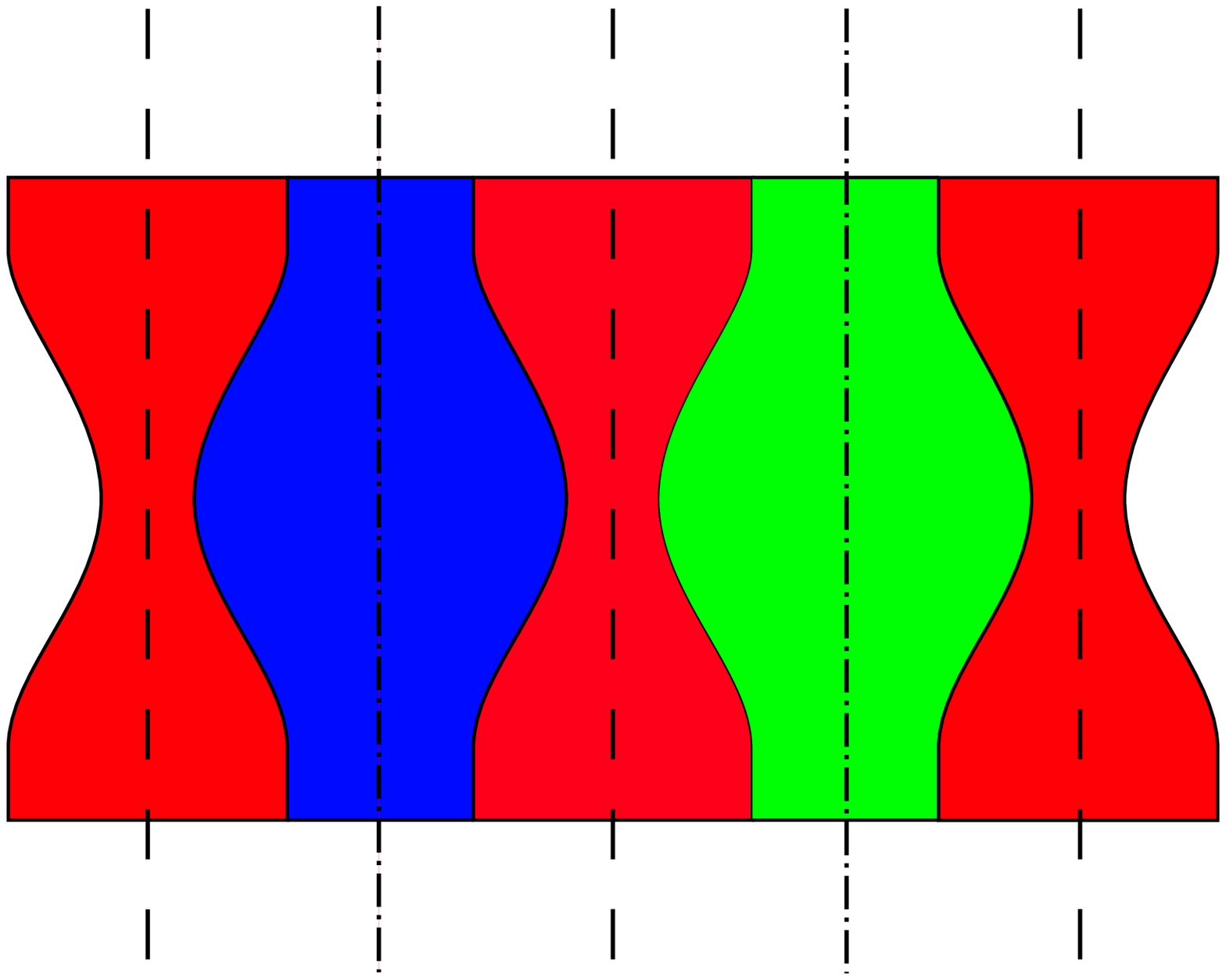}
     \label{Figure17a}
     }
  \subfigure[]{
      \includegraphics[height=4cm]{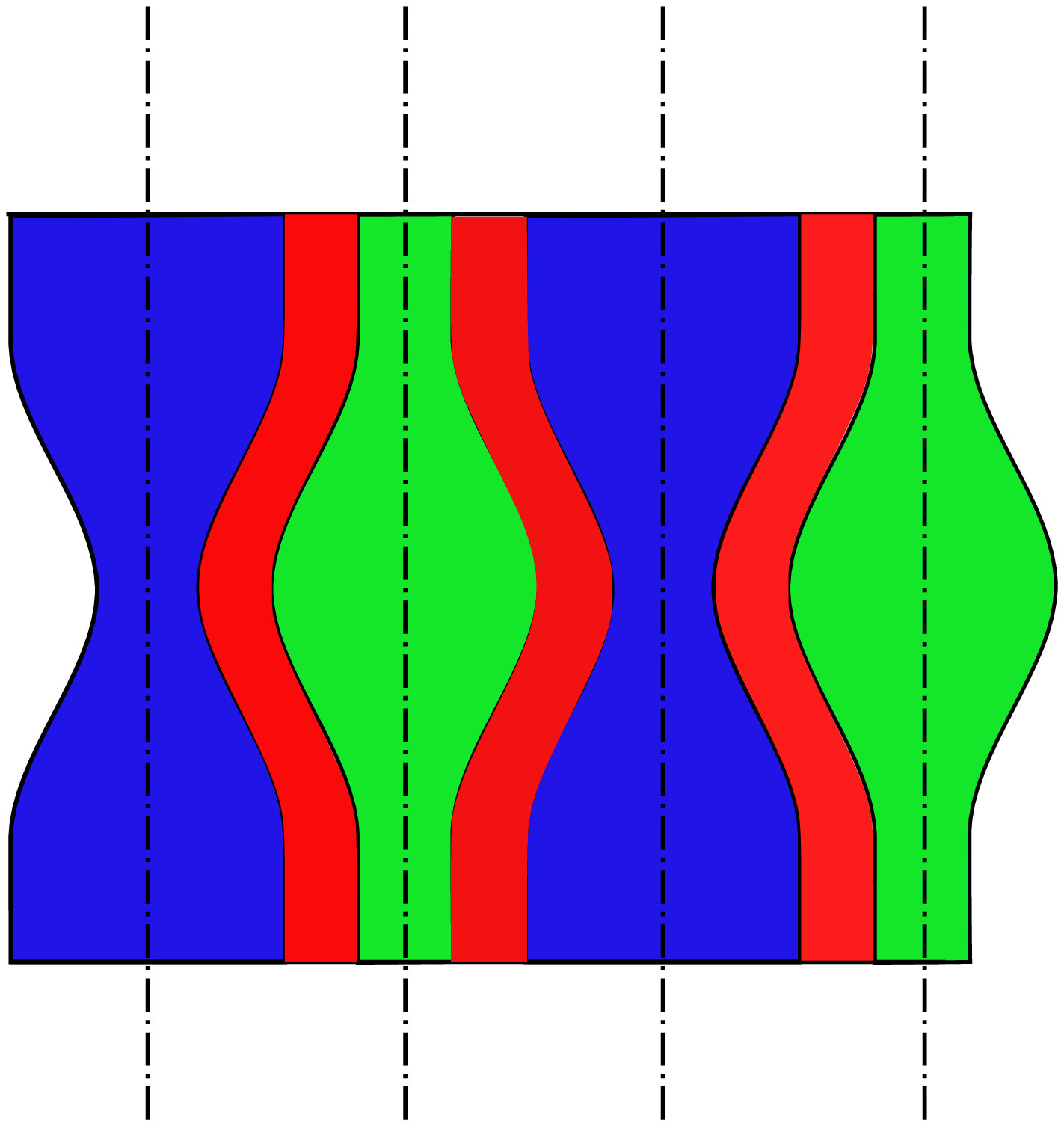}
     \label{Figure17b}
     }
   \caption{(Color Online) Predictions of oscillatory modes for the $\alpha \beta \alpha \gamma$ arrangement, reminiscent of the 1-$\lambda$-O mode (a) and 2-$\lambda$-O mode (b) in binary eutectics.}
\label{Figure17}
 \end{center}
\end{figure}

\begin{figure}[htbp]
 \begin{center}
  \subfigure[]{
      \includegraphics[height=5cm]{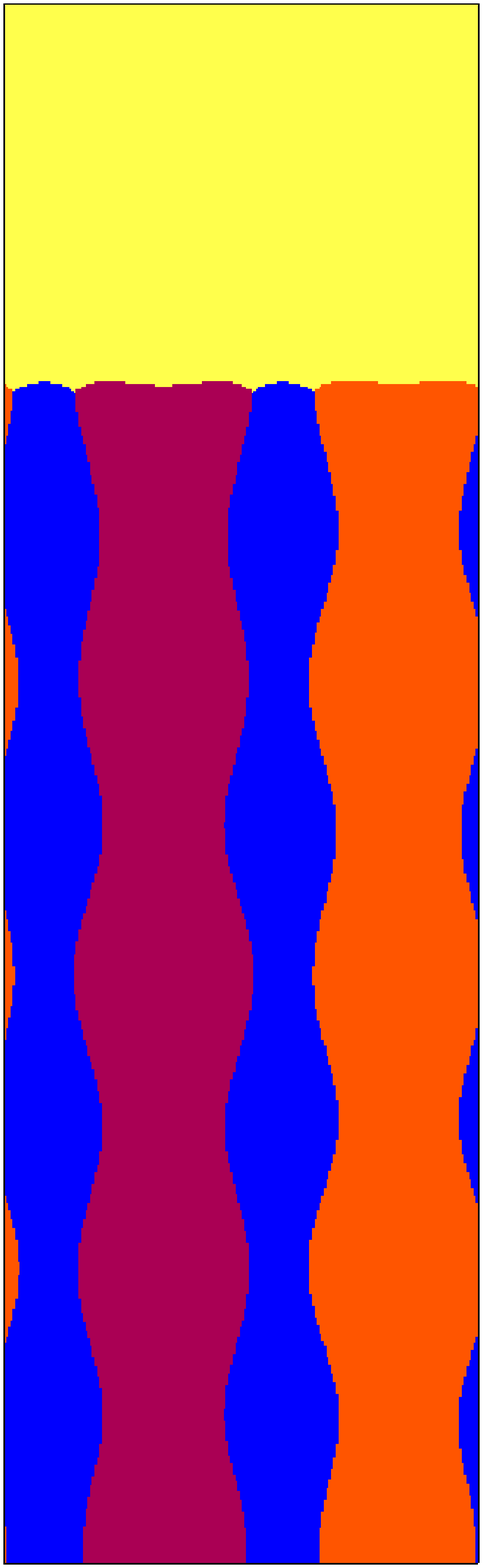}
      \label{Figure18a}
     }
  \subfigure[]{
      \includegraphics[height=5cm]{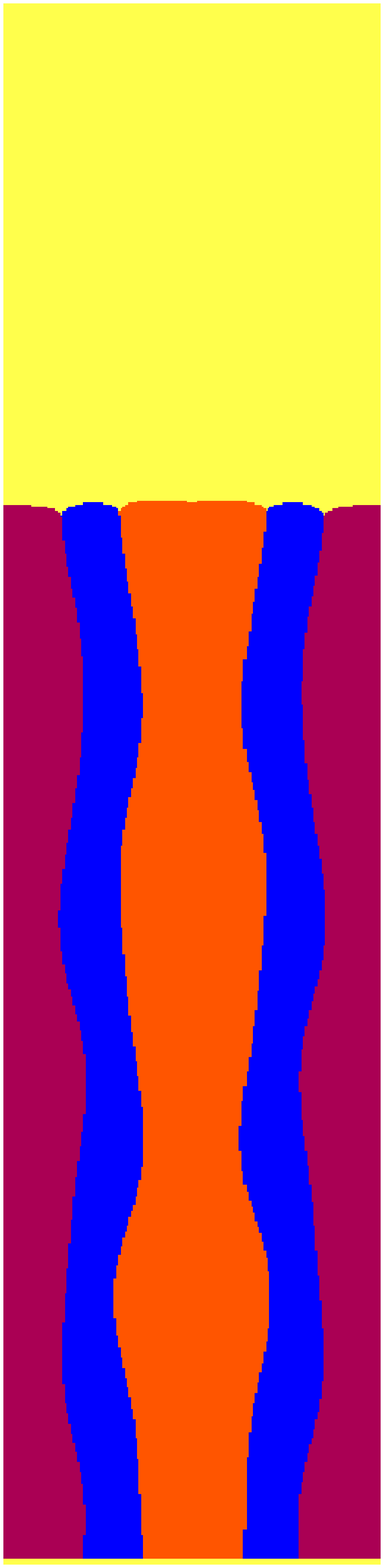}
      \label{Figure18b}
     }
  \subfigure[]{
      \includegraphics[height=5cm]{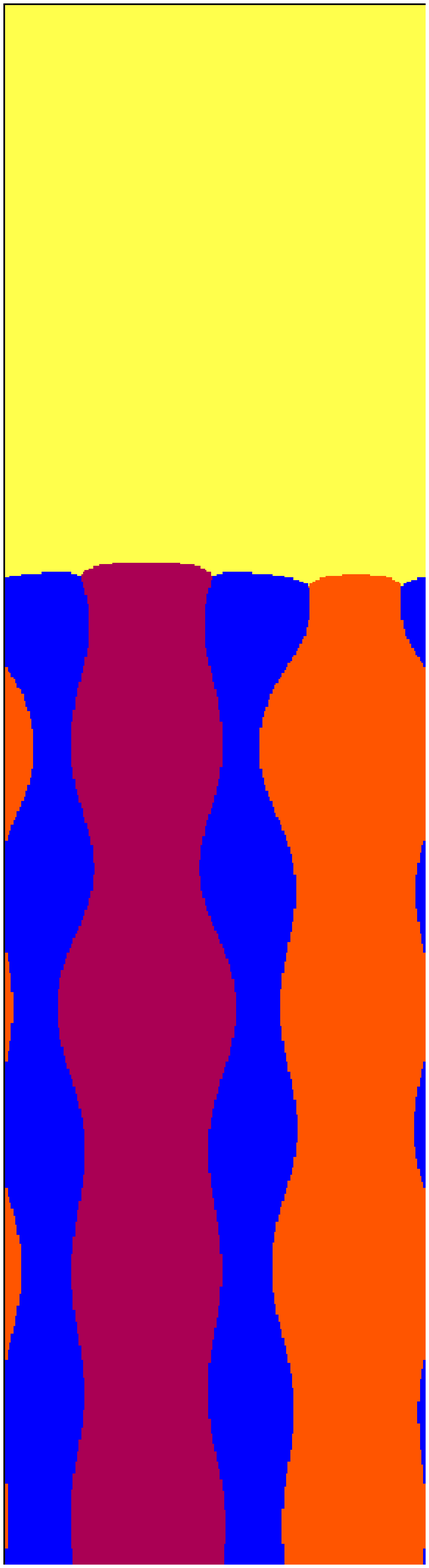}
      \label{Figure18c}
     }
\caption{(Color Online) Simulations of oscillatory modes of the $\alpha \beta \alpha \gamma$ configuration. The modes in (a) and (b) show resemblance to the 1-$\lambda$-O and 2-$\lambda$-O oscillatory modes of binary eutectics, respectively. 
Additionally, other modes (c) can also be observed, depending on the 
initial conditions. While we observe (a) at the eutectic concentration, 
(b) and (c) are modes at off-eutectic concentrations $\vc=(0.32,0.34,0.34)$. 
The spacings are (a) $\lambda=201$, (b) $\lambda=174$ (c) $\lambda=210$.}
\label{Figure18}
 \end{center}
\end{figure}

Concerning the oscillatory instabilities at large spacings, it is
useful to consider again the symmetry elements. For this cycle,
there are two real symmetry planes in the steady-state pattern that run
through the centers of the $\beta$ and $\gamma$ lamellae. Note that
these symmetries would exist even for unsymmetric phase diagrams and
unequal surface tensions. Therefore, by analogy with binary eutectics,
one may expect oscillatory modes that simply generalize the 1-$\lambda$-O
and 2-$\lambda$-O modes of binary eutectics, see Figure \ref{Figure17a}.
Indeed, for our simulations at the eutectic concentration, we retrieve 
the 1-$\lambda$-O type oscillation, figure \ref{Figure18a} as 
in our hypothesis (figure \ref{Figure17a}). 

This oscillatory instability 
can be quantitatively monitored by following the lateral positions 
of the solid-solid interfaces with time. More specifically, we 
extract the width of the $\beta$ phase as a function of 
the growth distance $z$. This is then fitted with a damped sinusoidal 
wave of the type $A_{0}+A\exp(-Bz)\cos((2\pi z/L)+D)$. The damping 
coefficient $B$ is obtained from a curve fit and plotted 
as a function of the spacing $\lambda$. The onset of the instability
is characterized by the change in sign of the damping coefficient.

For the off-eutectic concentration we get two modes (figure \ref{Figure18}).
While (figure \ref{Figure18b}) corresponds well to our 
hypothesis to the 2-$\lambda$-O type oscillation (figure \ref{Figure17b}), 
we also observe another mode as shown in figure \ref{Figure18c},
which combines elements of the two modes: both the width and the lateral
position of the $\alpha$ lamellae oscillate.

\subsection{Lamella elimination instability}

For the $\alpha\beta\alpha\gamma$ cycle, there is also a new
instability, which occurs for low spacings. We find that all 
spacings below the minimum undercooling spacing, as well as 
some spacings above it, are unstable with respect to 
lamella elimination: the system evolves to the 
$\alpha \beta \gamma$ arrangement by eliminating one of the
$\alpha$ lamellae, both at eutectic and off-eutectic concentrations. 
The points plotted to the left of the minimum in Figure \ref{Figure16b} 
are actually unstable steady states that can be reached only when the 
simulation is started with strictly symmetric initial conditions and the 
correct volume fractions of the solid phases.

This instability can actually be well understood using our theoretical
expressions. As already mentioned before, when we consider the cycle 
$\alpha \beta \alpha \gamma$ at the eutectic concentration with a lamella 
width configuration $(\xi , 1/3 , 1/3 -\xi, 1/3 )$, the global average
front undercooling attains a minimum for the symmetric pattern $\xi=1/6$.
However, the global front undercooling is not the most relevant information
for assessing the front stability. More interesting is the
undercooling of an {\em individual} lamella, because this can give
information about its evolution. More precisely, consider the undercooling
of one of the $\alpha$ lamellae as a function of $\xi$. If the undercooling
increases when the lamella gets thinner, then the lamella will fall further
behind the front and will eventually be eliminated. In contrast, if the
undercooling decreases when the lamella gets thinner, then the lamella will 
grow ahead of the main front and get larger. A similar argument has been 
used by Jackson and Hunt for their explanation of the long-wavelength 
elimination instability \cite{JH}.
It should be pointed out that the new instability found here is not a 
long-wavelength instability, since it can occur even when only one unit cell
of the cycle is simulated.

\begin{figure}[tbph]
\begin{center}
\includegraphics[width=8cm]{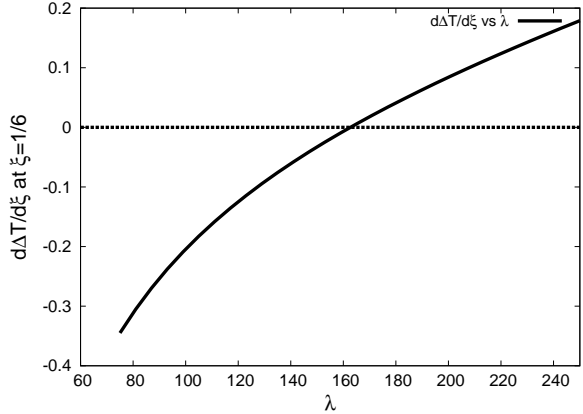}
\caption{Plot of $\partial \Delta T/\partial \xi$, taken at $\xi=1/6$
versus $\lambda$ for the $\alpha_{1} \beta \alpha_{2} \gamma$ cycle,
where $\Delta T$ is the undercooling of the $\alpha_1$ lamella and $\xi$ 
its width (relative to $\lambda$), calculated by our analytical 
expressions in the volume fraction configuration.
$(\xi , 1/3 , 1/3 -\xi, 1/3 )$ at V=0.01. The cycle is predicted
to be unstable to lamella elimination if $\partial \Delta T/\partial \xi<0$.
The $\lambda$ at which $\partial \Delta T/\partial \xi$ changes sign is the critical point beyond which the $\alpha \beta \alpha \gamma$ arrangement is stable with respect to a change to the sequence $\alpha \beta \gamma$ through a lamella elimination.}
\label{Figure19}
\end{center}
\end{figure}

Following the above arguments, we have calculated the growth temperature
of the first $\alpha$ lamella as a function of $\xi$ using the general
expressions in Eq.~\ref{abag-1lamella}. In Figure \ref{Figure19}, we plot 
the variation of $\partial \Delta T/\partial \xi$ at $\xi = 1/6$, 
as a function of $\lambda$. The point at which 
$\partial \Delta T/\partial \xi$ becomes positive then indicates the 
transition to a stable $\alpha\beta\alpha\gamma$ cycle. This criterion
is in good agreement with our simulation results. This argument can
also be generalized to more complicated cycles (see below).

\clearpage

\subsection{Longer cycles}
Let us now discuss a few more complicated cycles. The simple 
cycles we have simulated until now were such that during stable 
coupled growth the widths of all the lamellae corresponding to 
a particular phase were the same. This changes starting from
period $M=5$, where the configuration $\alpha \beta \alpha \beta \gamma$ 
is the only possibility (up to permutations). If we consider this
cycle at the eutectic concentration and note the configuration of lamella
widths as $(\xi,1/3-\xi,1/3-\xi,\xi,1/3)$ and compute the average front 
undercooling by our theoretical expressions, we find that the minimum
occurs for $\xi$ close to $0.12$. 
In addition, for this configuration, 
the undercooling of any asymmetric configuration (permutation of widths of lamellae) 
is higher than the one considered above.
If we rewrite symbolically this configuration as 
$\alpha_1,\beta_2,\alpha_2,\beta_1,\gamma$, it is easy to see
that this configuration has two symmetry axes of the same kind
as discussed in the preceding subsection: mirror reflection and
exchange of the phases $\alpha$ and $\beta$. One of them runs
along the interface between $\beta_2$ and $\alpha_2$, and the
other one in the center of the $\gamma$ lamella.

\begin{figure}[htpb]
\begin{center}
\subfigure[]{
\includegraphics[height=5cm]{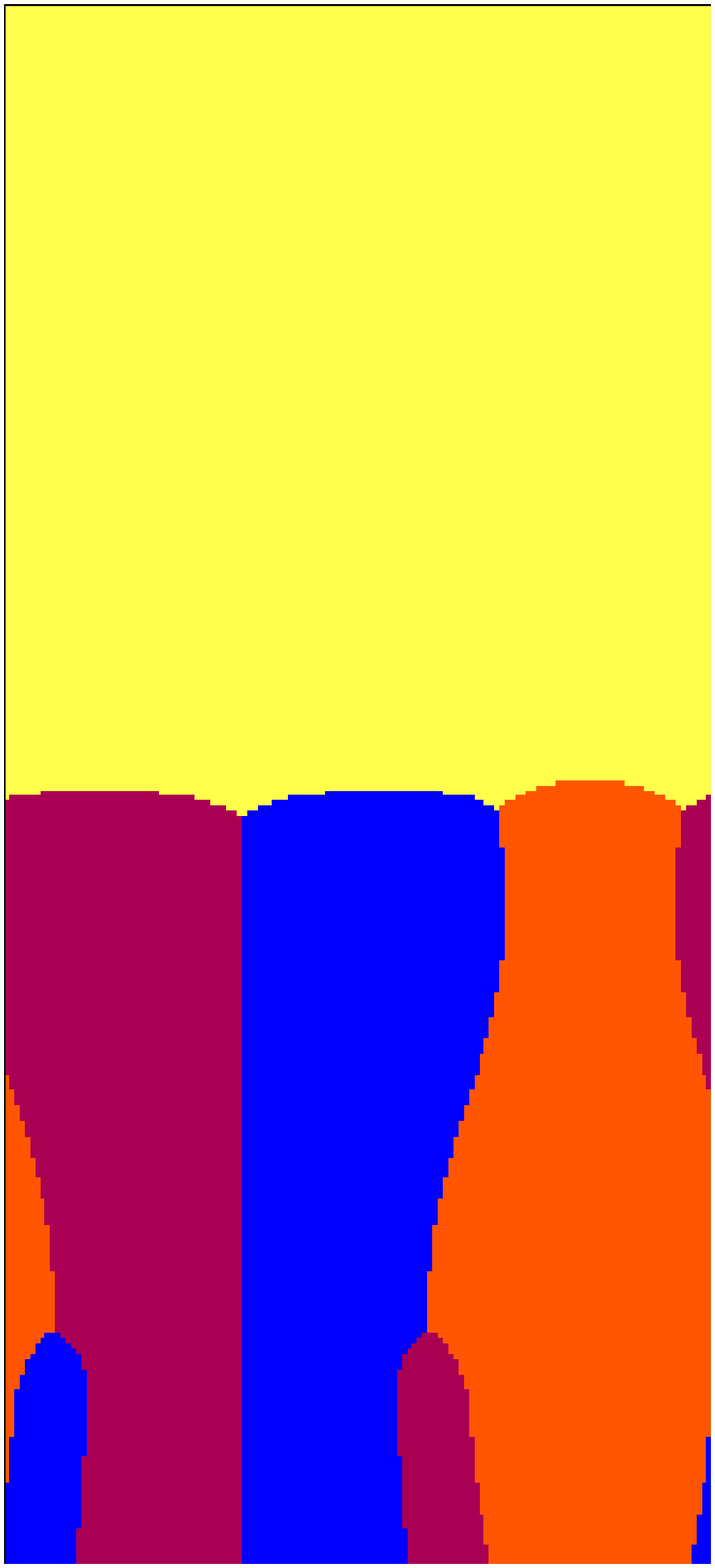}
\label{Figure20a}
}
\subfigure[]{
\includegraphics[height=5cm]{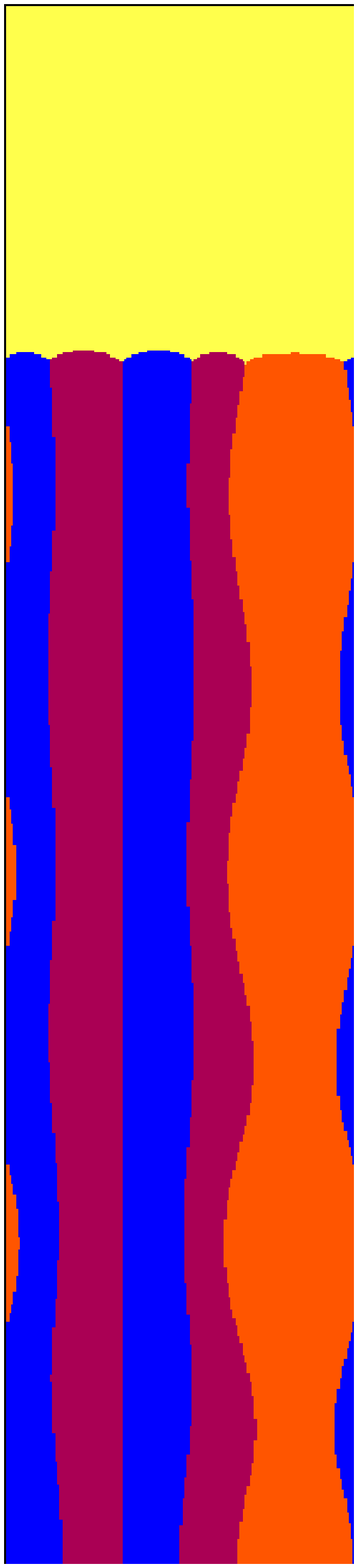}
\label{Figure20b}
}
\subfigure[]{
\includegraphics[height=5cm]{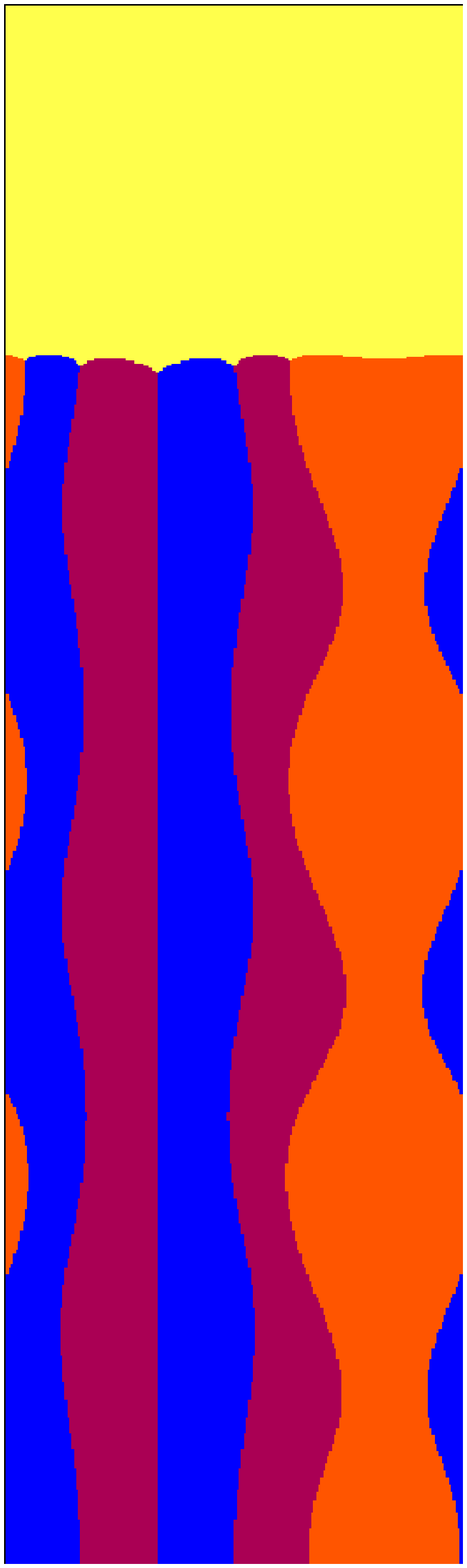}
\label{Figure20c}
}
\caption{(Color Online) Simulations at spacings $\lambda=135$ in (a), $\lambda = 150$ in (b) and $\lambda=180$ in (c), starting from an initial configuration of $\alpha\beta\alpha\beta\gamma$. There is no spacing for which the $\alpha\beta\alpha\beta\gamma$ develops into a stable lamellar growth front. Smaller spacings switch to the $\alpha\beta\gamma$ arrangement while the larger spacings exhibit oscillatory instabilities in both the width and the lateral positions of the lamellae.}
\label{Figure20}
\end{center}
\end{figure}

Not surprisingly, our simulation results confirm the importance
of this symmetry. The volume fractions in steady-state growth
are close to those that give the minimum average front undercooling, see 
Figs. \ref{Figure20b} and \ref{Figure20c}. 

Additionally,
we observe oscillations in the width of the largest $\gamma$ phase
and oscillations in the widths and the lateral position of the 
smaller lamellae of the $\alpha$ and $\beta$ phases, while the
interface between the larger $\alpha$ and $\beta$ phase remains
straight, such that the combination of all the $\alpha$ and $\beta$
lamellae oscillates in width as one ``composite lamella''. Thus, 
the symmetry elements of the underlying steady state are preserved 
in the oscillatory state.

For smaller spacings, this configuration is unstable, and 
the sequence changes to the $\beta\alpha\gamma$ arrangement as
shown in Figure \ref{Figure20a} by two successive lamella 
eliminations. It is noteworthy 
that we did not find any unstable sequence which switches to the 
$\alpha\beta\alpha\gamma$ arrangement, which again can be 
understood from the presence of the symmetry. Indeed, a 
symmetrical evolution would result in a change to a configuration 
$\alpha_1\beta_1\gamma$ or $\beta_2\alpha_2\gamma$, but precludes 
the change to a configuration of period length $M=4$.

\begin{figure}[htbp]
\begin{center}
\subfigure[]{
 \includegraphics[width=8cm]{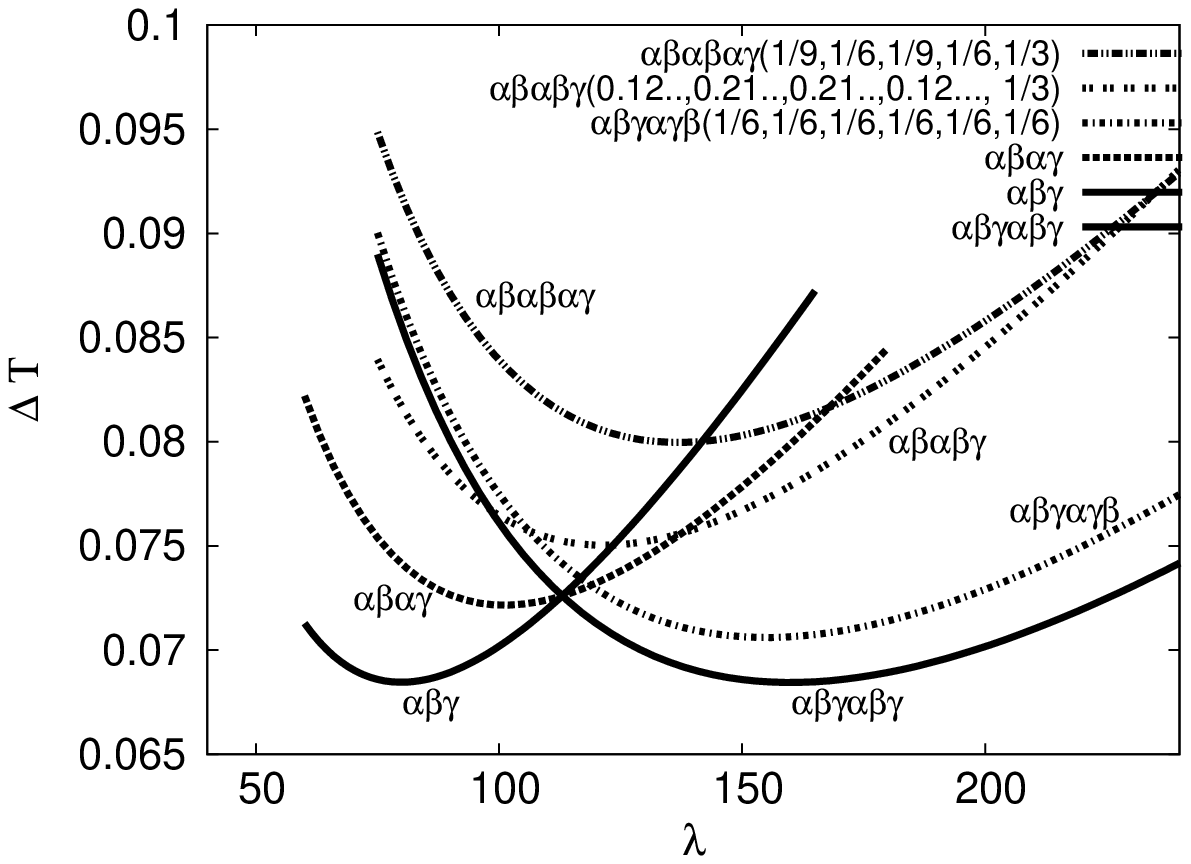}
\label{Figure21a}
}
\subfigure[]{
\includegraphics[width=8cm]{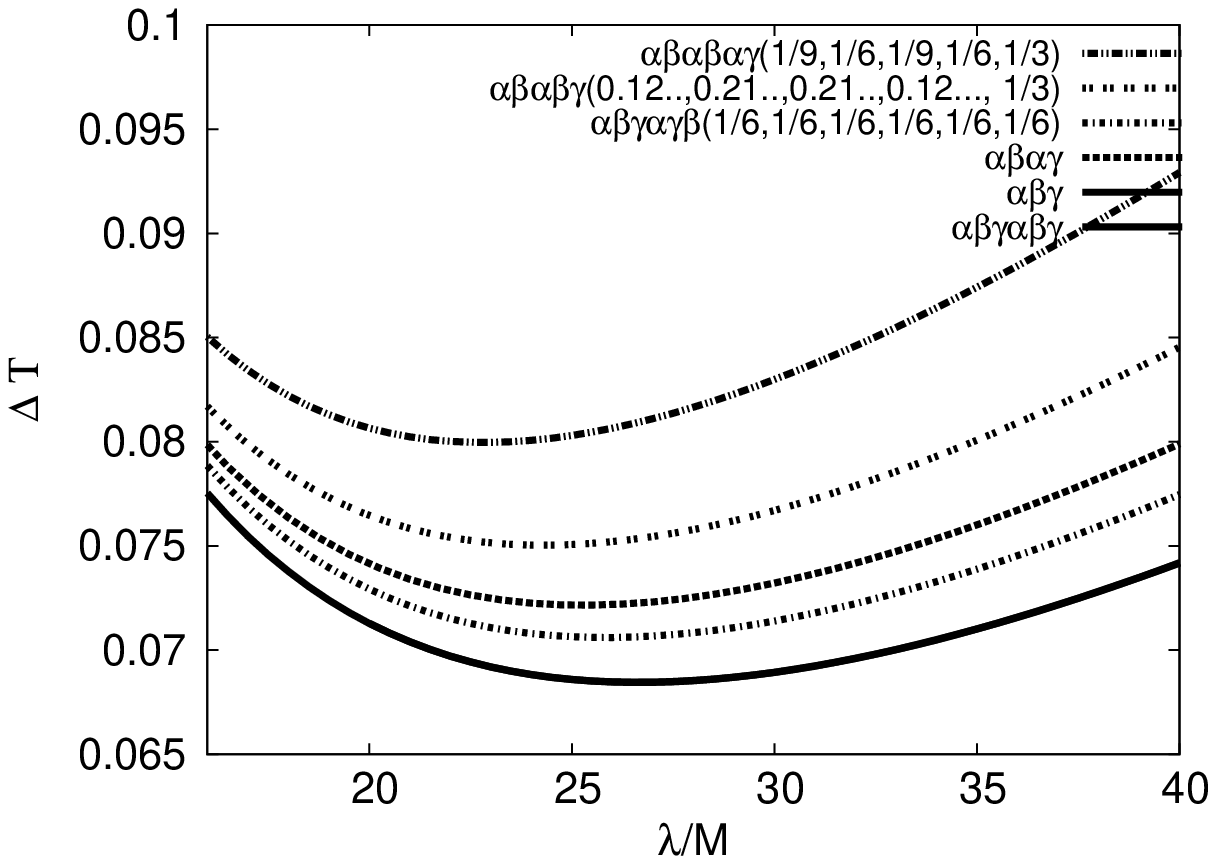}
\label{Figure21b}
}
\caption{(a)Synopsis of the theoretical predictions for the undercooling versus spacing of possible arrangements between period length $M=3$ to $M=6$, i.e. starting from $\alpha \beta \gamma$ to $\alpha \beta \gamma \alpha \beta \gamma$. (b) Same plot but with the lamellar repeat distance $\lambda$ scaled with the period length M. The variation among the arrangements is purely a result of the variation of the solutal
undercooling as can be infered from the discussion in Sec. \ref{theoretical_discussion}.}
\label{Figure21}
\end{center}
\end{figure}

Going on to cycles with period $M=6$, the first arrangement we 
consider is $\alpha_1 \beta_1 \alpha_2 \beta_2 \alpha_1 \gamma$, where 
we name the lamellae for eventual discussion and ease in description
according to the symmetries. Indeed, this arrangement has two exact
mirror symmetry planes in the center of the $\alpha_2$ and the $\gamma$
phases. We find that, if we calculate the average interface undercooling curves 
by varying the widths of individual lamella with the constraint 
of constant volume fraction, by choosing different $\xi$, in the width 
configuration $(\xi,1/6,1/3-2\xi,1/6,\xi,1/3)$, the average 
undercooling at the growth interface is minimal for the configuration 
$(1/9,1/6,1/9,1/6,1/9,1/3)$. This arrangement has the highest 
undercooling curve among the arrangements we have considered, shown
in Figure \ref{Figure21a}. 

\begin{figure}[tbph]
 \subfigure[]{
\includegraphics[height=3cm]{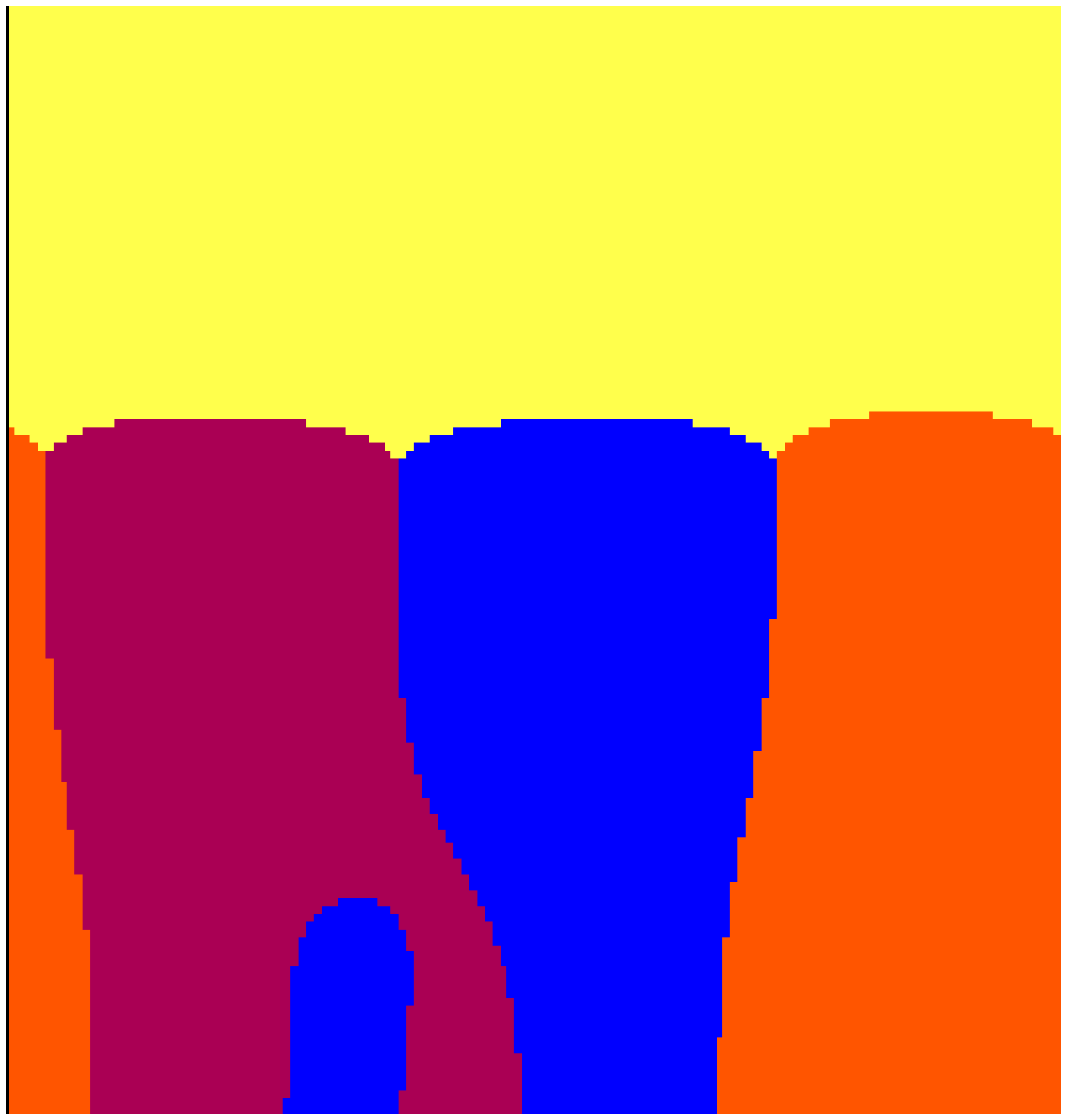}
\label{Figure22a}
}
\subfigure[]{
\includegraphics[height=3cm]{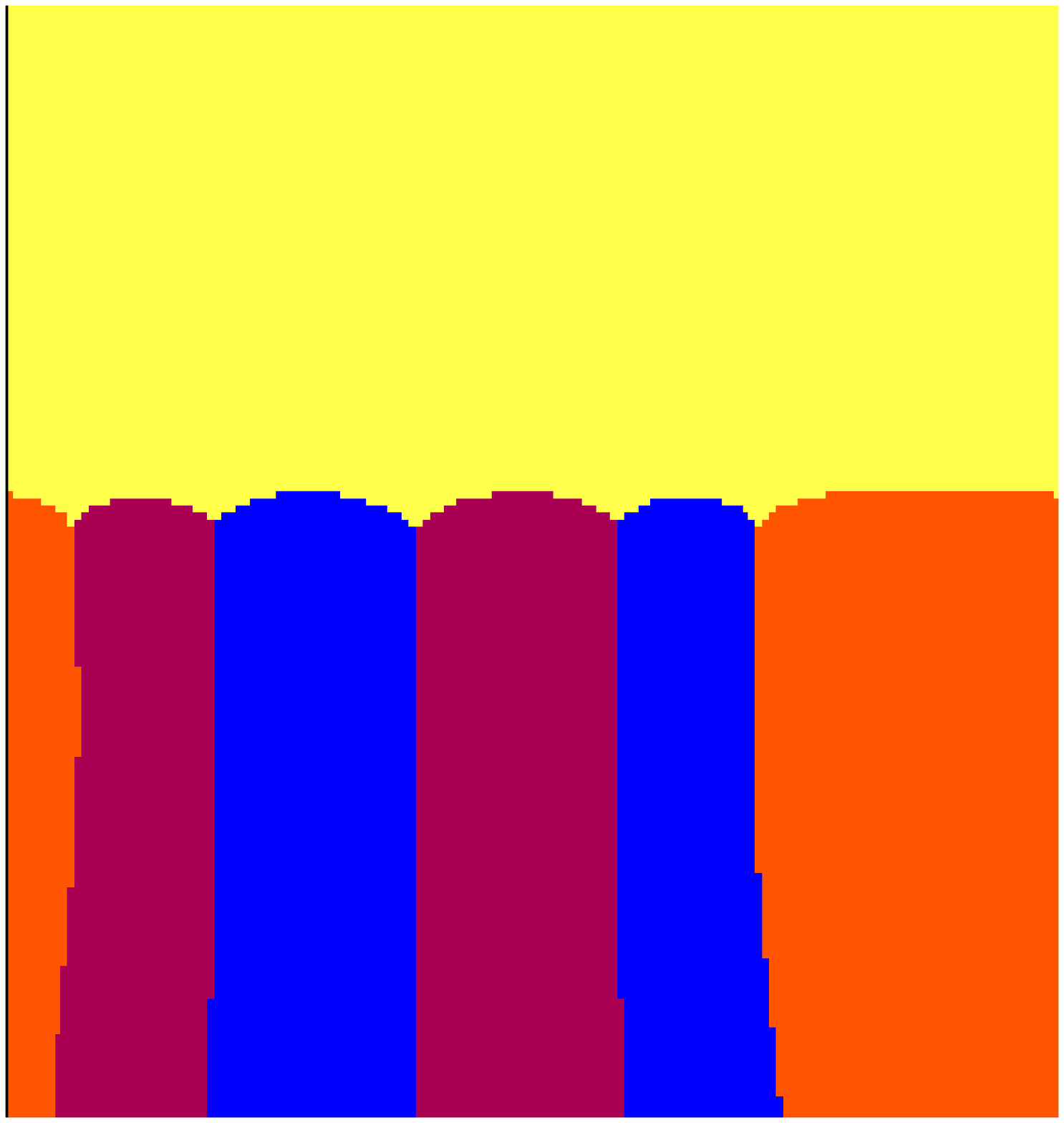}
\label{Figure22b}
}
\subfigure[]{
\includegraphics[height=3cm]{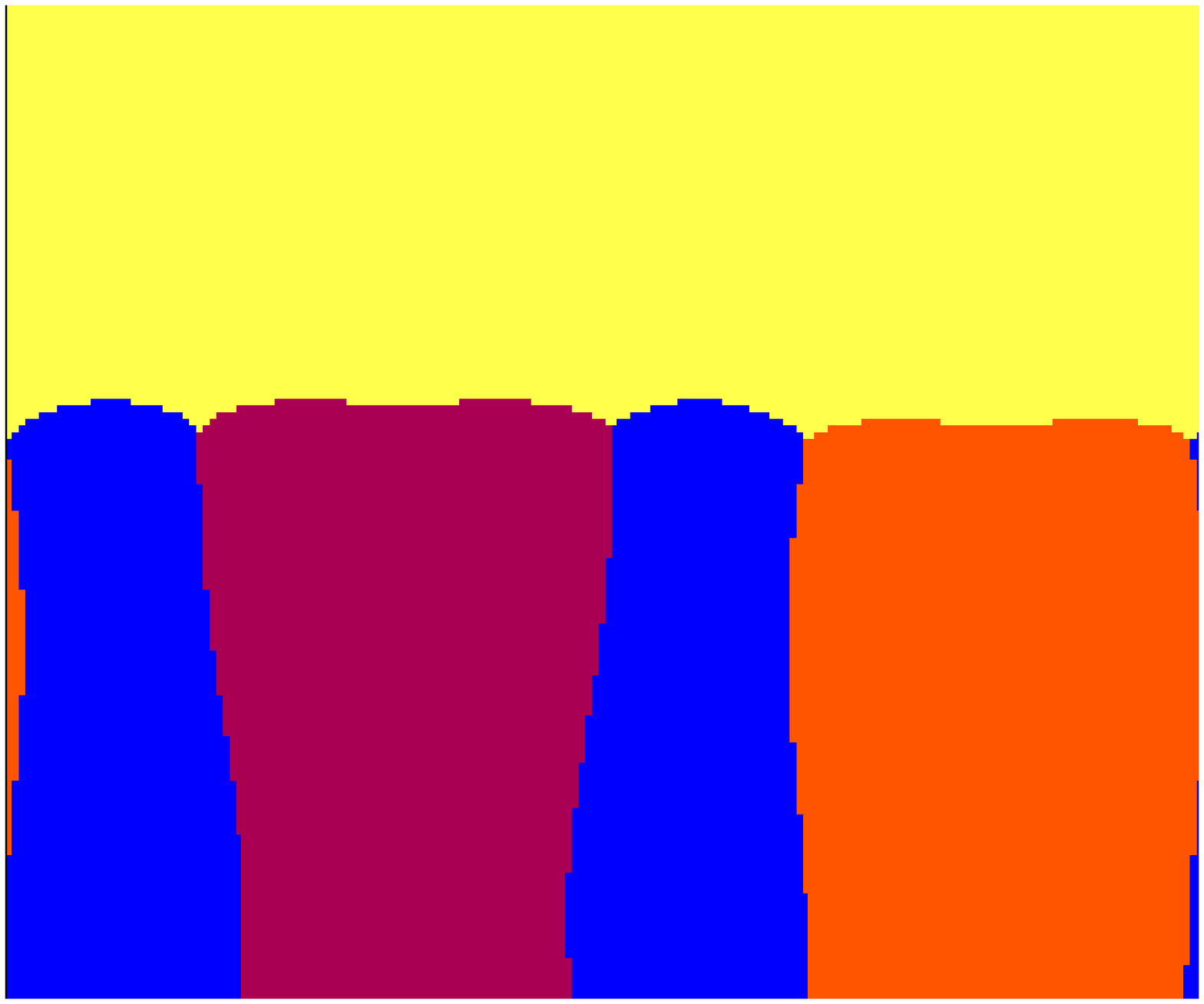}
\label{Figure22c}
}
\subfigure[]{
\includegraphics[height=3cm]{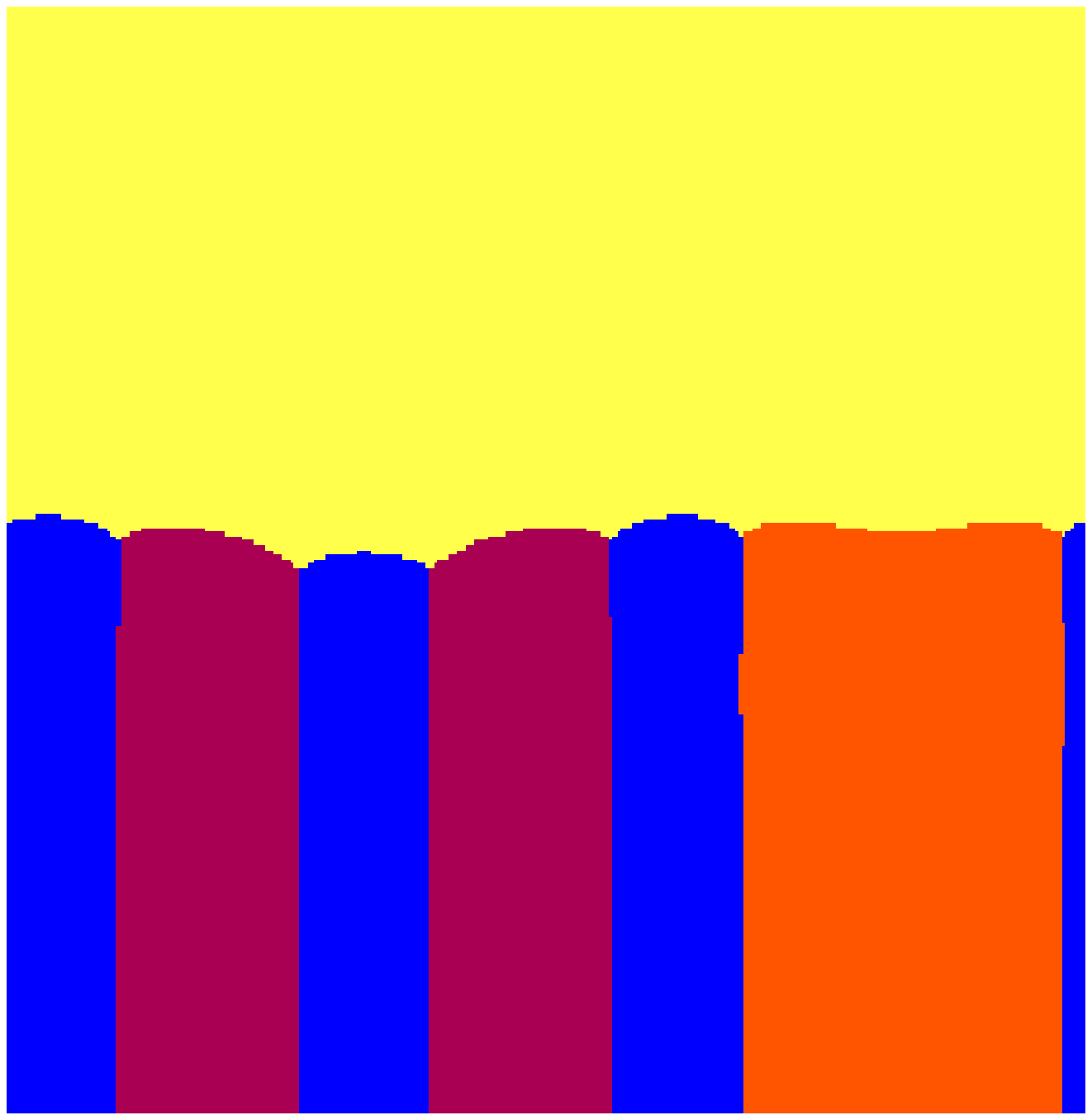}
\label{Figure22d}
}
\caption{(Color Online) Simulations starting from the arrangement $\alpha \beta \alpha \beta \alpha \gamma$ for spacings $\lambda=135$ in a), $\lambda=150$ in b), $\lambda=180$ in c) and $\lambda=240$ in d).}
\label{Figure22}
\end{figure}

It also has a very narrow range of stability, 
and we could isolate only 
one spacing which exhibits stable growth for $\lambda = 240$, 
Figure \ref{Figure22d}. Unstable arrangements near the 
minimum undercooling spacing evolve to the $\alpha \beta \gamma$ 
arrangement, Figure \ref{Figure22a}, while for other 
unstable configurations we obtain the arrangements in 
Figure \ref{Figure22b} and Figure \ref{Figure22c} 
as the stable growth forms corresponding to $\lambda=150$ and 
$\lambda=180$ respectively.

Apart from the (trivial) period-doubled arrangement 
$\alpha \beta \gamma \alpha \beta \gamma$, another possibility
for $M=6$ is $\alpha \beta \gamma \alpha \gamma \beta$ with a volume 
fraction configuration $(1/6,1/6,1/6,1/6,1/6,1/6)$. Simulations 
of this arrangement show that there exists a reasonably large range 
of stable lamellar growth, and hence we could make a comparison 
between simulations and the theory. We find similar agreement between
our measurements and theory as we did previously for the arrangements
$\alpha\beta\gamma$ and $\alpha\beta\alpha\gamma$. The plot in 
Figure \ref{Figure21} shows the theoretical predictions of all the
arrangements we have considered until now. 

\subsection{Discussion}

It should by now have become clear that there exists a large
number of distinct steady-state solution branches, each of
which can exhibit specific instabilities. In addition, the
stability thresholds potentially depend on a large number of 
parameters: the phase diagram data (liquidus slopes, coexistence
concentration), the surface tensions (assumed identical here),
and the sample concentration. Therefore, the calculation of a
complete stability diagram that would generalize the one for
binary eutectics of Ref.~\cite{Sarkissian} represents a formidable
task that is outside the scope of the present paper. Nevertheless,
we can deduce from our simulations a few guidelines that can be
useful for future investigation.

Lamellar steady-state solutions can be grouped into three
classes, which respectively have (I) equal number of lamellae
of all three phases (such as $\alpha\beta\gamma$ and 
$\alpha\beta\gamma\alpha\gamma\beta$), (II) equal number
of lamellae for two phases (such as $\alpha\beta\alpha\gamma$),
and (III) different numbers of lamellae for each phase.

For equal global volume fractions of each phase (as in
most of our simulations), class III will have the narrowest 
stability ranges because of the simultaneous presence of very 
large and very thin lamella in the same arrangement, which 
make these patterns prone to both oscillatory and lamella
elimination instabilities. 

Any cycle in which a phase appears more than once can transit
to another, simpler one by eliminating one lamella of this phase.
This lamella instability always appears for low spacings below
a critical value of the spacing that depends on the cycle.
The possibility of a transition, however, depends also on the
symmetries of the pattern. For instance, the arrangement 
$\alpha \beta \alpha \beta \alpha \gamma$, if unstable, 
can transform into the $\alpha \beta \gamma$, 
$\alpha \beta \alpha \gamma$ or the $\alpha \beta \alpha \beta \gamma$
arrangements, while for an arrangement $\alpha \beta \alpha \beta \gamma$, 
it is impossible to evolve into the $\alpha \beta \alpha \gamma$ 
arrangement if the symmetry of the pattern is preserved by the
dynamics. 

For large spacings, oscillatory instabilities occur and can lead
to the emergence of saturated oscillatory patterns of various 
structures. The symmetries of the steady states seem to determine 
the structure of these oscillations, but no thorough survey of
all possible nonlinear states was carried out.

\section{Some remarks on pattern selection}
\label{sec_selection}
\begin{figure}[htbp]
\begin{center}
\subfigure[]{
 \includegraphics[width=8cm]{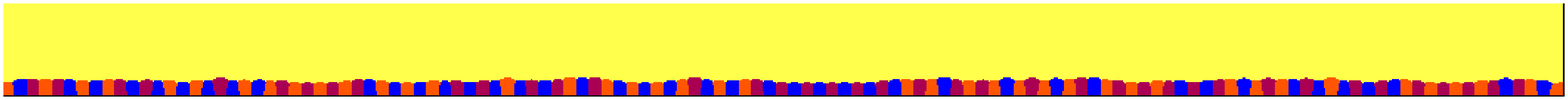}
 \label{Figure23a}
}
\subfigure[]{
 \includegraphics[width=8cm]{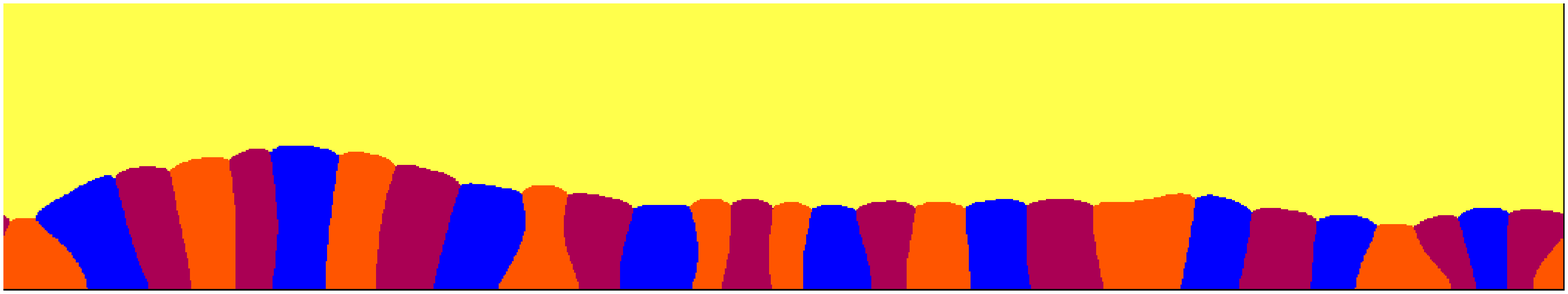}
 \label{Figure23b}
}
\caption{(Color Online) Two snapshots of 2D dynamics in a large system. Isothermal simulations are started from a random configuration in (a) where the probability of occurrence of each phase is $1/3$, which is also the global concentration in the liquid.  The temperature of the system is T=0.785 and the concentration of the liquid is the eutectic concentration. A slowly changing pattern with a non-planar front is achieved. Some lamellae are eliminated, but no new lamellae are created.}
\label{Figure23}
\end{center}
\end{figure}

Up to now, we have investigated various regular periodic patterns and
their instabilities. The question which, if any, of these different 
arrangements, is favored for given growth conditions, is still open. 
From the results presented above, we can already conclude that this
question cannot be answered solely on the basis of the 
undercooling-vs-spacing curves. Indeed, we have shown that by
appropriately choosing the initial conditions, any stable configuration
can be reached, regardless of its undercooling. This is also consistent
with experiments and simulations on binary eutectics \cite{Akamatsu,Parisi10}.
To get some additional insights on what happens in extended systems, 
we conducted some simulations of isothermal solidification where the 
initial condition was a random lamellar arrangement. More precisely, 
we initialize a large system with lamellae of width $\lambda=25$
and choose a random sequence of phases such that two neighboring
lamellae are of different phases as shown in Figure \ref{Figure23a}. 
The global probabilities of all the phases are $1/3$, 
which corresponds to the eutectic concentration,
and the temperature is set to $T=0.785$.

Under isothermal growth conditions, one would expect that, at a given 
undercooling, the arrangement with highest local velocity would be the 
one that is chosen. However, in order for the front to adopt this pattern, 
a rearrangement of the phase sequence is necessary. In our simulations, 
we find that lamella elimination was possible (and indeed readily occurred). 
In contrast, there is no mechanism for the creation of new lamellae in 
our model, since we did not include fluctuations that could lead to 
nucleation, and the model has no spinodal decomposition that could lead to the 
spontaneous formation of new lamellae, as in Ref.~\cite{Plapp+02}. 
As a result, some of the lamellae became 
very large in our simulations, which led to a non-planar growth front,
as shown in Figure \ref{Figure23b}. No clearcut periodic pattern emerged,
such that our results remain inconclusive.

We believe that lamella creation is an important mechanism required
for pattern adjustment. In 2D, nucleation is the only possibility for
the creation of new lamellae. In contrast, in 3D, new lamellae can also
form by branching mechanisms without nucleation events, since there
are far more geometrical possibilities for two-phase 
arrangements \cite{Moulinet,Walker}. Therefore, we also conducted a
few preliminary simulations in 3D. 

\begin{figure}
 \begin{center}
  \subfigure[]{
   \includegraphics[width=4cm]{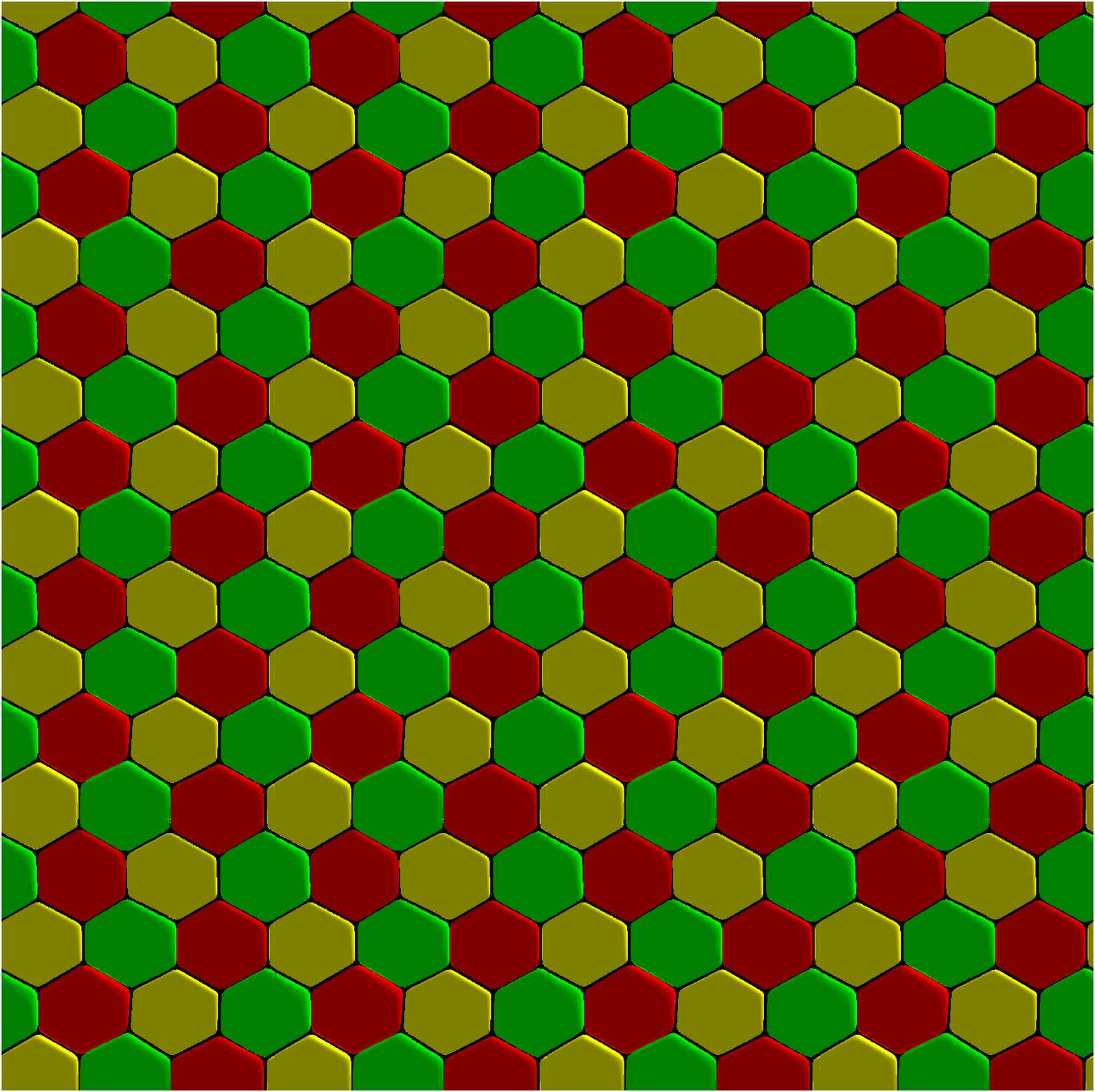}
  \label{Figure24a}
  }
  \subfigure[]{
   \includegraphics[width=4cm]{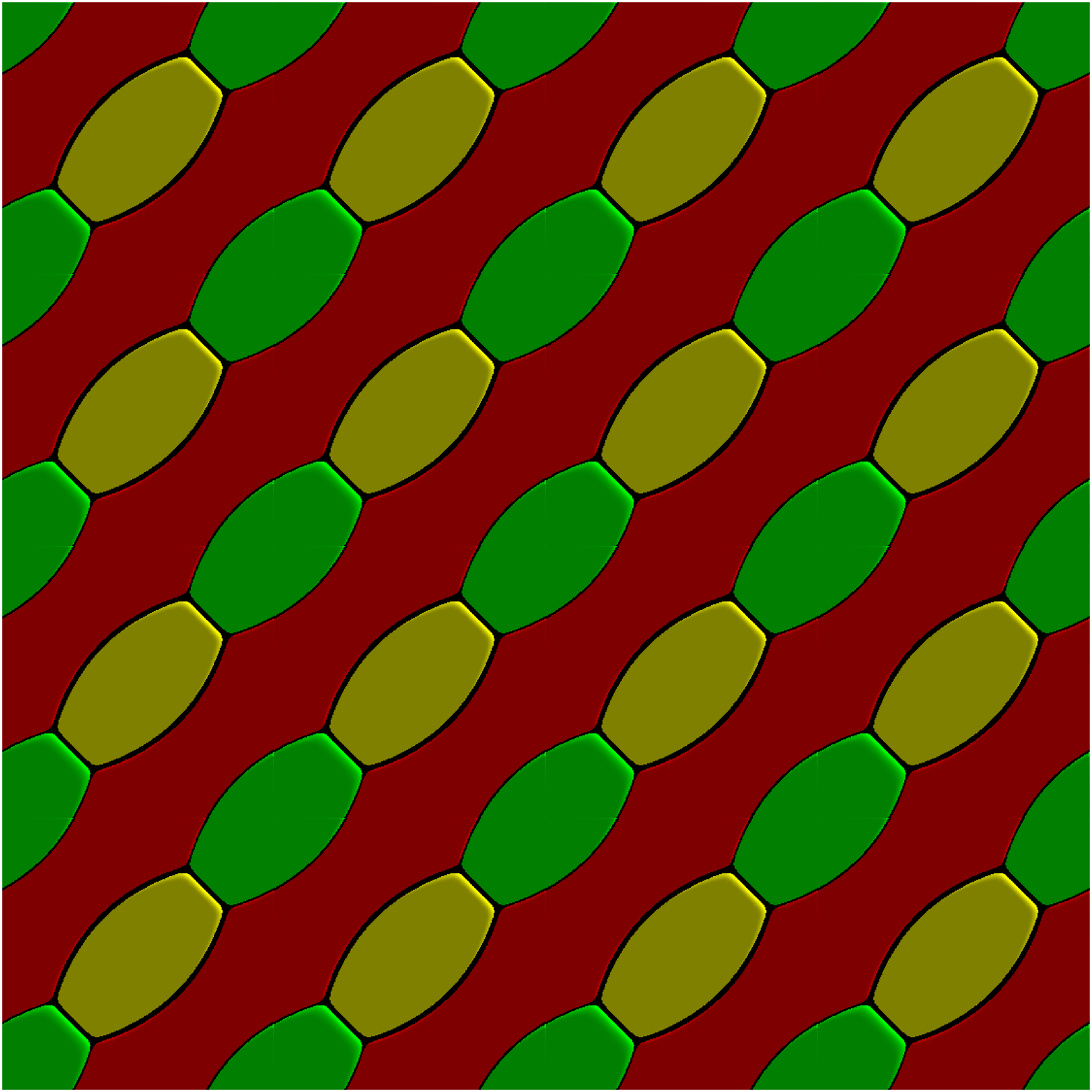}
  \label{Figure24b}
  }
  \subfigure[]{
   \includegraphics[width=4cm]{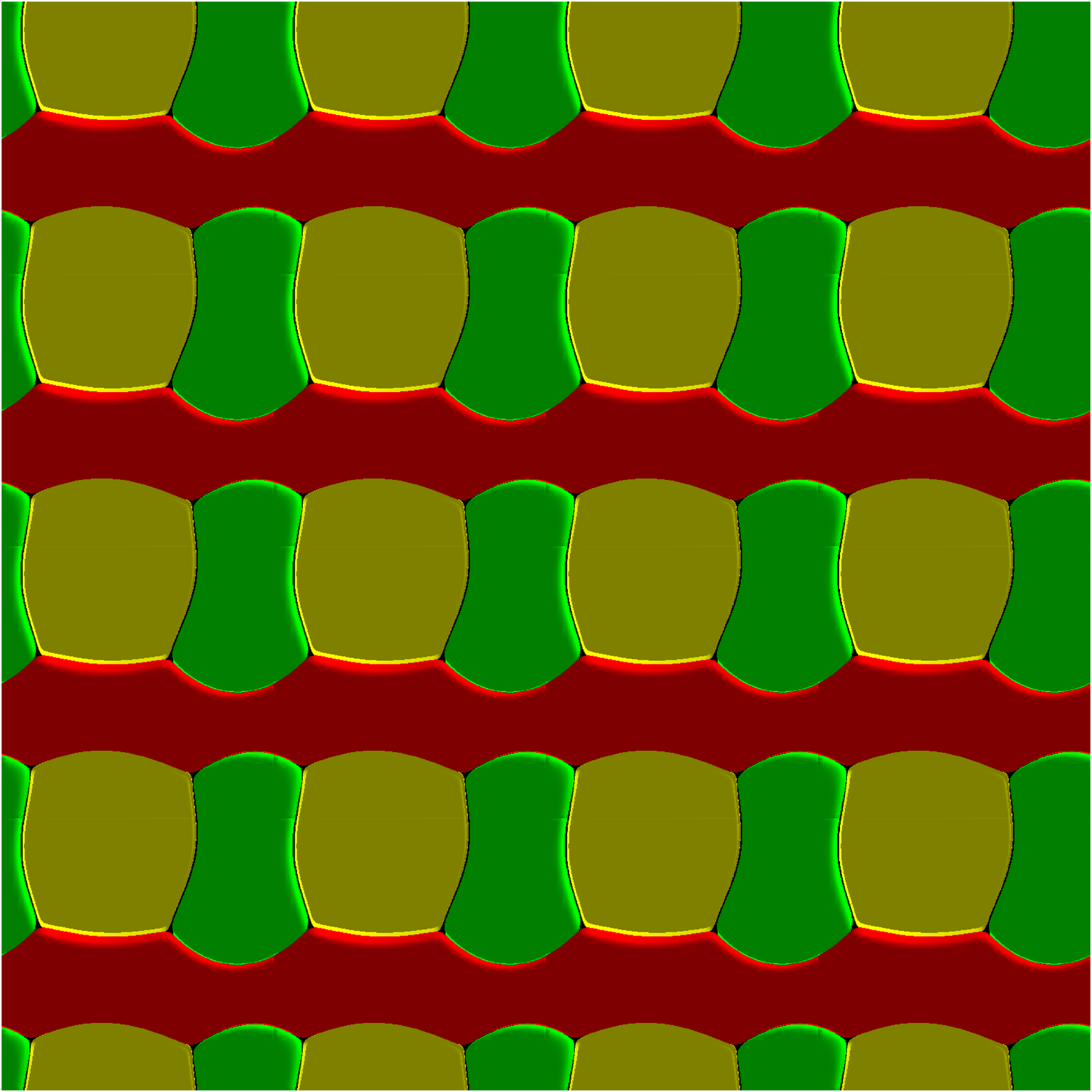}
   \label{Figure24c}
 }
\caption{(Color Online) Cross-sections of patterns obtained in three-dimensional directional solidification. In each picture, the simulation unit cell is tiled in a 4$\times$4 array to get a better view of the pattern. The pattern in (a) was started from a random configuration and evolved to a perfectly hexagonal pattern (at the eutectic composition for a symmetric phase diagram). At an off-eutectic concentration, starting with two isolated rods of $\alpha$ and $\beta$ phase, the result shown in (b) is one of the possible structures, while with an asymmetric phase diagram at the eutectic concentration, we get a regular brick structure (c) from a random initial condition.}
\label{Figure24}
\end{center}
\end{figure}

The cross sections of the simulated
systems are $150 \times 150$ grid points for results in 
Figs. \ref{Figure24}(a) and (c), and $90 \times 90$ grid points 
for the system Figure \ref{Figure24}(b). The longest run 
took about 7 weeks on 80 processors, for the simulation of the pattern 
in Figure \ref{Figure24}(a). This long simulation time is due to the 
fact that the pattern actually takes a long time to settle 
down to a steady state; the total solidification distance was of the 
order of 800 grid points. The other simulations required less time to 
reach reasonably steady states. The patterns shown in Figs. \ref{Figure24}
(a) and (c) start from random initial conditions 
(very thin rods of randomly assigned phases), the former with the 
symmetric phase diagram used previously, the latter with a slightly 
asymmetric phase diagram constructed with the changed parameters listed
below,
\begin{minipage}{16cm}
 \parbox{9cm}{
\begin{displaymath}
L_{i}^{\alpha} = \left(\begin{array}{llll}
       & A         & B          &  C       \\
\alpha & 2.0       & 1.0        &  1.0     \\ 
\beta  & 1.0       & 2.0        &  1.0     \\
\gamma & 1.0       & 1.0        &  2.0
\end{array}\right)
\end{displaymath}
}
\parbox{5cm}{
\begin{displaymath}
T_{i}^{\alpha} = \left(\begin{array}{llll}
       & A       & B       &  C       \\
\alpha & 1.0     & 0.59534 &  0.63461 \\ 
\beta  & 0.59534 & 1.0     &  0.63461 \\
\gamma & 0.59534 & 0.59534 &  1.0
\end{array}\right).
\end{displaymath}
}
\end{minipage}
 
The picture of Figure \ref{Figure24}(b) corresponds to a pattern resulting of a simulation
which is started with two isolated rods of $\alpha$ and $\beta$ in a matrix of $\gamma$,
with an off-eutectic concentration of $\vc=(0.3,0.4,0.3)$.

As shown in Figure \ref{Figure24}, many different steady-state patterns 
are possible in 3D. Not surprisingly, the type of pattern seen in the 
simulations depends on the concentration and on the phase diagram. Patterns 
very similar to Figure \ref{Figure24}(c) have recently been observed in 
experiments in the Al-Ag-Cu ternary system \cite{Ratke}. It should be stressed 
that our pictures have been created by repeating the simulation cell four 
times in each direction in order to get a clearer view of the pattern. 
This means that in a larger system, the patterns might be less regular.
Furthermore, we certainly have not exhausted all possible
patterns. A more thorough investigation of the 3D patterns
and their range of stability is left as a subject for future work.
\section{Conclusion and outlook}
In this paper, we have generalized a Jackson-Hunt analysis for arbitrary 
periodic lamellar three-phase arrays in thin samples, and used 2D phase-field
simulations to test our predictions for the minimum undercooling spacings 
of the various arrangements. For the model used here the value of 
the interface kinetic coefficient cannot be determined, which leads to some
incertitude on the values of the undercooling, but this does not influence our 
principal findings. When the correct values of the surface free energy
(that take into account additional contributions coming from the chemical
part of the free energy density) are used for the comparisons with the
theory, we find good agreement for the minimum undercooling spacings for
all cycles investigated. Moreover, we find that, as in binary eutectics, all 
cycles exhibit oscillatory instabilities for spacings larger than some critical
spacing. The type of oscillatory modes that are possible are determined
by the set of symmetry elements of the underlying steady state. 

We have repeatedly made use of symmetry arguments for a classification
of the oscillatory modes. In certain cases, the symmetry is exact and
general, which implies that the corresponding modes should exist for
arbitrary phase diagrams and thus be observable in experiments. For 
instance, the mirror symmetry lines in the middle of the $\alpha$
lamellae in the $\alpha\beta\alpha\gamma$ arrangement exist even for
non-symmetric phase diagrams and unequal surface tensions, and hence 
the corresponding oscillatory patterns and their symmetries should be
universal. In other cases, we have used a symmetry element which is specific
to the phase diagram used in our simulations: a mirror reflection, 
followed by an exchange of two phases. For a real alloy, this symmetry 
obviously can never be exactly realized because of asymmetries in the 
surface tensions, mobilities, and liquidus slopes, and therefore some of 
the oscillatory modes found here might not be observable in experiments. 
However, their occurrence cannot be completely ruled out without a 
detailed survey, and we expect certain characteristics to be quite
robust. For instance, we have repeatedly observed that two neighboring 
lamellae of different phases can be interpreted as a ``composite lamella'' 
that exhibits a behavior close to the one of a single lamella in a binary 
eutectic pattern. Such behavior could appear even in the absence of
special symmetries, and thus be generic.

Furthermore, a new type of instability (absent in binary eutectics) was found, where a 
cycle transforms into a simpler one by eliminating one lamella. We interpret 
this instability, which occurs for small spacings, through a modified 
version of our theoretical analysis. It is linked to the existence of
an extra degree of freedom in the pattern if a given phase appears more 
than once in the cycle. We have not determined the full stability diagram 
that would be the equivalent of the one given in Ref.~\cite{Sarkissian} for
binary eutectics, because of the large number of independent parameters 
involved in the ternary problem.

We have made a few attempts to address the question of pattern selection,
with inconclusive results both in 2D and 3D. In 2D, the process of pattern
adjustment was hindered by the absence of a mechanism for lamella creation,
and in 3D the system sizes that could be attained were too small. Based on
the findings for binary eutectics, however, we believe that there is no
pattern selection in the strong sense: for given processing conditions,
the patterns to be found may well depend on the initial conditions and/or
on the history of the system. This implies that the arrangement with 
the minimum undercooling may not necessarily be the one that emerges 
spontaneously in large-scale simulations or in experiments.

The most interesting direction of research for the future is certainly a 
more complete survey of pattern formation in 3D and a comparison to 
experimental data.
To this end, either the numerical efficiency of our existing code has
to be improved, or a more efficient model that generalizes the model
of Ref.~\cite{Plapp} to ternary alloys has to be developed.


\section{Acknowledgements}

This work was financially supported by three sources: Centre National d'Etudes Spatiales (France), 
CCMSE (Center for Materials Science and Engineering) funded by the state
of Baden-W\"{u}ttemberg, Germany and the European Fond for Regional Development (EFRE), and
the DFG (German Research Foundation -project number NE822/14-1).

\bibliography{main_aps}
\end{document}